\documentclass[a4paper,12pt]{article}
\usepackage{graphicx}
\usepackage{amssymb}
\usepackage{epstopdf}
\usepackage{bm}
\usepackage[a4paper, total={6.5in, 9in}]{geometry}
\usepackage{hyperref}
\usepackage{xcolor}
\usepackage{dsfont}
\usepackage{eurosym}
\usepackage{hyperref}
\usepackage{tikz}
\usepackage{amsmath}
\usepackage{stmaryrd}
\usepackage{caption}
\usepackage{subcaption}

\usepackage{algorithm,algorithmic}
\usepackage[normalem]{ulem}

\begin{document}
\title{Modeling metaorder impact \\with a Non-Markovian Zero Intelligence model}

\author{$\mathrm{Adele \ Ravagnani}^\mathrm{1},  
	\  \mathrm{Fabrizio \ Lillo}^\mathrm{1,2}$ \\  
	$^\mathrm{1}$\small{\emph{Scuola Normale Superiore, Pisa, Italy}}\\
    $^\mathrm{2}$\small{\emph{Dipartimento di Matematica, Universit\`a di Bologna, Italy}}
}

\date{\today}

\maketitle

\abstract{Devising models of the limit order book that realistically reproduce the market response to exogenous trades is extremely challenging and fundamental in order to test trading strategies. We propose a novel explainable model for small tick assets, the Non-Markovian Zero Intelligence, which is a variant of the well-known Zero Intelligence model. The main modification is that the probability of limit orders' signs (buy/sell) is not constant but is a function of the exponentially weighted mid-price return, representing the past price dynamics, and can be interpreted as the reaction of traders with reservation prices to the price trend. With numerical simulations and analytical arguments, we show that the model predicts a concave price path during a metaorder execution and to a price reversion after the execution ends, as empirically observed. We analyze in-depth the mechanism at the root of the arising concavity, the components which constitute the price impact in our model, and the dependence of the results on the two main parameters, namely the time scale and the strength of the reaction of traders to the price trend.

}
\vspace{0.5cm} 

\section{Introduction}
In electronic markets, trade intentions of market participants are handled by limit order books (LOBs). They are collections of buy and sell orders, which are matched on a price-time priority basis, with a continuous-time double auction mechanism. The resulting dynamics are complex, extremely fast, they lead to a self-organized price formation process and summarize the interactions between a diversity of agents with different objectives and trading strategies. In order to reproduce these dynamics and their statistical properties (so-called \textit{stylized facts}), researchers have been engaged in this challenging modeling task since the early 2000s, resulting in a vast amount of papers \cite{konark_review_2024}. Obtaining a realistic simulator of the LOB dynamics is fundamental both for academic and practical interests. It allows to gain theoretical insight about financial markets and their functioning. On the other hand, it allows to precisely test trading strategies before they are used in real markets. These tests are usually performed by relying on the so-called \textit{backtesting}, which consists in simulating the performance of the trading strategies by relying on historical data, under several scenarios. However, even though the backtesting approach is commonly adopted, it is inaccurate: contrary to employing LOB simulators, it does not allow to consider the market response to each action of the strategy we aim to test. Motivated by these issues, we focus on devising a novel explainable LOB model that reproduces, as realistically as possible, the market response when exogenous orders are submitted. Explainability is fundamental to us since we aim to shed light on the mechanisms at the root of market impact dynamics.

\begin{figure}
    \centering
    \includegraphics[width=0.6\linewidth]{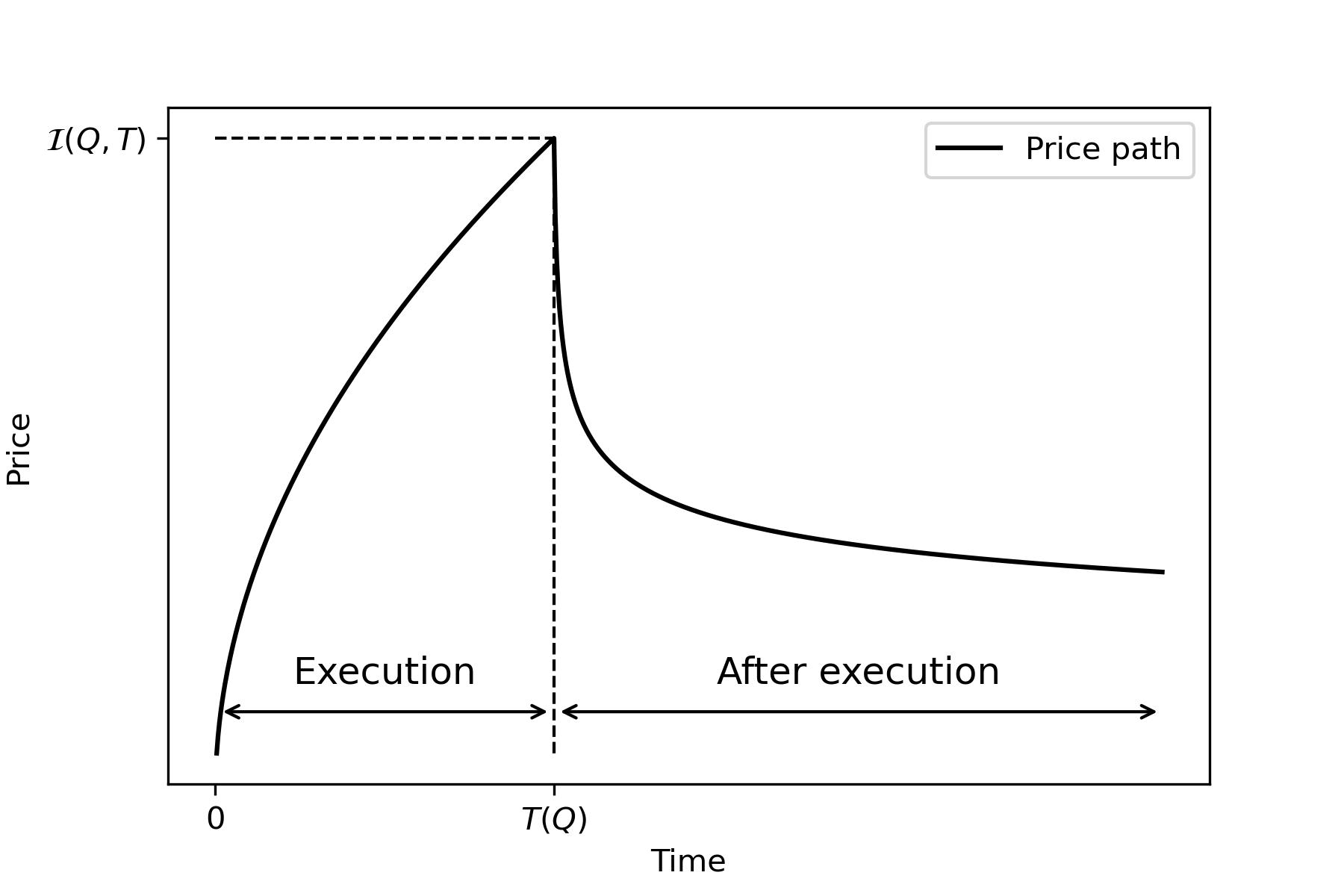}
    \caption{Expected average shape of the market impact during and after the execution of a buy metaorder. $Q$ is the metaorder size (in shares), $T$ is its duration.}
    \label{fig_meta_order_impact_th}
\end{figure}

In this paper we are mostly interested in the execution of large orders, the  so-called \textit{metaorders}. As it is known since the seminal Kyle paper \cite{kyle1985}, it is optimal to split the metaorder into small \textit{child market orders} that are executed progressively. This allows to limit the \textit{market impact}, which is the positive correlation between the sign of an incoming order and the subsequent price change, and which represents the main source of trading costs for medium and large size investors. In a LOB simulator, the desired behavior of a metaorder execution's responsiveness is its concavity, as it is empirically observed and studied in several papers \cite{bouchaud_book_2018}. Concerning the behavior of the price at the end of the metaorder execution, several works empirically and theoretically studied this issue. It emerged that, following the peak price reached at the end of the execution, a reversion occurs, with a decay and converging to a plateau \cite{moro_2009, brokman2015, zarinelli_2015, bucci2018}. A graphical representation of the expected average shape of the market impact during and after the execution of a buy metaorder is provided in Fig. \ref{fig_meta_order_impact_th}. The Non-Markovian Zero Intelligence LOB model we propose in this paper, aims to reproduce the empirically observed market impact shape. Our model is an extension of a well-established simulator, the so-called Zero Intelligence model \cite{daniels_2003, smith_2003, farmer_2005}, which leads to a linear market impact when executing a metaorder.

\subsection{Related literature}\label{sub_section_literature}
Since the first studies of the early 2000s, a vast amount of research articles \cite{konark_review_2024} has been devoted to the challenging task of modeling the dynamics of the Limit Order Book (LOB). Following \cite{konark_review_2024}, four broad categories of LOB simulators can be identified: Point Processes, Agent-Based, Deep Learning Based and Stochastic Differential Equations Based models. Among the Point Processes models, the Zero Intelligence simulators \cite{bouchaud_2002, smith_2003, cont_2010, abergel_2011} consider orders arrival as independent Poisson processes. The Zero Intelligence model \cite{daniels_2003, smith_2003, farmer_2005} (aka Santa Fe model) which is extended in this paper, belongs to this class of models. Despite the simplicity of the Zero Intelligence simulators, they allow to reproduce several stylized facts, such as the spread distribution. However, especially given the unrealistic assumption of orders' independence, when metaorders are executed by employing these simulators, the market impact results to be linear in time and in addition, other important properties as the inter-arrival times and cross-excitation behavior, cannot be replicated. In order to overcome this major limitation, Poisson models with variable order intensity, as \cite{hult_2010, huang_2013, lu_2018b}, were proposed. For instance, the Queue-Reactive model \cite{huang_2013} considers intensities which depend on the current state of the order book. An impressive achievement of this model, which makes it stand out among Point Process simulators, is its ability to reproduce the concavity of the market impact against both time and volume.
However, this model still lacks of endogenously excited order flow. This issue motivated the development of another set of simulators: the Hawkes Process based, as \cite{toke_2010, zheng_2014, bacry_2016, lu_2018a, mounjid_2019, lee_2022}. They are fully explainable and able to correctly reproduce relevant microstructural properties as the volatility clustering and the Epps effect. However, their calibration is not straightforward and the choice of the kernel can be crucial for the model's performance.

A different approach is employed by Agent Based simulators \cite{paddrik_2012, byrd_2019, belcak_2020}. The LOB dynamics result from the interaction of several categories of investors (e.g. informed, uninformed, high frequency, trend followers, mean reverters, noise and algorithmic traders) and each of them is modeled differently. Some of these models, as \cite{byrd_2019, belcak_2020}, return a concave market impact. On the other hand, two main drawbacks of this class of simulators can be outlined: the modeling of the different types of investors' behavior is mostly heuristic and the simulation of this plethora of market agents leads to high computational costs.

Given the high complexity and nonlinearity of the LOB's dynamics, in the last years, machine learning has been increasingly employed in order to devise LOB simulators. The architectures proposed in the literature are mostly based on Long Short-Term Memory
Networks, Recurrent Neural Networks, and Generative Adversarial Networks, e.g. \cite{li_2020, shi_2021, shi_2022, coletta_2022, cont_2023a, nagy_2023, kumar_2024}. These models produce satisfying performances in the reproduction of the stylized facts. Nonetheless, if we consider the refined and recent contribution \cite{cont_2023a}, the extended study about a metaorder execution which is performed by the authors, highlights the inability of the model to capture the well-known concavity. Moreover, the models based on machine learning lack of explainability, they sensitively depend on several hyperparameters that need to be calibrated carefully and on an enormous amount of parameters.

Parsimony and explainability are instead the features of the Stochastic Differential Equations Based models as \cite{lakner_2016, huang_2017, hambly_2020, cont_degond_2023}. They model the LOB evolution as a set of differential equations and allow to focus on the steady state of the dynamics. Despite their mathematical tractability, an explicit solution cannot be derived by most of these models. Moreover, they highly depend on the prior assumptions they use and the resulting LOB simulators require high computational costs.

The majority of the models we revised above, manage to reproduce the dynamics of \textit{large-tick} assets. These stocks have a spread which is most of the times equal to the tick size and dense LOBs, i.e. all levels of the LOB are occupied. This is in contrast with \textit{small-tick} assets, which are such that the mean spread is much larger than the tick size and so, their LOB is sparse. Consequently, small-tick assets dynamics are more complex and challenging to reproduce. 

\subsection{Contributions of the paper and outline}
The aim of this article is to propose an extension of the Zero Intelligence (ZI) model \cite{daniels_2003}, that is able to reproduce the concavity of the market impact during the execution of a metaorder and the subsequent price reversion after its end. 
As we have seen in Subsection \ref{sub_section_literature}, in the literature, models of the LOB which are more sophisticated than the ZI, have been proposed. Nonetheless, we focus on this Point Process model since, even if the analytical treatment of the full model is extremely difficult, several relations can still be derived by adopting some assumptions. Moreover, this model can simulate the dynamics of both small-tick and large-tick assets. Our focus is on the former class, since their modeling is little explored in the literature. Additionally, the computational costs of the ZI model are contained and, most importantly, its explainability allows to provide insights into the mechanism underlying market impact, which is a major objective of this work.

Our model places itself as the small-tick assets counterpart of the Queue-Reactive (QR) model \cite{huang_2013}, which was devised for large-tick assets. The QR model well reproduces the concavity of the market impact, it is Markovian and the probability of the orders' arrival at a given queue depends on the queue's size. If, as in our case, the focus is on small-tick assets, considering orders' arrivals which depend on the queues' sizes is unreasonable: the LOBs for this class of assets usually have queues with small sizes, often equal to the minimum size of a single order. Therefore, in order to devise an LOB model that can reproduce the concavity of the market impact, we propose a non-Markovian variant of the ZI model. The idea is that given a limit order, its sign is sampled with a probability that is a function of an indicator of the past price trend. Financially, this choice links our model to the Locally Linear Order Book model \cite{donier_2015} since it can be interpreted by considering that in real markets, long term investors have \textit{reservation prices} and their decisions to submit or not submit limit orders are driven by these values and the current price trend. In our model, the reaction of these traders occurs with a time scale that corresponds to the memory we employ in the evaluation of the price dynamics, and its strength is tuned by a parameter that enters the definition of the probability that a limit order is a sell/buy. Throughout the paper, we study how our results and market impact dynamics depend on these two parameters.

The paper is organized as follows. Section \ref{sec_LOB_overview} provides a brief overview of limit order books. In Section \ref{section_sf}, the Zero Intelligence model is revised. Section \ref{section_modifiedZI} presents our non-Markovian version of the model with theoretical insights and numerical results. Finally, conclusions are drawn in Section \ref{section_conclusion}. In the Appendices, details about the data employed for parameters' estimation and the approach adopted are provided; also, some figures, which are explained in the main text, are reported.

\section{Limit Order Books}\label{sec_LOB_overview}
\begin{figure}
    \centering
    \includegraphics[width=0.5\linewidth]{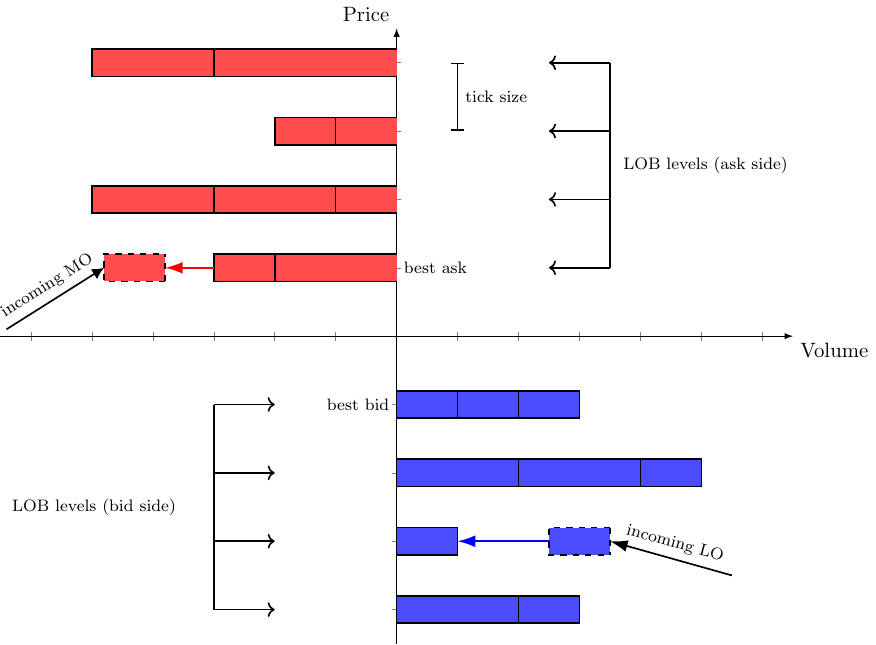}
    \caption{Graphical illustration of the Limit Order Book.}
    \label{fig_lob}
\end{figure}

In LOBs several types of events can occur. A \textit{buy limit order} of $Y$ shares at price $y$ (multiple of a given monetary value, called the \textit{tick size}, that is set by the regulators for each exchange) represents the intention of an agent to buy $Y$ shares at a price that is lower than or equal to $y$. On the other hand, a \textit{sell limit order} of $Z$ shares at price $z$ represents the intention of an agent to sell $Z$ shares at a price that is greater than or equal to $z$. Buy and sell limit orders (LOs) constitute the bid and ask side of the LOB respectively and they are separated by the \textit{bid-ask spread}, that is the distance between the highest and lowest prices of the bid and ask sides (\textit{best bid} and \textit{best ask}). Instead, the mean of these two prices is the so-called \textit{mid-price}, which is usually employed as a proxy of the asset price. Agents can also cancel their limit orders and submit \textit{market orders} (MOs), which are instantaneously executed at the best price. Therefore, a trade occurs when there are a buyer and a seller that agree on a price. We note that in the majority of LOBs, the order book matching is done on a price/time priority basis (\textit{First In First Out}): for a given price level, the first order to be executed is the first in the queue i.e. the first which has been posted. A graphical representation of the LOB for a given stock at a given time is provided in Fig. \ref{fig_lob}.

\section{The Zero Intelligence model and its limitations}\label{section_sf}
The ZI model \cite{daniels_2003, smith_2003, farmer_2005} was designed with the aim of being as analytically tractable as possible. It allows to simulate the LOB by relying on an extremely simple description: LOs, cancellations and MOs are considered as three independent Poisson processes, with rates that are independent of prices and sizes. Each order has unitary size and bid or ask sides are equally probable, as well as the price levels. Therefore, only four parameters (the LO rate $\lambda$, the MO rate $\mu$, the cancellation rate $\delta$ and the unitary volume $q_0$) need to be estimated from empirical data and this can be performed straightforwardly via elementary computations. We adopt the approach illustrated in \cite{bouchaud_book_2018}, that we present in Appendix \ref{sec_parameters_estimation}.

Despite its simplicity, the Zero Intelligence (ZI) model reproduces fairly well the mean and the distribution of the bid-ask spread (it slightly underestimates the mean spread), and the increasing
mean volume profile near to the best quotes. On the other hand, it does not capture the decrease in mean volume in limit order book's deeper levels; it overestimates volatility for small-tick assets and underestimates it for large-tick assets. Most importantly, this LOB simulator leads to a flat response function (as we see in the next subsection) and profitable market-making strategies. Contrary to real markets where the market orders' signs are strongly and positively autocorrelated, the order flow in the ZI model is uncorrelated. This makes the impact of a market order to be instantaneous and permanent.

\subsection{A linear model of market impact}\label{subsec_ZI_impact}
One important limitation of the ZI model is that it provides a linear model of market impact, both as a function of time and of the metaorder size. Indeed, the response function, or lag-$\tau$ impact, is constant. This function is defined as 
$$\mathcal{R}(\tau) = \langle(m_{t + \tau} - m_{t})\epsilon_t \rangle_{t, MO},$$ where $m_t$ is the mid-price before the market order at time $t$, $\epsilon_t$ is the sign of the market order at time $t$ and $\langle\cdot \rangle_{t, MO}$ represents the mean over all market orders. In real markets, the response function is concave, positive, increasing and saturates for large $\tau$ while in the ZI model, as reported in \cite{bouchaud_book_2018}, it is well approximated by:
\begin{equation}\label{eq_response_th}
    \mathcal{R}(\tau) \simeq\frac{1}{2}\mathbb{P}(q_{best} = 1)\langle x_1\rangle \equiv k 
\end{equation}
where $q_{best}$ is the volume (in units of $q_0$) at the best quote, $x_1$ is the first gap size i.e. the difference between the best bid/ask and the second price level and $\langle \cdot \rangle$ denotes its average. For small-tick assets the probability that $q_{best}$ is unitary is close to $1$ and so, the response function is about half the mean first gap size. 

Consequently, since in the ZI model the impact of a single trade is time-independent and equal to $k$, if we consider a metaorder that is executed in $[0, T]$ with trading speed $v_t$ and total volume $Q$, the mid-price impact is
\begin{equation}\label{linear_impact}
    m_T - m_{0} = \int_0^T k v_t dt = k Q
\end{equation}
where $m_t$ is the mid-price at time $t$. 

We confirm these predictions by performing several simulations of the ZI calibrated on real data. We use the LOBSTER dataset\footnote{\href{https://lobsterdata.com/}{lobsterdata.com}}
for the small-tick asset Tesla on January 5, 2015, which has a tick-size equal to $0.01 \$$. After the pre-processing\footnote{For details about the LOBSTER data and the pre-processing steps we performed, see Appendix \ref{appendix_lobster}.}, we end up with $170,416$ events. The average mid-price and its standard deviation are: $209.62 \$$ and $1.44 \$$. The mean spread is $16.71$ ticks. Limit orders constitute the $49.90\%$ of the market events, cancellations the $50.09 \%$ and market orders the $0.01 \%$. The parameters of the ZI are estimated\footnote{Code related to this calibration step and the models employed in the following is available at: \href{https://github.com/adeleravagnani/non-markovian-zero-intelligence-lob-model}{github.com/adeleravagnani/non-markovian-zero-intelligence-lob-model}} as illustrated in Appendix \ref{sec_parameters_estimation} obtaining: $\lambda =0.0131$, $\mu =0.0441$, $\delta =0.1174$, $q_0=101$.

\begin{figure}
    \centering
    \includegraphics[scale = 0.5]{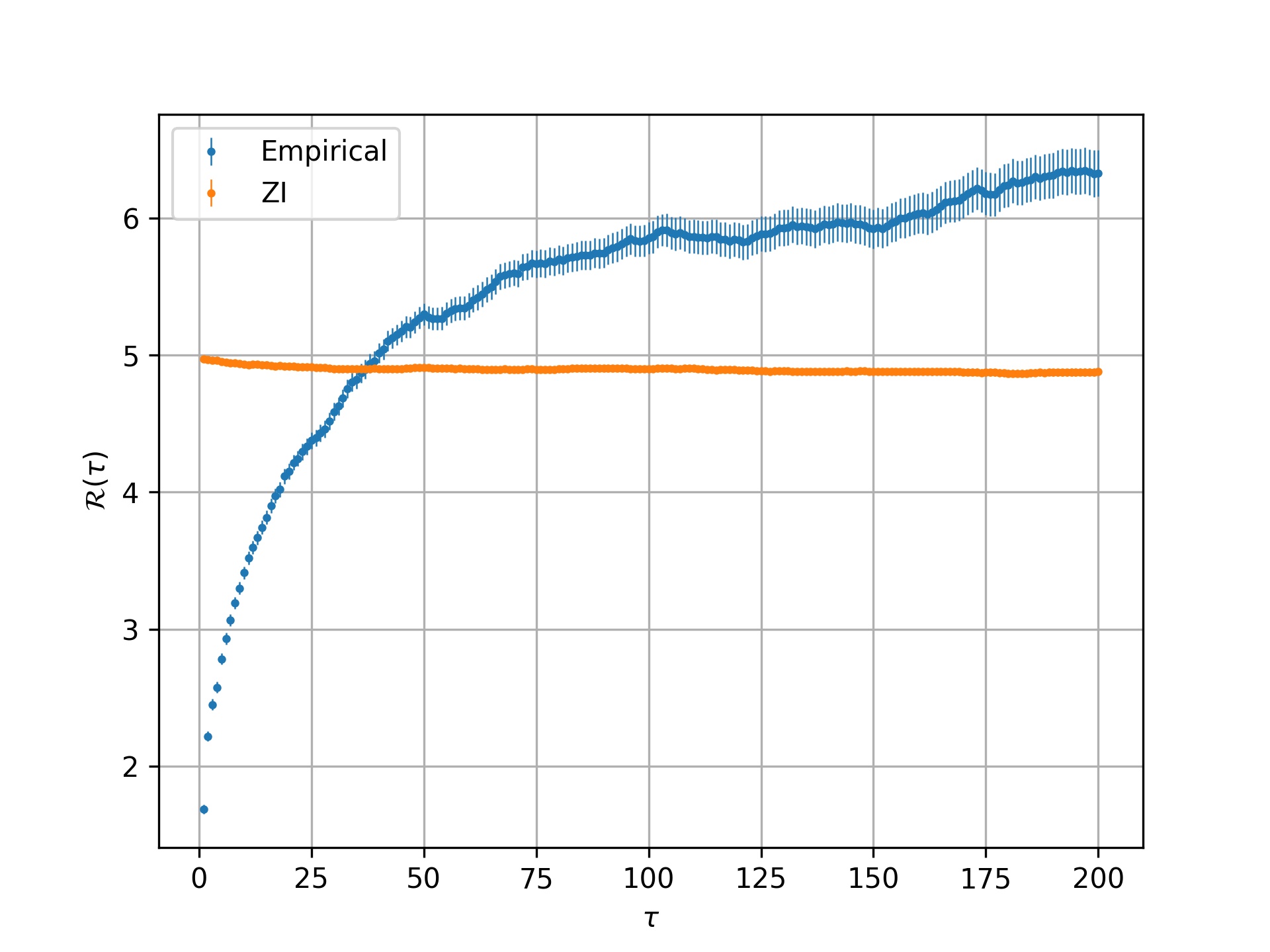}
    \caption{Comparison between the response function $R(\tau) = \langle (m_{t + \tau} - m_t)\epsilon_t \rangle_{t, MO}$ (in ticks) computed on empirical data and on data simulated with the ZI model.
    }
    \label{fig_ZI_response}
\end{figure}

After the estimation of the ZI model, we run several simulations (without any metaorder execution) to estimate the response function ${\mathcal R}(\tau)$ and the parameter $k$ characterizing the impact. Fig. \ref{fig_ZI_response} shows the response function obtained from the simulations of the ZI model together with its empirical counterpart. As expected, the two curves are very different since the ZI model predicts a constant response function, while real data shows a slow increase.

The estimated average value of the response function of the ZI model is equal to $ \hat{k} = (4.917 \pm 0.002)$ ticks. The theoretical value predicted by Eq. \eqref{eq_response_th} is $ \hat{k}^{th} = (5.107\pm 0.02$) ticks which is slightly larger than $\hat{k}$ but in overall agreement with it.

We now consider the price trajectory during a metaorder execution. We simulate a calibrated ZI model and superimpose the execution of a buy metaorder with constant speed and one unit, i.e. $q_0$, executed for each child MO.  This last choice is motivated by the fact that in the ZI model, each order has unitary size. In order to investigate how the results depend on the trading speed and the total volume, we carry out two experiments. (1) We set the total volume equal to $Q = 100$ (in units of $q_0$)  and we vary the number of events between two child orders that we denote as \textit{trading interval} $\Delta$ (in event time), which is the inverse of the trading speed $v$ ($\Delta = 1/v$); (2) we set the trading interval $\Delta = 50$ and vary the total volume executed $Q$. All the results are averaged over $200$ simulations.

\begin{figure}
    \includegraphics[scale = 0.5]{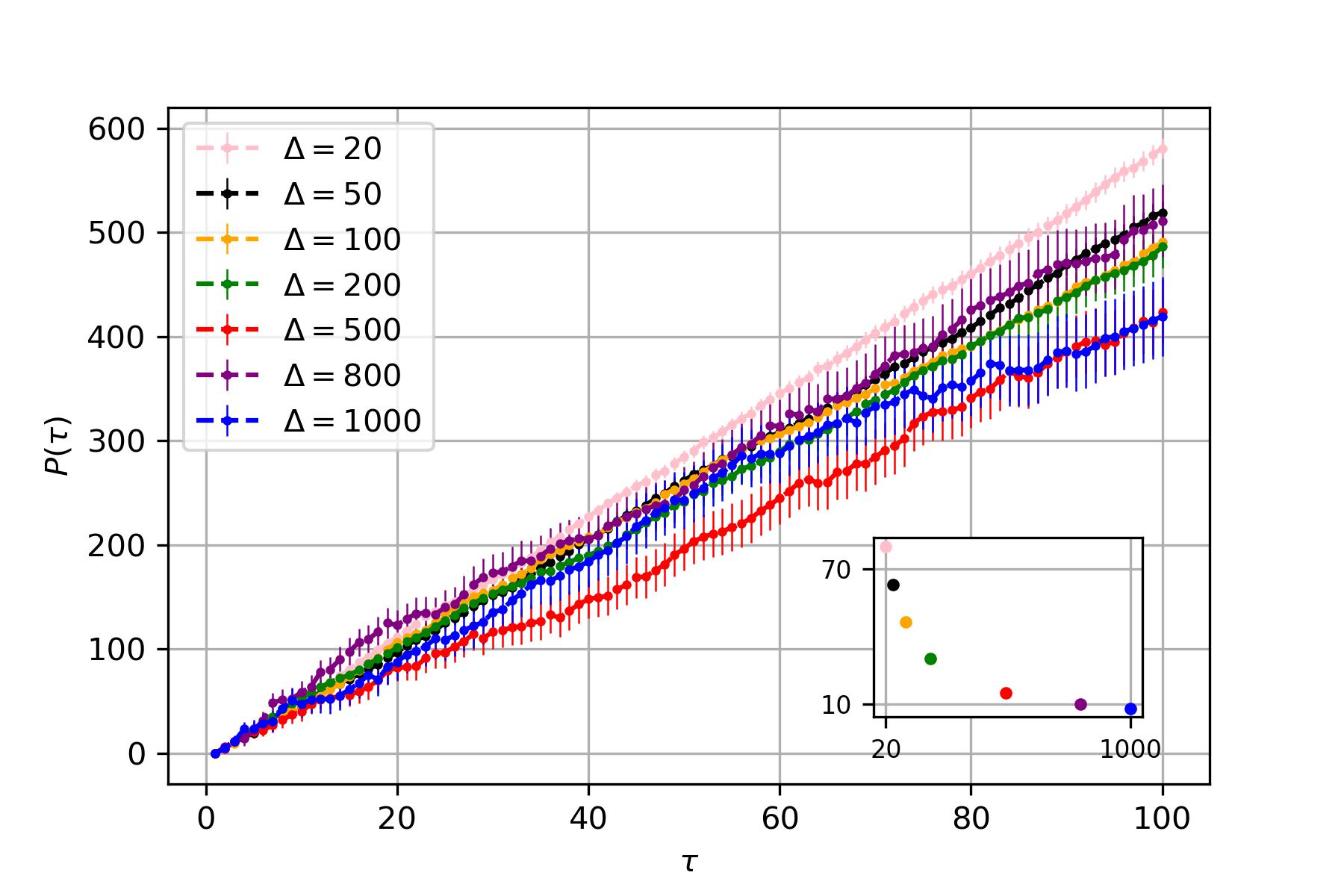}
    \includegraphics[scale = 0.5]{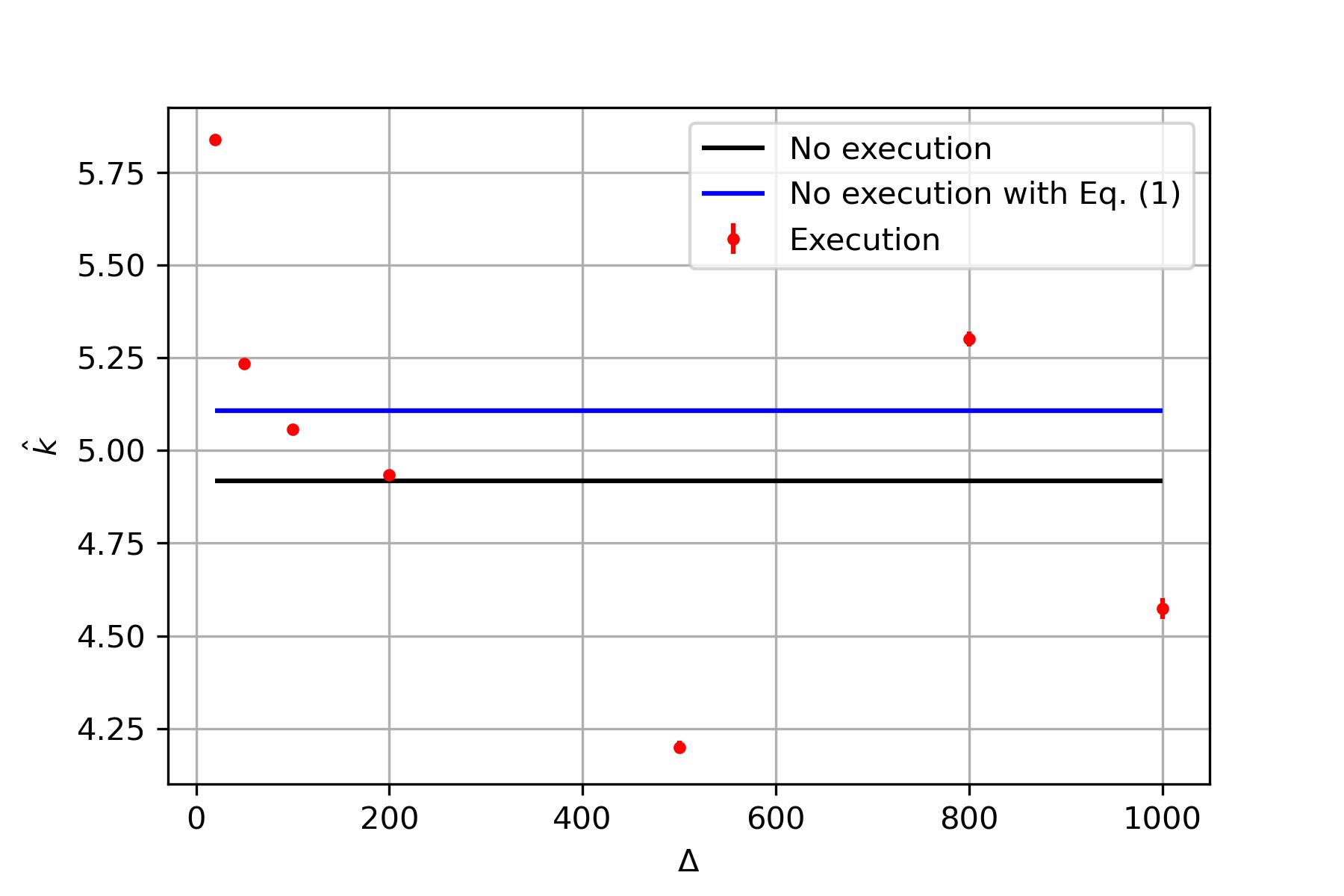}
    \caption{ZI model. On the left, we plot the average path of the mid-price increment versus the number of the child MOs of the metaorder. More precisely, the plotted quantity is: $P(\tau) = \langle (m_{(\tau-1)(\Delta + 1)} - m_0)\epsilon \rangle_{sim}$ with $\tau=1, \ldots, Q$. The metaorder has total volume $Q= 100$, buy direction i.e. $\epsilon = +1$ and different values of trading interval $\Delta$ (in event time). The inset plot represents the participation rate as a function of the trading interval. On the right, we plot the estimated slope of $P(\tau)$. The black line is the coefficient of the single trade response function obtained on simulated data without metaorder execution. The blue line is the coefficient obtained on simulated data without any metaorder execution and by employing Eq. \eqref{eq_response_th}.}
    \label{fig_response}
\end{figure}
\begin{figure}
\includegraphics[width=.5\linewidth]{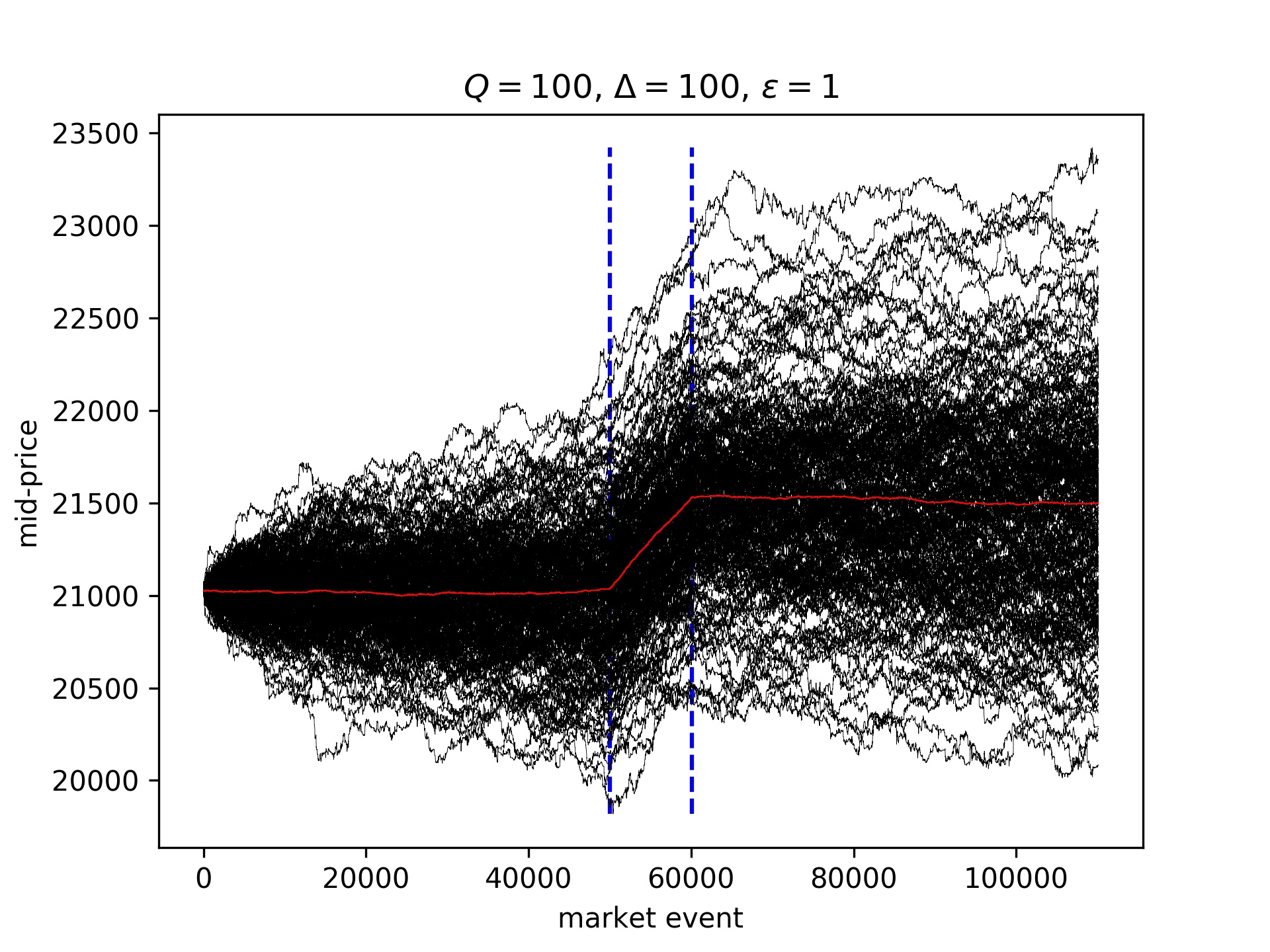}
\includegraphics[width=.5\linewidth]{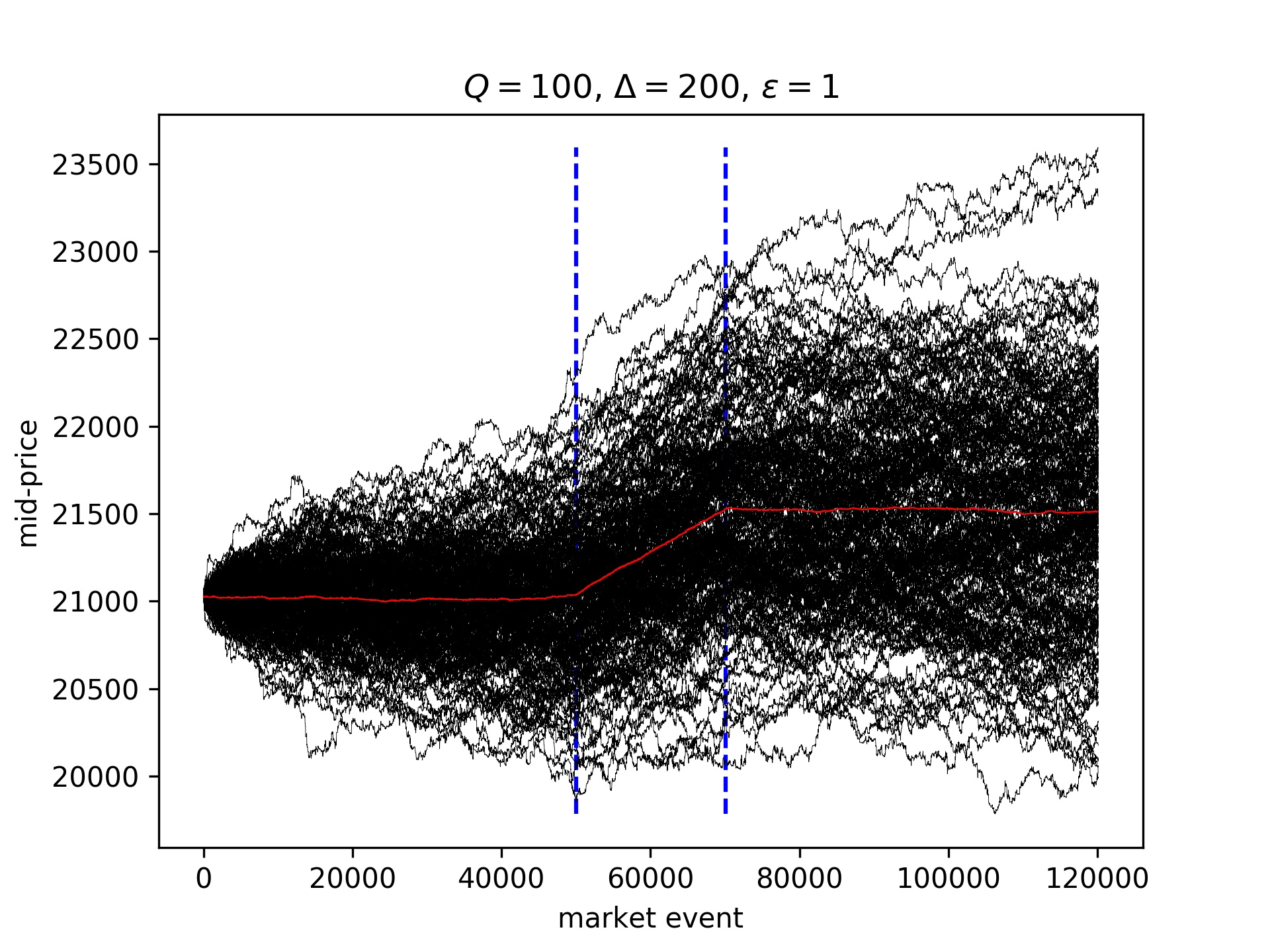}

    \caption{ZI model. Mid-price path for an execution with total volume $Q= 100$, buy direction and different values of trading interval $\Delta$ (in event time). Each black line corresponds to a simulation. The red line is the mean path. The vertical dashed blue lines are the beginning and the end of the metaorder execution.}
    \label{fig_mid_price_paths_trading_interval_1}
\end{figure}
\begin{figure}
    \centering
    \includegraphics[scale = 0.5]{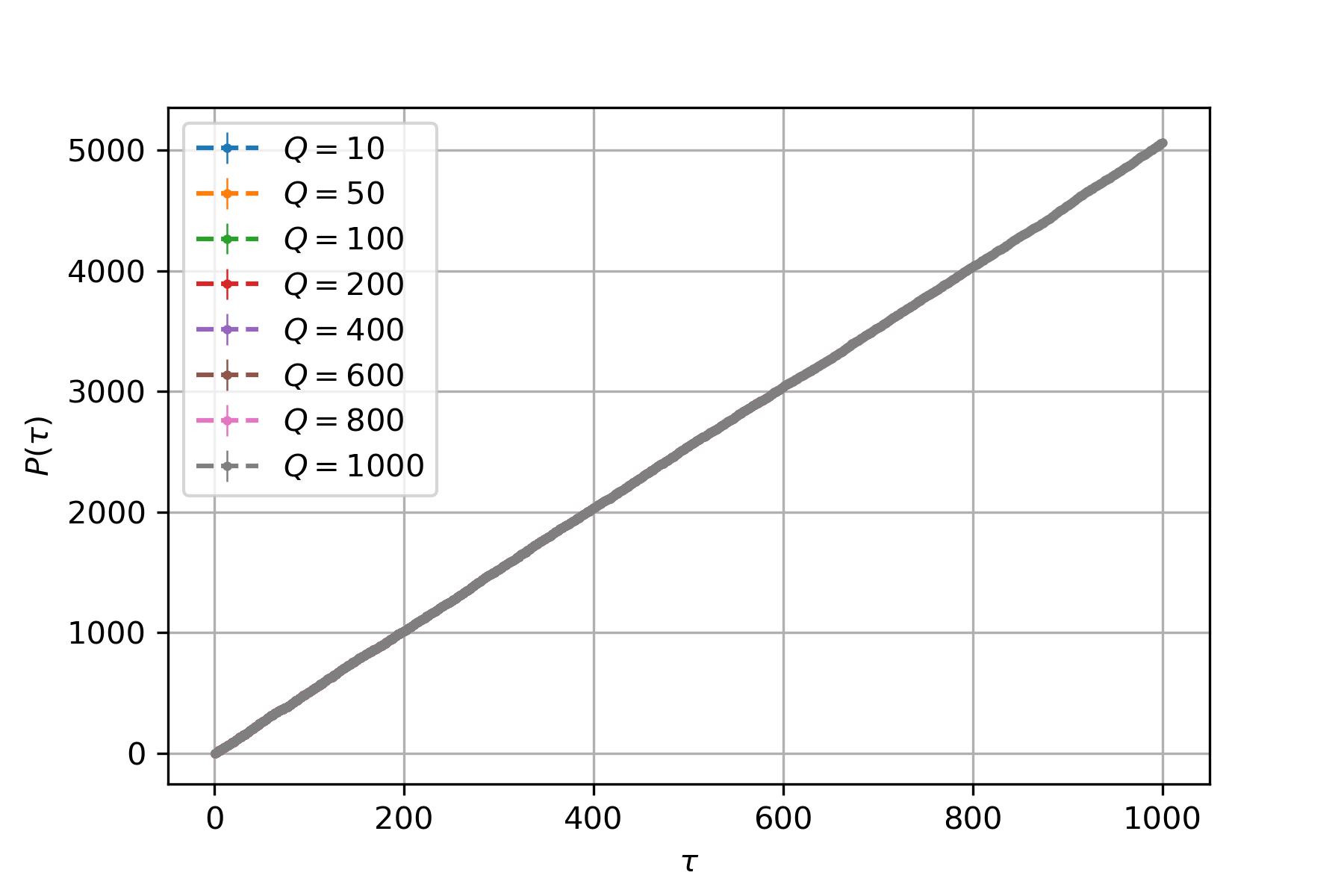}
    \includegraphics[scale = 0.5]{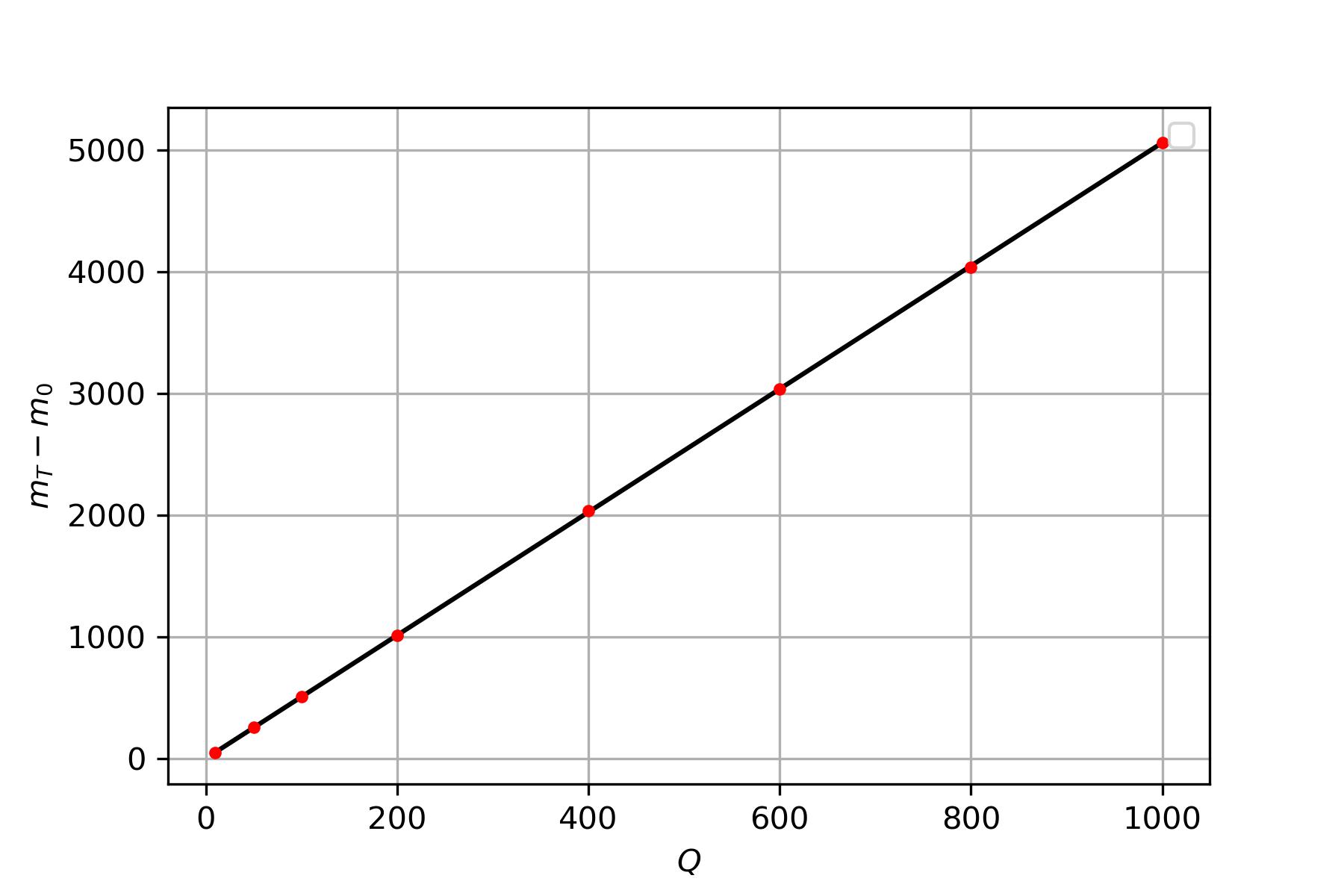}
    \caption{ZI model. Path of the mid-price difference versus the number of the child MOs of the metaorder. It is defined as: $P(\tau) = \langle (m_{(\tau-1)(\Delta + 1)} - m_0)\epsilon \rangle_{sim}$ with $\tau=1, \ldots, Q$. The metaorder has trading interval $\Delta = 50$ (in event time), buy direction i.e. $\epsilon = +1$ and different values of total volume $Q$ (left). Mid-price difference $m_T - m_{0}$ (in ticks) as a function of the total volume $Q$. The black line is the fitted one with slope equal to $(5.063 \pm 0.011)$ ticks (right).}
    \label{fig_response_shares}
\end{figure}

{\bf Experiment 1.}  
The left panel of Fig. \ref{fig_response} shows the estimated path of the mid-price increment $m_{(\tau - 1)(\Delta + 1)}-m_0$ as a function of $\tau = 1, \ldots, Q$ that is the number of child orders of the metaorder executed in $[0, T]$ with $T = (Q-1)(\Delta + 1)$. Intuitively, if we define the \textit{participation rate} as $100*Q/N_{MO}^{ex}$ where $N_{MO}^{ex}$ is the number of MOs executed by all market in the period of the metaorder execution, this variable increases as the trading interval decreases (see inset plot in the left panel of Fig. \ref{fig_response}). The right panel shows the fitted slopes as a function of the trading interval. This illustrates the equality between the permanent impact parameter and the constant response function of the ZI model $\hat{k}$, which is estimated by running the ZI model without any execution of a metaorder (black line in the right plot of Fig. \ref{fig_response}). Although the estimated coefficients are not perfectly constant, they fluctuate quite close to the theoretical value $\hat{k}$.

Fig. \ref{fig_mid_price_paths_trading_interval_1} shows the mid-price paths before, during, and after the executions. In this case the time on the abscissa is the market event time. By focusing on the average path (red line), we observe that during the execution there is a linear increase in the mid-price while before and after the execution, the mid-price remains approximately constant at the level reached by the last trade of the metaorder.

{\bf Experiment 2.} Results are reported in Fig. \ref{fig_response_shares}. As expected, the larger the total volume, the stronger the increase of the mid-price during the execution of the metaorder. By fitting the difference between the mid-price at the end and at the beginning of the execution, i.e. $m_T - m_{0}$, as a function of $Q$ and referring to Eq. \eqref{linear_impact}, we obtain a coefficient equal to $(5.063 \pm 0.011)$ ticks. This is consistent with $\hat{k}$ estimated without any execution and the coefficients estimated with a metaorder execution and different trading intervals.

In conclusion, the numerical investigation of the ZI model shows the main limitations in the behavior of the price dynamics during and after the execution of a metaorder. As seen, during execution, the price increases linearly with time, and after the end of the metaorder, the price remains constant. These properties are at odds with what observed empirically (displayed schematically in Fig. \ref{fig_meta_order_impact_th}). The next Section introduces a modified ZI model which provides more realistic price patterns.  

\section{The Non-Markovian Zero Intelligence model}\label{section_modifiedZI}
In the ZI model, each event is simulated without considering any information about past events and, as we previously recalled, this makes the order flow uncorrelated and the impact constant. Inspired by \cite{fosset}, we propose a non-Markovian model which is analogous to the ZI, except for one aspect. Contrary to the ZI model where the LOs' signs are picked randomly with equal probability, we set
\begin{equation}\label{eq_probability_sellLO}
    \mathbb{P}(\text{sell LO at time } t|\text{LO at time } t) = \frac{1}{1 + \exp(-\alpha \bar{R_t})}
\end{equation}
where $\bar{R_t}$ is the exponentially weighted mid-price return and $\alpha$ is an intensity parameter ($\alpha > 0$). The former represents the past price trend and it is defined as 
\begin{equation}\label{eq_def_EWMA_return}
    \bar{R_t} = \sum_{s= 1}^t e^{-\beta(t - s)}r_{s},
\end{equation}
where $\beta > 0$ is an inverse characteristic time, $r_s = m_{s} - m_{s-1}$, $ m_{s}$ is the mid-price before event $s$ and the initial time is $s=0$ with a given $m_0$\footnote{We note that Eq. \eqref{eq_def_EWMA_return} could be interpreted as an operative definition in our model. On the other hand, we can state a more formal definition that ensures stationarity: $\bar{R_t} = \sum_{s= -\infty}^t e^{-\beta(t - s)}r_{s}$.}. It is important to stress that the specific functional forms of the probability in Eq. \eqref{eq_probability_sellLO} and of the indicator of the past price trend in Eq. \eqref{eq_def_EWMA_return} could be generalized to alternative forms.

Our choice can be justified by considering that in real markets, long term investors have \textit{reservation prices}: they are not willing to buy (sell) at prices that are higher (lower) than a given value. These values are usually motivated by considerations that rely on fundamental analysis. Consequently, the decisions to submit or not submit LOs by these traders are driven by their observation of the price path. For instance, let us consider a trader who wants to sell $Z$ shares with a reservation price equal to  $12.5\$$, the best bid is $10\$$ and the best ask is $12\$$. Then, if the price shows an upward trend, the trader may decide to submit a LO at $12.5\$$ in order to gain time priority and ensure she sells as soon as the price reaches her target. In the framework of our model, in order to understand whether the price has an upward or downward trend, traders have a memory that is dictated by the parameter $\beta$. On the other hand, $\alpha$ corresponds to the intensity of trading reaction of these investors to the price trends. This interpretation connects our model to the Latent Liquidity Theory introduced in \cite{toth_2011} and in particular, to the Locally Linear Order Book (LLOB) model \cite{donier_2015}. This reaction-diffusion model is grounded on the idea that most of the liquidity in the book remains latent i.e. it is not displayed in the book, until the market price becomes close to the reservation price. In other words, this latent liquidity represents trading intentions of long term investors who have reservation prices (to buy or to sell) that are updated in time because of price changes, incoming news, etc. However, contrary to our model, the LLOB is Markovian and the adjustment of the reservation price is independent from the market dynamics, except for the mechanism that triggers the liquidity to become visible. 

An alternative interpretation of our model is related with \textit{contrarian} investment strategies \cite{chan1988, lakonishok1994}: investors buy past losers and sell past winners. This trading behavior is opposed to \textit{momentum} trading and it is widely explored in the literature. Several works, as \cite{odean1998, grinblatt2000, badrinath2002,dehaan2011, che2018,baltzer2019,grinblatt2020, bradrania2023}, provide empirical evidences that the contrarian behavior is commonly adopted, especially among individual investors. It is driven by the mean-reversion paradigm and partially rooted in the \textit{disposition effect} i.e. the investors' tendency to hold on to losers stocks and sell the winners.

We notice that if $\alpha = 0$, we recover the standard ZI model i.e. independently of the exponentially weighted mid-price return, sell and buy LOs are equally probable. 

Finally, the dynamics of the LOB, following the Non-Markovian Zero Intelligence (NMZI) model, can be simulated according to Algorithm \ref{alg_ZI}, displayed in Appendix \ref{sec_parameters_estimation}. We stress that the non-Markovianity of our model is due to the introduction of the time-varying probability that a LO is a sell: given a state of the LOB, each time we sample a LO, the evolution of the LOB and so, its next state depends on previous states and in particular, on past mid-prices. 

In the following, the Section is organized as follows. In Subsection \ref{subsec_intuition_model}, we provide an overview of the dynamics of the mid-price that we obtain and the intuition of the mechanism which is at the root of the market impact concavity in our simulator. During the execution, the mid-price path is first concave and then, it reaches a stationary regime where it converges to a straight line. After the execution, it exponentially decays. We analyze in depth each component of this behavior. In order to do so, in Subsection \ref{sec_equivalent_descriptions} we start to focus on the behavior of $\bar{R}_t$, we state two different ways to compute it and derive a condition that links these two descriptions. In Subsection \ref{subsec_stationary_regime}, we focus on the stationary regime of the mid-price path. We derive a relation between the stationary values of $\bar{R}_t$ and the slope of the mid-price trajectory. Then, in Subsection \ref{sec_master_equations} we derive two master equations describing the evolution of the mid-price and the spread in the time windows between two child MOs. In Subsection \ref{subsec_impact_components}, we analyze the interplay between the impact components that leads to the concavity in our model. Finally, in Subsection \ref{sec_after_execution}, we focus on the mid-price path after the execution of the metaorder ends and we show that the decay of the mid-price is exponential.

\subsection{A glimpse of market impact dynamics in our model}\label{subsec_intuition_model}

\begin{figure}
\includegraphics[width=.5\linewidth]{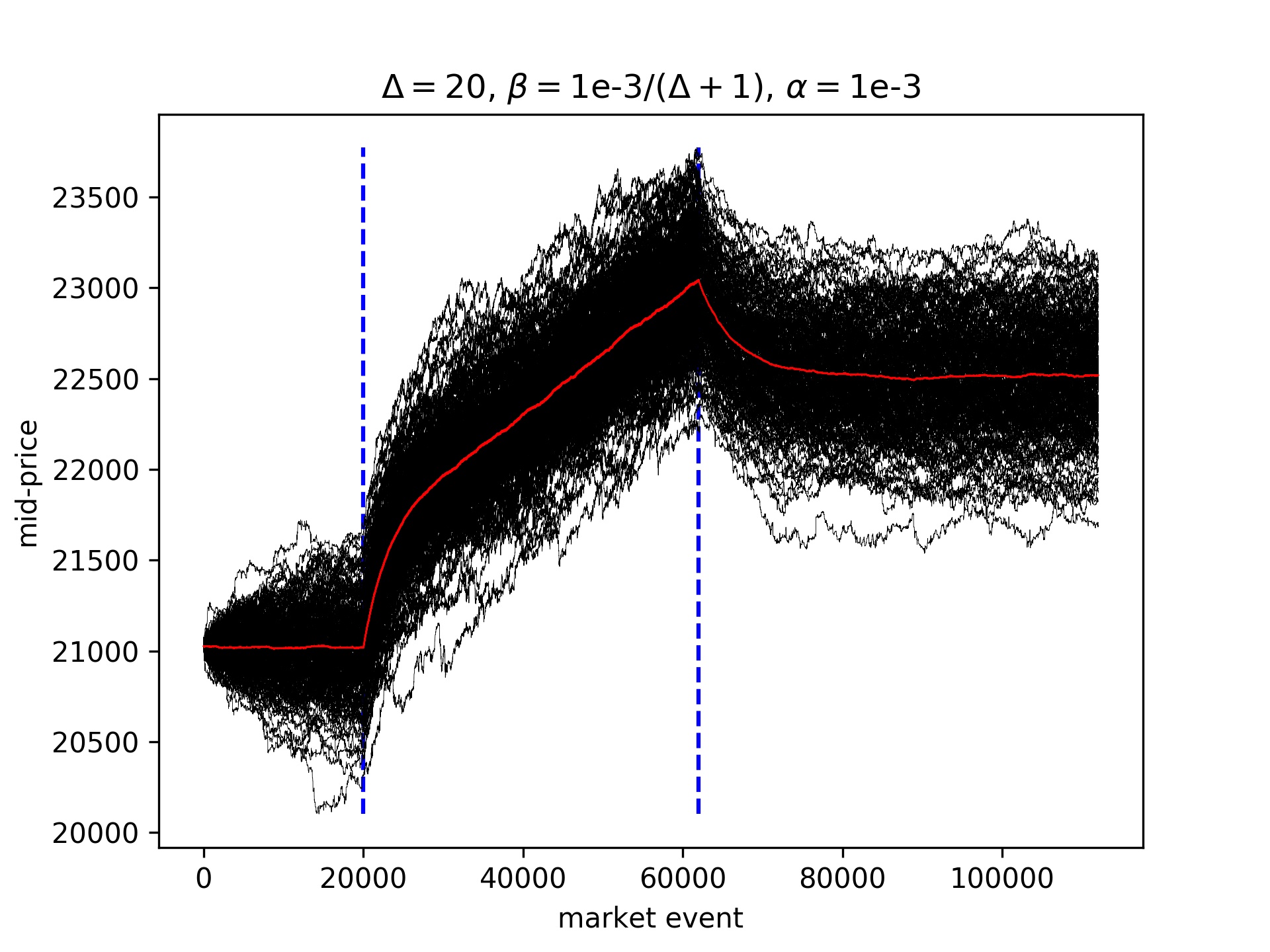}
\includegraphics[width=.5\linewidth]{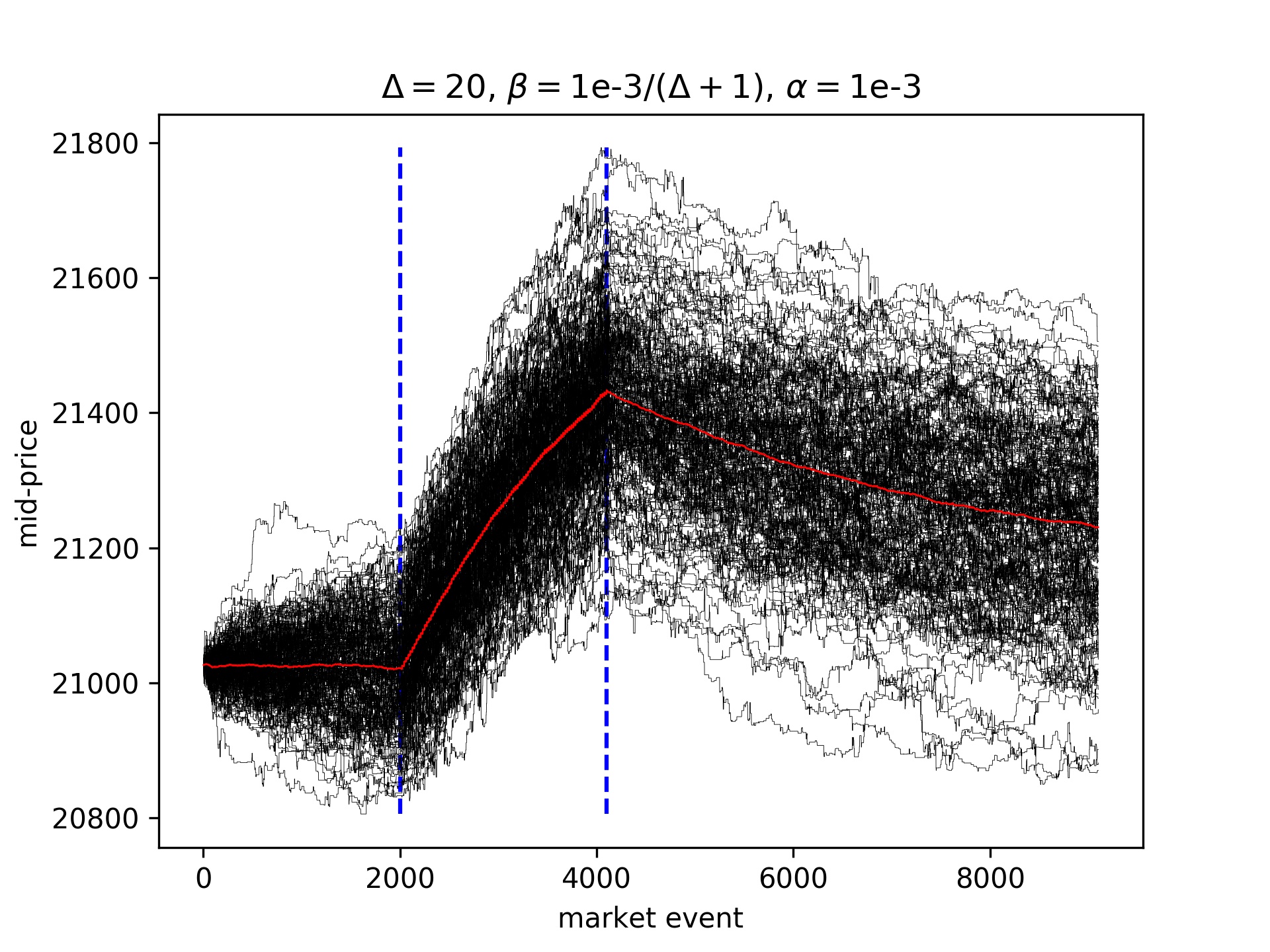}

    \caption{NMZI model. Mid-price path for an execution with total volume $Q= 2,000$ (left) and $Q= 100$ (right), buy direction and $\Delta = 20$. Each black line corresponds to a simulation. The red line is the mean path. The vertical dashed blue lines are the beginning and the end of the metaorder execution.}
    \label{fig_mid_price_paths_trading_interval_mZI}
\end{figure}

\begin{figure}
    \centering
    \includegraphics[width=0.5\linewidth]{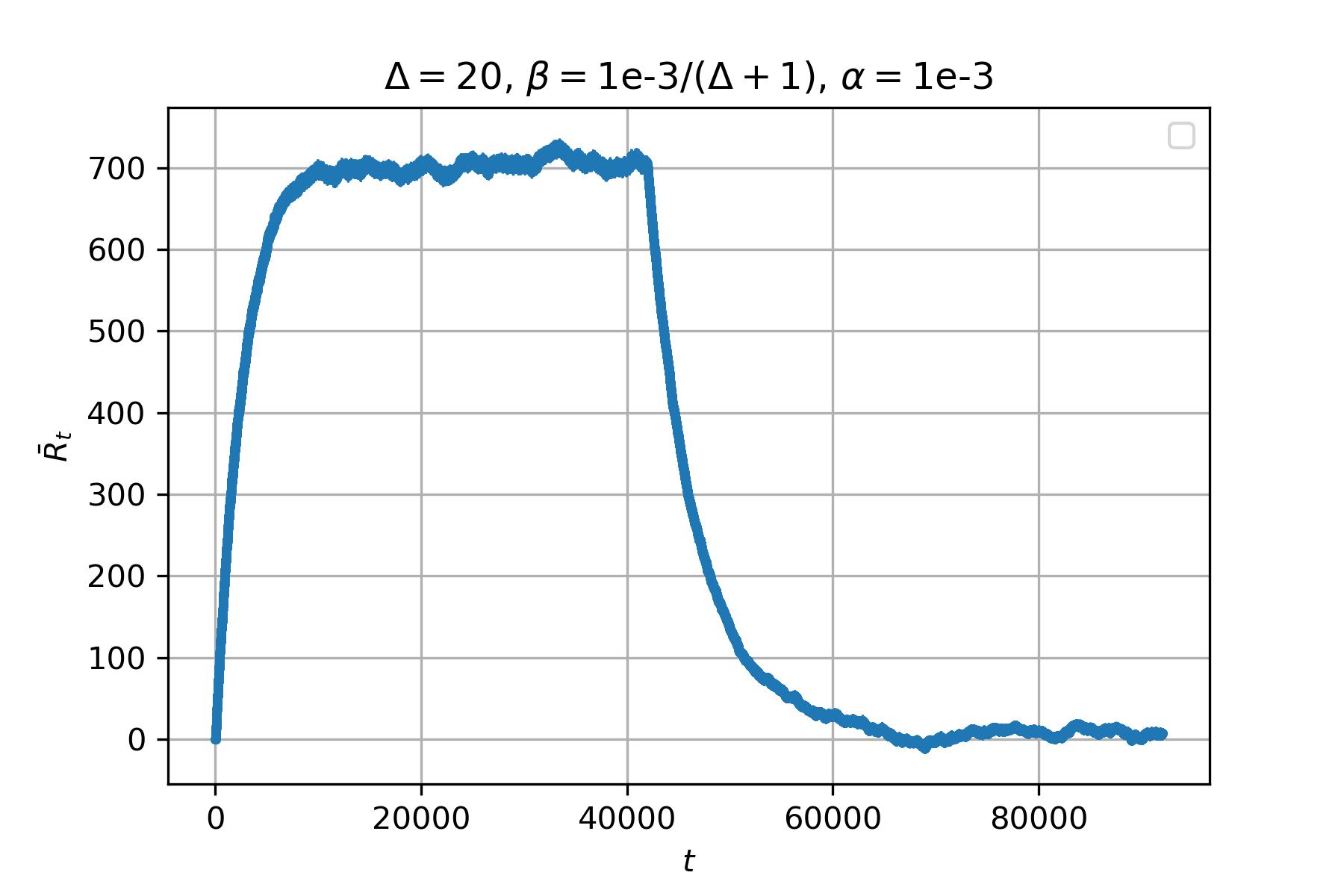}\includegraphics[width=0.5\linewidth]{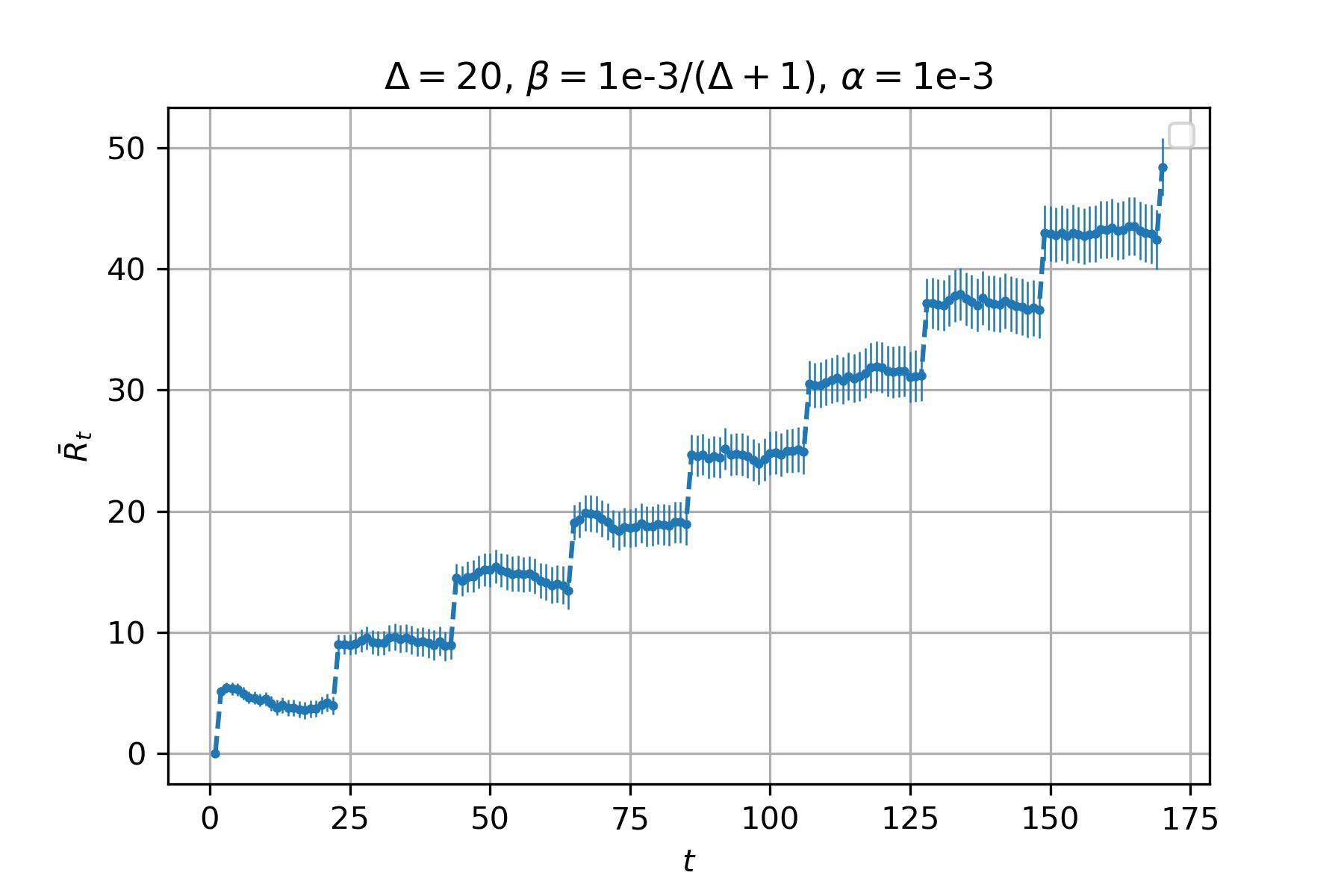}
    \caption{NMZI model. Evolution of $\bar{R}_t$ ($t$ denote the market events since the beginning of the metaorder execution) when a metaorder with total volume $Q= 2,000$, buy direction and $\Delta = 20$ is executed. In the left panel, all the evolution (during the execution of the metaorder and after its ending) is plotted while on the right, a zoom on the evolution until the $9th$ child MO is provided.}
    \label{fig_exp_weighted_return_zoom}
\end{figure}

In order to grasp the outcome of our model, let us start to consider the parameters $\beta = 10^{-3}/(\Delta + 1)$, $\alpha = 10^{-3}$ and $\Delta = 20$. We simulate the LOB dynamics\footnote{The parameters $\lambda, \mu, \delta, q_0$ of the NMZI model are the same of the ZI model we estimate in Subsection \ref{subsec_ZI_impact}. Similarly to what we do for the ZI model, in all simulations of the NMZI model we perform, we execute a buy metaorder with constant speed and $1$ unit executed for each child MO. All results are averaged over $200$ simulations. We note that without any execution of metaorders, the response function we obtain is flat and equal to ($4.867\pm 0.05$) ticks while, the corresponding value according to Eq. \eqref{eq_response_th} turns out to be equal to $(5.114\pm0.002)$ ticks.} with the execution of a buy metaorder of $Q=2,000$ (in units of $q_0$) and in the left panel of Fig. \ref{fig_mid_price_paths_trading_interval_mZI}, the mid-price paths before, during and after the execution of the metaorder are plotted. Before the execution starts, the mid-price is approximately constant. During the execution, two regimes can be distinguished: first, the price path is concave (denoted in the following \textit{concave regime} and it is more evident in the right panel of Fig. \ref{fig_mid_price_paths_trading_interval_mZI} where we represent the mid-price evolution when a metaorder of size $Q=100$ is executed) and then, it increases linearly in the so-called \textit{stationary regime}. Finally, after the execution ends, the reversion to a lower price value occurs. This behavior is mirrored by the exponentially weighted mid-price return whose dynamics during and after the execution of a metaorder with volume $Q=2,000$ is represented in the left panel of Fig. \ref{fig_exp_weighted_return_zoom}. By construction, $\bar{R}_t$ is a measure of the past price trend.

\begin{figure}
    \centering
    \includegraphics[width=0.7\linewidth]{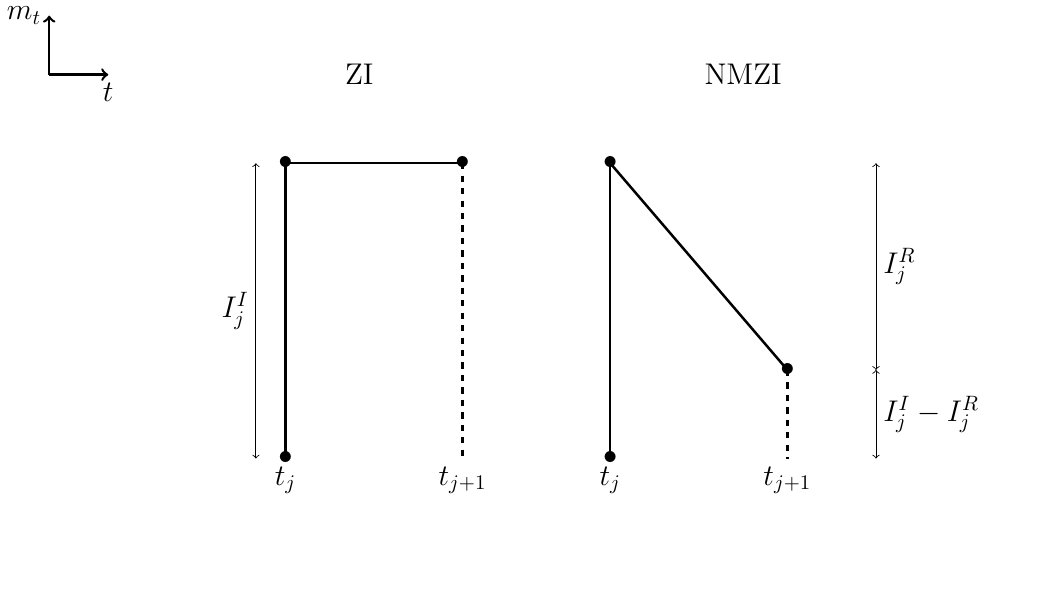}
    \caption{Stylized representation of the mid-price path in the intervals between two child MOs in the ZI and in the NMZI models. $t_j$ and $t_{j+1}$ are the times of the child MOs $j$ and $j+1$. $I_j^I$ and $I_j^R$ are the immediate and reversion components of the price impact of the child MO $j$; $I_j^I-I_j^R$ is its overall impact.}
    \label{stylized_fig_impact_components}
\end{figure}

We start to provide the intuition that explains this behavior in our model. In the next subsections, we will deepen it and analyze in detail each component of the mid-price trajectory. Let us consider a metaorder which is composed of $Q$ child MOs, each of them of unitary size, submitted every $\Delta$ steps and with the same sign e.g. buy. When the first child order of the metaorder is executed, its \textit{immediate price impact} is about equal to half the first gap size in the ask side. This is equivalent to Eq. \eqref{eq_response_th} with $\mathbb{P}(q_{best} = 1) \simeq 1$ that holds for small-tick assets (this is the class of assets under our interest). For the standard ZI model, in the time interval until the next child MO, the mid-price remains stable. This is not the case for our model which leads to a price reversion. In Fig. \ref{stylized_fig_impact_components}, we provide a stylized representation of the mid-price path between two child MOs in the ZI and in the NMZI models. In our model, the immediate impact component is such that $\bar{R_t}$ increases and so does the probability of sell LOs over buy LOs. This triggers more sell LOs and consequently, more sell LOs in the spread, which make the mid-price decrease on average: this is the \textit{reversion component of the impact} in the interval between two child MOs. As we see in the following and it is represented in the right panel of Fig. \ref{fig_exp_weighted_return_zoom}, $\bar{R_t}$ increases after each child MO and it is approximately constant within each interval between two child MOs. In this way, the entity of the reversion increases as more child MOs are executed and the imbalance between sell and buy LOs increases, making the first gap size in the ask side likely to decrease. This would cause a reduction of the immediate price impact component. However, in the following we see that the variation of the immediate impact component is marginal. To sum up, in the concave regime, we have two components in the price impact of every child MO: an immediate contribution which mildly varies in time and a reversion which increases in time. The difference between these two components is the so-called \textit{overall impact of the single child MO} and as we see in the following, it decreases as new child MOs are executed, making the concavity in the mid-price trajectory appear. 

At a certain point a transition to the stationary regime occurs since $\bar{R}_t$ becomes constant and so does the probability that a LO is a sell. Finally, the decay after the execution ends is a consequence of the reversion mechanism we have in the mid-price trajectory after a child MO.

\subsection{The behavior of $\bar{R}_t$ and two equivalent descriptions for its computation}\label{sec_equivalent_descriptions}
In the preceding subsection we qualitatively explained the mechanisms of our model. Now, we go deeper and, as a first step, we focus on the behavior of $\bar{R}_t$ in the concave regime: we aim to justify the stepwise characteristic shown in the right panel of Fig. \ref{fig_exp_weighted_return_zoom}. In order to do so, we introduce a different way to compute $\bar{R}_t$, namely by using only child MOs' times and in this setting, by construction, $\bar{R}_t$ increases after each child MO of a buy metaorder and it is constant within each interval between two child MOs. Therefore, by showing the conditions which lead to the equality between the two different ways to compute $\bar{R}_t$, it follows that even if we compute $\bar{R}_t$ by using all market events' times, it is approximately a step function as represented in the right panel of Fig. \ref{fig_exp_weighted_return_zoom}. This highlights that $\bar{R_t}$ reflects the price dynamics due to the metaorder execution and this is included in the sampling of the signs of the LOs. 

Let us denote as (i) the description for which $\bar{R}_t$ is computed by using all market events' times and as (ii) the description for which $\bar{R}_t$ is computed by using only child MOs' times. They are approximately equivalent for a suited rescaling and a condition on the exponent $\beta$, which enters the definition of $\bar{R}_t$ in Eq. \eqref{eq_def_EWMA_return}. If $t$ denotes the market events' times and the child MOs are executed at $t_j = \Delta j + j - 1$ with $j=1, \ldots, Q$, the starting time is $0$ and $t_Q = T$, we have the following relations:
\begin{equation}\label{eq_EWMAreturns_descriptions}
    \begin{split}
       \text{description (i): }\bar{R_t}^{(1)} &= \sum_{s= 1}^t e^{-\beta_1(t - s)}(m_{s} - m_{s-1})\\
       \text{description (ii): } \bar{R_t}^{(2)} &= \sum_{s= 1}^{t} \mathbb{I}\Bigg[\frac{s-\Delta - 1}{\Delta + 1} \in \mathbb{N}_0\Bigg] e^{-\beta_2(\frac{t^+ - s}{\Delta + 1})}(m_{s} - m_{s-\Delta - 1})\\
       &\text{where } t^+ = max_{s \leq t}\Bigg\{ s: \frac{s-\Delta - 1}{\Delta + 1} \in \mathbb{N}_0\Bigg\}.
    \end{split}
\end{equation}
When $t$ is the time just after a child MOs, the outputs of these relations are approximately equivalent if $\beta_2 = (\Delta + 1)\beta_1$ and $\beta_1\ll 1/\Delta$. We observe that $\beta_2$ represents the characteristic time of the description (ii) therefore, requiring $\beta_2 \ll (\Delta + 1)/\Delta$ amounts at including in the computation of $\bar{R}_t$ the price dynamics which cover a large number of past child MOs, exploiting the memory of the model. Let us consider an example with $\Delta = 2$. The first $5$ child MOs occur for market times $2, 5, 8, 11, 14$ (the starting time is $t=0$). If $t= 15$, the term in $\bar{R}_{15}^{(2)}$ related to the 4th child MO is $(m_{12}- m_9)e^{-\beta_2}$. In $\bar{R}_{15}^{(1)}$, the corresponding contribution is: $(m_{12}- m_{11})e^{-3\beta_1} + (m_{11}- m_{10})e^{-4\beta_1} + (m_{10}- m_9)e^{-5\beta_1}$ which can be rewritten as
\begin{equation*}
    (m_{12}- m_{9}e^{-2\beta_1})e^{-3\beta_1} - m_{11}e^{-3\beta_1}(1 - e^{-\beta_1}) -m_{10}e^{-4\beta_1}(1 - e^{-\beta_1}).
\end{equation*}
Focusing on the first term, comparing it with the corresponding one of $\bar{R}_{15}^{(2)}$ and setting $\beta_2 = (\Delta + 1)\beta_1$, their difference is
\begin{equation*}
    m_{9}e^{-(\Delta + 1)\beta_1}(1 - e^{-\Delta\beta_1}) 
\end{equation*}
and $(1 - e^{-\Delta\beta_1})\simeq 0$ if $\beta_1\ll 1/\Delta$. Similarly if $\beta_1\ll 1$, the intermediate terms $m_{10}e^{-3\beta_1}(1 - e^{-\beta_1})$ and $m_{9}e^{-4\beta_1}(1 - e^{-\beta_1})$ are about 0. On the other hand, if $t$ are times between child MOs (in the example above $t=0,1, 2, 4, 5, 7, 8, 10, 11, 13, 14, \ldots$), the exponentially weighted mid-price return in the second description remains constant within the trading interval: $\bar{R}_3^{(2)} = \bar{R}_4^{(2)} = \bar{R}_5^{(2)}$, $\bar{R}_6^{(2)} = \bar{R}_7^{(2)} = \bar{R}_8^{(2)}, \ldots$. This is in contrast with the first description where for instance, we have
\begin{equation*}
    \begin{split}
        \bar{R}_{17}^{(1)} 
        &= \bar{R}_{16}^{(1)}e^{-\beta_1} + (m_{17} - m_{16}) =\\ 
        &= \bar{R}_{15}^{(1)}e^{-2\beta_1} + e^{-\beta_1}(m_{16} - m_{15}) + (m_{17} - m_{16})\\
       \bar{R}_{17}^{(2)} &= \bar{R}_{16}^{(2)} = \bar{R}_{15}^{(2)} . 
    \end{split}
\end{equation*}
If $\beta_1 = \beta_2/(\Delta + 1)\ll 1/\Delta$ i.e. the memory to evaluate the price trend is much larger than the trading interval (necessary condition to fully exploit the non-Markovianity of our model), and the mid-price differences within the trading interval $\Delta$ are negligible with respect to the ones due to the child MOs, 
the reversion of $\bar{R}_t^{(1)}$ within the trading intervals is marginal and overall, the two descriptions are approximately equivalent. 

Additionally, we observe that for a buy metaorder, $\bar{R}_{t_{j+1}}^{(1/2)} > \bar{R}_{t_j}^{(1/2)}, \ j=1, \ldots Q - 1$ i.e. each child MO makes the mid-price increase since $\bar{R}_t$ is approximately constant between two child MOs and the immediate impact of a buy MO is positive and about half the first gap in the ask side (for small-tick assets).

We notice that the economically relevant description is (i): the price trend indicator $\bar{R}_t$ should be computed by relying on all market events' times since \textit{a priori} the child MOs' times are not known. 

\subsection{The stationary regime}\label{subsec_stationary_regime}
We showed the behavior of $\bar{R}_t$ in the concave regime however, at some point this quantity stabilizes to a stationary value, as represented in the left panel of Fig. \ref{fig_exp_weighted_return_zoom}. In the framework of description (ii), Eq. \eqref{eq_EWMAreturns_descriptions} can be rewritten in the following recursive way:
\begin{equation}\label{eq_recursive_EWMA_returns}
    \begin{split}
        \bar{R}_{t_j+1}^{(2)} &= r_{t_j + 1} + e^{-\beta_2} \bar{R}_{t_j-\Delta}^{(2)}  \\ 
    \end{split}
\end{equation} 
where $r_{t_{j+1}} = m_{t_{j+1}} - m_{t_j - \Delta}$ is the mid-price change due to child MO $j$. Given what we showed in the preceding subsection, the price trend indicator is non-decreasing during the execution of a buy metaorder and we can identify the \textit{stationary regime} as the set of times $\{t \in  [t^*, T]: \  \bar{R}_t \simeq \bar{R}_{t-1} \simeq \bar{R}^* > 0\}$. From Eq. \eqref{eq_recursive_EWMA_returns}, it follows that in this regime also $r_t\rightarrow r^*$ and the exponentially weighted mid-price return and the mid-price return are approximately proportional:
\begin{equation}\label{eq_proportionality}
    \bar{R}^* \simeq \frac{r^*}{1-e^{-\beta_2}}.
\end{equation} 

Let us note that the mid-price return can also be seen as the slope of the mid-price trajectory. Therefore, after a regime in which the mid-price trajectory is concave, it stabilizes to a straight line with slope $r^*$.

We point out that Eq. \eqref{eq_recursive_EWMA_returns} expresses an implicit link between $\bar{R}_t$ and the intensity parameter $\alpha$. As we show in Subsection \ref{sec_equivalent_descriptions}, descriptions (i) and (ii) to compute $\bar{R}_t$ are equivalent under suited conditions. If we employ description (ii), $\bar{R}_t$ is a step function: after each child MO, it is dampened by $\gamma$ and increases of $m_{t_j + 1} - m_{t_j - \Delta}$ where $t_j$ denote the child MOs' times. The quantity $m_{t_j + 1} - m_{t_j - \Delta}$ is the mid-price difference between two child MOs and it is linked to $\alpha$ through the time-varying probability that a LO is a sell. Indeed, as we see in Subsection \ref{subsec_impact_components}, $\alpha$ affects the magnitude of the impact between two child MOs.

\subsubsection{Numerical results}

\begin{figure}
    \includegraphics[width=0.5\linewidth]{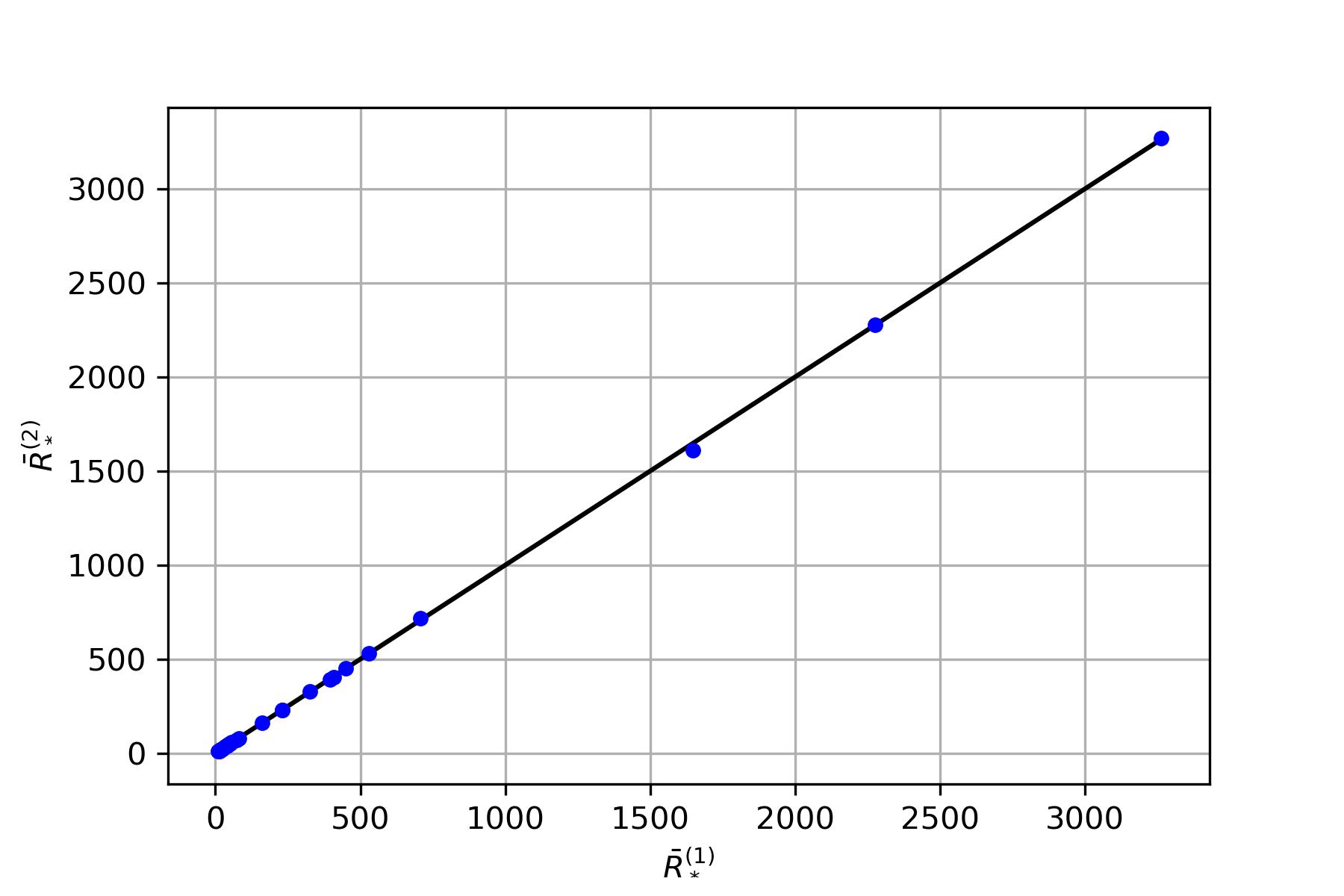}
    \includegraphics[width=0.5\linewidth]{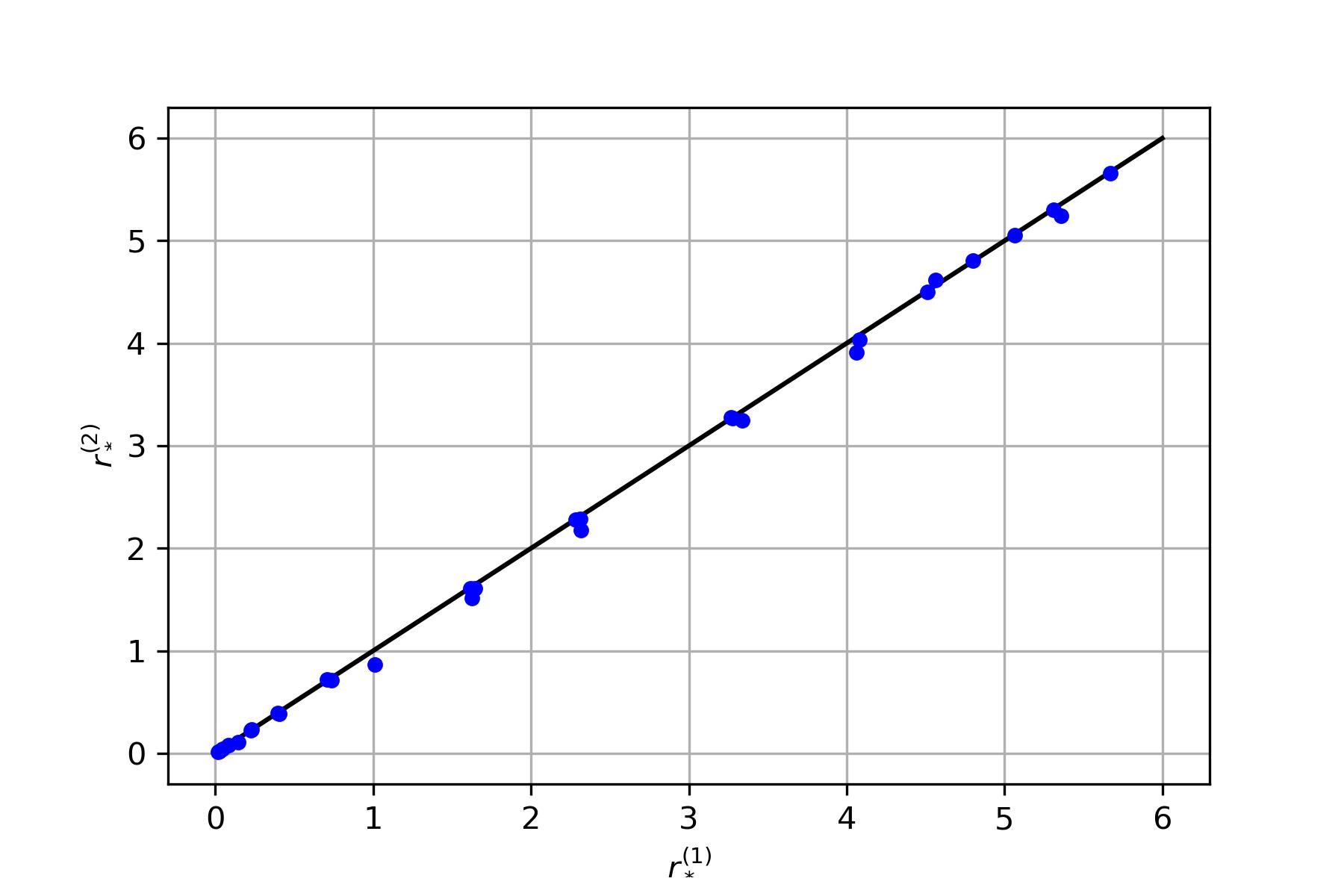}
    \caption{NMZI model. Comparison between the stationary values of $\bar{R}_t$ (left) and $r_t$ (right) obtained with the two different descriptions. The superscript (1) refers to the description for which the exponential weighted mid-price return $\bar{R}_t$ is computed by using all market events' times; the superscript (2) refers to the description for which $\bar{R}_t$ is computed by using only child MOs' times and $\beta_2 = \beta_1(\Delta + 1)$. The dark line is the bisector of the first quadrant.}
    \label{equivalent_descriptions}
\end{figure}

\begin{figure}
    \centering
    \includegraphics[width=0.5\linewidth]{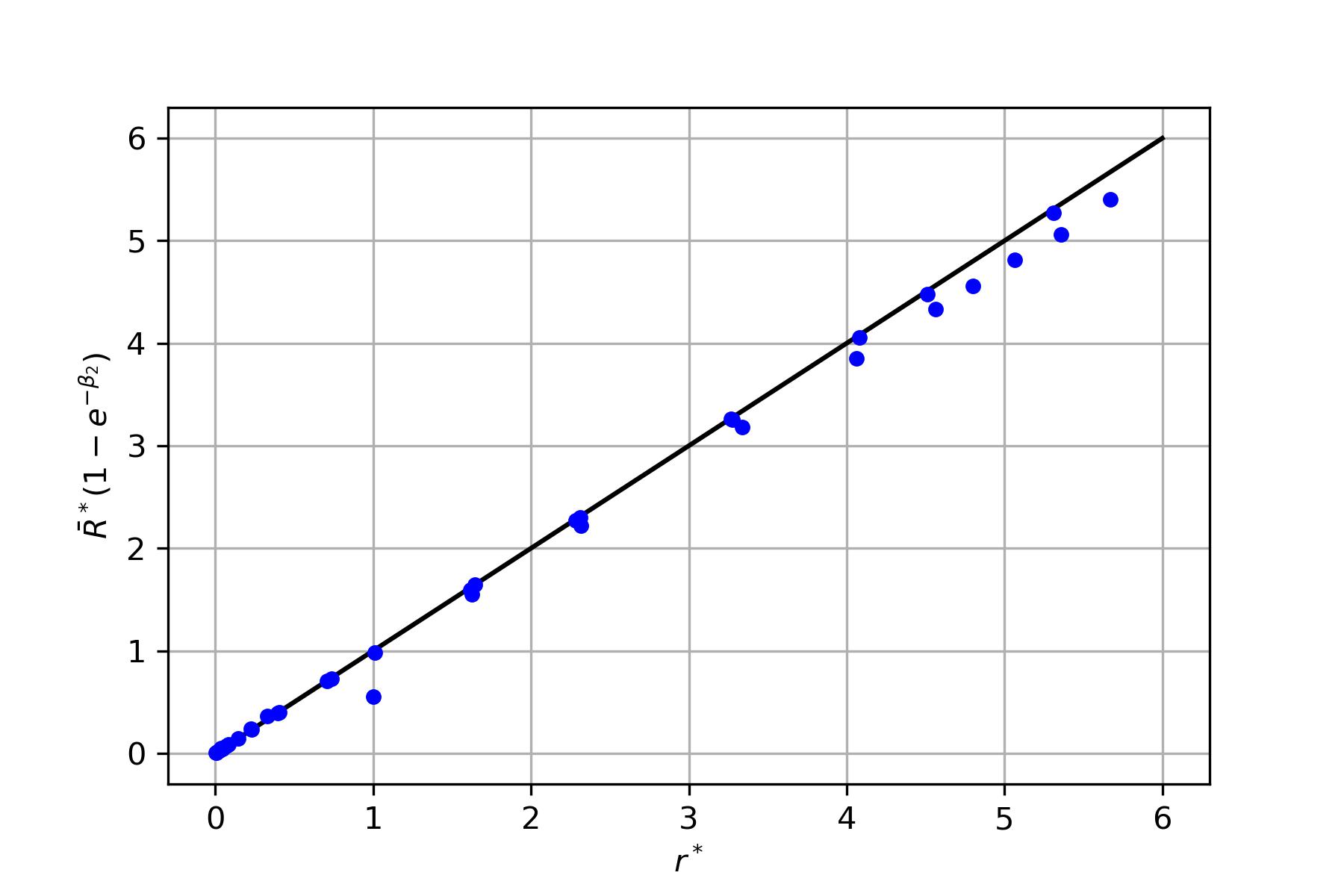}
    \caption{NMZI model. Check of the proportionality relation between the stationary values of $r_t$ and $\bar{R}_t$. The dark line is the bisector of the first quadrant.}
    \label{proportionality_stationary_values}
\end{figure}

We verify numerically the findings we illustrated so far. We compare descriptions (i) and (ii). Several combinations of the parameters $\beta_2, \alpha, \Delta$ are considered: $\beta_2 = [10^{-3}, 10^{-2}, 10^{-1}]$, $\alpha = [10^{-4}, 10^{-3}, 10^{-2}, 10^{-1}]$, $\Delta = [20, 50, 100]$. Let us recall that $\beta_2$ is the parameter which enters the definition of the exponentially weighted mid-price return $\bar{R}_t$ under description (ii). Therefore, in order to compare the two descriptions, given $\beta_2$, the corresponding parameter of the description (i) i.e. $\beta_1$, is set equal to $\beta_2/(\Delta + 1)$. 

We determine the values $\bar{R}^*$, $r^*$ and $t^*$ in the following way. We consider $\bar{R}_t$ during the metaorder's execution and fit it with an exponential function. More precisely, in order to obtain more trustworthy estimations, we fit the set of points $\{\bar{R}_{t_j}\}$ where $t_j, \ j=1, \ldots, Q$ are the child MOs' times. In this way, we only consider the values of $\bar{R}_t$ at the times of each child MO, we eliminate the step behavior and make the fit easier. The functional form of the fit is of the type: $f(j) = a - be^{-cj}$ and the stationary value $\bar{R}^*$ corresponds to the parameter $a$. Similarly, the beginning of the stationary regime in terms of the number of the child MOs is identified as $\tau^* = -\log(\sigma_a/b)/c$ where $\sigma_a$ is the error of the estimated parameter $a$ i.e. $\tau^*$ is the time such that $be^{-cj} = \sigma_a$. Given that $t_j=(\Delta + 1)j -1, \ j=1, \ldots, Q$, the value $t^*$ is equal to $(\Delta + 1)\tau^* - 1$. Finally, a linear fit is run on the mid-price path which corresponds to child MOs that occur at times after $\tau^*$. The slope of this liner fit is the stationary value of the mid-price difference $r^*$.

In Fig. \ref{equivalent_descriptions}, we compare the stationary values $\bar{R}^*$ and $r^*$, obtained with the two descriptions. They lie around the bisector, highlighting the equivalence between the two descriptions given $\beta_1 = \beta_2/(\Delta + 1)$. 

This analogy does not only concern the stationary values but it also holds for the whole evolution, as shown in Fig. \ref{fig_ewmareturns_mZI} in Appendix \ref{appendix_ModifiedZI} which displays the evolution in time of the exponentially weighted mid-price return $\bar{R}_t$ for $\alpha = 10^{-3}$, two different choices of trading interval $\Delta$ i.e. $\Delta = 20, \ 100$, and of the parameter $\beta_2$ i.e. $\beta_2 = 10^{-2}$, $10^{-3}$. We observe that, as expected, $\bar{R}_t$ stabilizes to an equilibrium value after a concave regime. 
Similarly, Fig. \ref{fig_response_mZI} in Appendix \ref{appendix_ModifiedZI} represents the mid-price paths as a function of the number of the child MOs. They exhibit a concave increase followed by a linear behavior, which mirror the increasing and constant behavior of $\bar{R}_t$ respectively.  

Now, we aim to investigate the role that the parameters $\beta_2$ and $\alpha$ play in the findings related to the stationary regime. We consider the results obtained by relying on the economically relevant description (i) i.e. $\bar{R}_t$ is computed by using all market events. In Figures \ref{fig_heatmap_Rstar_mZI}-\ref{fig_heatmap_mathcalRstar_mZI}-\ref{fig_heatmap_tstar_mZI} in Appendix \ref{appendix_ModifiedZI}, we show heat maps representing the stationary values $\bar{R}^*$, $r^*$ and $\tau^*$ for several parameters' combinations ($\beta_2 = [10^{-3}, 10^{-2}, 10^{-1}]$, $\alpha = [10^{-4}, 10^{-3}, 10^{-2}, 10^{-1}]$, $\Delta = [20, 50, 100]$).

{\bf Role of $\boldsymbol\beta_2$.} For fixed values of $\Delta$ and $\alpha$, as $1/\beta_2$ increases so do $\bar{R}^*$ and $\tau^*$. This means that considering a longer memory in the evaluation of the price trend leads to a longer concave regime and so, to a better reproduction of the behavior of the market impact we observe in real markets. This strengthens the intuition we build our model on: including the price dynamics in the orders' sampling process of a LOB model is crucial to reproduce the market impact concavity. 

Concerning $r^*$, if $1/\beta_2$ increases, the stationary value of the mid-price differences decreases. Since $\bar{R}^*$ is larger for larger $1/\beta_2$, also the value of the probability that a LO is a sell in the stationary regime is larger and equivalently, the strength of posting sell LOs in response to an upward price trend. As we see in Subsection \ref{subsec_impact_components}, this implies that the reversion in the interval between two child MOs is stronger and the overall impact of each child MO in the stationary regime, that is $r^*$, is smaller.

{\bf Role of $\boldsymbol\alpha$}. For fixed values of $\Delta$ and $\beta_2$, as $\alpha$ decreases, $\bar{R}^*$ and $\tau^*$ increase. This can be explained by considering that the greater the reaction intensity of traders to price trends, the faster is the convergence of $\mathbb{P}(\text{sell LO at }t|\text{LO at }t)$ to the constant value $1$. A measure of the transition speed of this probability, is given by its first derivative with respect to $\bar{R}_t$, that is equal to $\alpha/4$ for $\bar{R}_t = 0$. The limited concave regime we observe for larger values of $\alpha$, is coherent with our idea that time varying sampling probabilities or similarly, decisions to submit LOs that vary in time and are driven by the price trend, lead to the emergence of the concavity in the mid-price path. 

On the other hand, as $\alpha$ increases, $r^*$ decreases. As we highlight before, larger values of $\alpha$ are accompanied with faster transition speeds to the stationary regime and to the value $1$ for the probability that a LO is a sell. Therefore, as we see in Subsection \ref{subsec_impact_components}, if $\alpha$ is larger, the reversion in the interval between two child MOs is stronger and the overall impact of each child MOs in the stationary regime is smaller.

Finally, we verify the proportionality relation between the exponentially weighted mid-price return and the response function, that we derived in Eq. \eqref{eq_proportionality}. We consider the results obtained by relying on description (i). The stationary values $r^*$ obtained for this description are compared to $\bar{R}^*(1 - e^{-\beta_2})$ and the results are reported in Fig. \ref{proportionality_stationary_values} for the same parameters' combinations considered in Fig. \ref{equivalent_descriptions}.  
A good agreement between $\bar{R}^*(1 - e^{-\beta_2})$  and $r^*$ is obtained.

From now on, the description under interest is (i), being the economically relevant one, as previously pointed out. Therefore, we intend $\beta_1$ for $\beta$ and $\beta_1(\Delta + 1)$ for $\beta_2$.

\subsection{The mid-price path between two child MOs}\label{sec_master_equations}
To understand the origin of the concavity of the price in the NMZI model, we present a model of the mid-price evolution in the intervals between two child MOs. Previously, we showed that $\bar{R}_t$ can be approximated by a constant function in the intervals between two child MOs. Therefore, given the recursive relation which links $\bar{R}_t$ and $r_t$, if $\bar{R}_t$ is constant so does $r_t$. Since $r_t$ expresses the mid-price difference, we expect that a linear function is likely to describe the price reversion in the intervals between two child MOs. In the following, we focus on the stationary regime where the probability that a LO is a sell is constant and we set it to $p_1 = 1/(1 + e^{-\alpha\bar{R}^*})$. By means of master equations, we study the evolution of the mid-price in the time windows between two child MOs. 

We need to consider that, given a state of the order book at time $t$ with mid-price $m = m(t)$\footnote{In this subsection we adopt this notation for the mid-price at time $t$. However, $m(t)$ is analogous to $m_t$.}, it can switch from the value $m$ to one of the values $m'= m + \frac{1}{2}, m + 1, \ldots, m + \frac{s-1}{2}$, where $s = s(t)$ is the spread value at time $t$, because of a buy LO inside the spread at time $t+1$. On the other hand, if at time $t+1$ we have a sell LO inside the spread, the mid-price decreases to $m'= m - \frac{1}{2}, m - 1, \ldots, m - \frac{s-1}{2}$. Sell (buy) MOs and buy (sell) cancellations make the mid-price decrease (increase) by half the first bid (ask) gap size. In the setting of Algorithm \ref{alg_ZI} (displayed in Appendix \ref{sec_parameters_estimation}), the first gap size can assume values between $1$ and $K -2$ where $K$ is the number of tick levels in the grid which represents the LOB. Therefore, since when LOs inside the spread occur, the value that the mid-price can assume depends on the spread, first we need to derive the master equation which describes the evolution of this quantity. Then, we will be able to derive the master equation related to the mid-price.

\subsubsection{Spread evolution}
We derive the master equation which describes the evolution of the spread in the intervals between two child MOs, and we show that it does not depend on the probability that a LO is a sell; therefore, it is also valid for the ZI model. This is due to the fact that the imbalance between sell and buy LOs does not impact the evolution of the spread indeed, the contribution of these two types of orders cancel out in the master equation. This is in contrast to the mid-price: in this case, the imbalance between sell and buy LOs is crucial in determining the reversion in the intervals between two child MOs.

In order to write the master equation describing the evolution of the spread between two child MOs, we need to consider that the spread can change in two ways. Given the spread $s(t)$ at time $t$, it can decrease to the values $1, 2, \ldots, s(t)-1$ if at time $t+1$ there is the placement of a LO inside the spread. On the other hand, the spread can increase to the values $s(t) +1, s(t) + 2, \ldots, s_{MAX}$ if at time $t+1$ a MO or a cancellation occurs and the first gap size is $x_1(t) = s(t+1) - s(t)$. In the setting of Algorithm \ref{alg_ZI}, the maximum value that the spread can assume is $s_{MAX} = K - 1$.

Consequently, the transition rates from spread $s = s(t)$ to spread values $s'=\{1, 2, \ldots, s-1\}$ are:
\begin{equation*}
    \omega_{s\rightarrow s'} = \frac{\lambda K}{\Gamma(t)} p_1 \frac{1}{K - b_1(t) - 1} + \frac{\lambda K }{\Gamma(t)}  (1 - p_1) \frac{1}{b_1(t) + s(t)}
\end{equation*}
where $b_1(t) \in [0, K - 1]$ is the best bid price at time $t$, $\Gamma(t) = \lambda K + 2\mu + \delta n_{orders}(t)$ and $n_{orders}(t) = n_{orders}^{ask}(t) + n_{orders}^{bid}(t)$ is the total number of orders in the book at time $t$. Let us observe that the factors $(K - b_1(t) - 1)^{-1}$ and $(b_1(t) + s(t))^{-1}$ refer to the probability of sampling a given price level for a sell/buy LO respectively. In the implementation of the model, the grid which represents the LOB is centered after each step therefore, we can approximate $(K - b_1(t) - 1) \simeq (b_1(t) + s(t)) \simeq K/2$. We obtain:
\begin{equation*}
    \omega_{s\rightarrow s'} = \frac{2\lambda}{\Gamma(t)}.
\end{equation*}

Similarly, the rates of going from spread values $s'=\{1, 2, \ldots, s-1\}$ to spread $s$ are:
\begin{equation*}\begin{split}
    \omega_{s'\rightarrow s} &= \Bigg(\frac{\mu}{\Gamma(t)}+ \frac{\delta n_{orders}(t)}{\Gamma(t)}  \frac{n_{orders}^{ask}(t)}{n_{orders}(t)} \frac{n_{orders}^{ask_1}(t)}{n_{orders}^{ask}(t)} \Bigg)\mathbb{P}(q_{best}^{ask}(t) = 1)\mathbb{P}\Big(s - s' = x_1^a(t)\Big) + \\
    \\&+ \Bigg(\frac{\mu}{\Gamma(t)} + \frac{\delta n_{orders}(t)}{\Gamma(t)}  \frac{n_{orders}^{bid}(t)}{n_{orders}(t)} \frac{n_{orders}^{bid_1}(t)}{n_{orders}^{bid}(t)}\Bigg)\mathbb{P}(q_{best}^{bid}(t) = 1)\mathbb{P}\Big(s - s' = x_1^b(t)\Big)
\end{split}
\end{equation*}
where $n_{orders}^{bid_1/ask_1}(t)$ are the numbers of orders at the best bid/ask price level at time $t$ and $x_1^{b/a}(t)$ are the first gaps in the bid/ask side at time $t$. For small-tick assets we can approximate $\mathbb{P}(q_{best}^{ask/bid}(t) = 1)\simeq 1 \ \forall t $ and therefore, $n_{orders}^{bid_1/ask_1}(t) \simeq 1 \ \forall t $. Additionally, we consider the probability of having the spread difference which coincides with the first gap size as a Kronecker delta: $$\mathbb{P}\Big(s - s' = x_1^{b/a}(t)\Big) = \mathbb{I}\Big(s - s' = x_1^{b/a}(t)\Big).$$
Therefore, the rates above can be simplified as 
\begin{equation*}
    \omega_{s'\rightarrow s} = \Bigg(\frac{\mu}{\Gamma(t)}+ \frac{\delta}{\Gamma(t)}\Bigg)\mathbb{I}\Big(s - s' = x_1^a(t)\Big) + \Bigg(\frac{\mu}{\Gamma(t)}+ \frac{\delta}{\Gamma(t)}\Bigg)\mathbb{I}\Big(s - s' = x_1^b(t)\Big).
\end{equation*}
Analogously, we have that the transition rates from spread $s$ to spread values $s'=\{s+1, s+2, \ldots, K-1\}$ are:
\begin{equation*}
    \omega_{s\rightarrow s'} = \Bigg(\frac{\mu}{\Gamma(t)}+ \frac{\delta}{\Gamma(t)}\Bigg)\mathbb{I}\Big(s' - s = x_1^a(t)\Big) + \Bigg(\frac{\mu}{\Gamma(t)}+ \frac{\delta}{\Gamma(t)}\Bigg)\mathbb{I}\Big(s' - s = x_1^b(t)\Big),
\end{equation*}
and the rates of going from spread values $s'=\{s+1, s+2, \ldots, K-1\}$ to the spread $s$:
\begin{equation*}
    \omega_{s'\rightarrow s} = \frac{2\lambda}{\Gamma(t)}.
\end{equation*}

Finally, let us define $ \mathbb{P}(s,t)$ as the probability of having spread $s$ at time $t$. The master equation for the spread evolution is:
\begin{equation}\label{master_eq_spread}
    \begin{split}
        \mathbb{P}(s,t + 1) - \mathbb{P}(s,t) &= \sum_{s'\neq s}\Big[\omega_{s'\rightarrow s} \mathbb{P}(s',t) - \omega_{s\rightarrow s'} \mathbb{P}(s,t)\Big] = \\
        &= \frac{\mu + \delta}{\Gamma(t)}\mathbb{P}\Big(s - x_1^a(t), t\Big) + \frac{\mu + \delta}{\Gamma(t)}\mathbb{P}\Big(s - x_1^b(t), t\Big) + \\
    &+ \frac{2\lambda}{\Gamma(t)}\sum_{s'= s +1}^{K -1} \mathbb{P}(s', t) - \mathbb{P}(s, t)\Big[\frac{2\lambda (s - 1) + 2\mu + 2\delta}{\Gamma(t)}\Big]
    \end{split}
\end{equation}
and therefore, we have that
\begin{equation*}
    \langle s(t) \rangle = \sum_{s'=1}^{K -1}s'\mathbb{P}(s',t).
\end{equation*}
We observe that the master equation \eqref{master_eq_spread} depends on three quantities which are functions of time: $\Gamma(t) = \lambda K + 2\mu + \delta n_{orders}(t)$, $x_1^a(t)$, $x_1^b(t)$. However, as simulations will show, the number of orders in the book $n_{orders}(t)$ as well as the first gap sizes $x_1^{b/a}(t)$ have a modest range of variability in the stationary regime, and if they are considered as constant, it negligibly impacts our results. Moreover, it is evident that the spread evolution does not depend on the probability of sell limit orders making this equation also valid for the standard ZI model.

For $t\to+\infty$, we expect that the spread reverts to the equilibrium value, which can be roughly determined as $\hat{s}$ such that:
\begin{equation}\label{spread_equi}
    \frac{\lambda K }{\Gamma(t)}\frac{(\hat{s} - 1)}{K/2} = \frac{2(\mu + \delta)}{\Gamma(t)} \iff \hat{s} = \frac{\mu + \delta}{\lambda} + 1.
\end{equation}
This corresponds to the value for which the contributions of LOs inside the spread and cancellations and MOs which are aggressive i.e. that lead to mid-price changes, are balanced out. It is the mean spread we expect to obtain by simulating both the ZI and the NMZI models, without any execution of metaorders.

\subsubsection{Mid-price evolution}
Similarly to what we did for the spread and by considering the observations made at the beginning of the subsection, we obtain the master equation describing the evolution of the mid-price in the intervals between two child MOs:
\begin{equation}\label{master_eq_midprice}
    \begin{split}
        \mathbb{P}(m,t + 1) - \mathbb{P}(m,t) &= \frac{2\lambda}{\Gamma} \sum_{j=1}^{s(t)-1}\Bigg[p_1\mathbb{P}\Big(m + \frac{j}{2},t\Big) + (1 - p_1)\mathbb{P}\Big(m - \frac{j}{2},t\Big)\Bigg] + \\
        &+\frac{\mu + \delta}{\Gamma}\Bigg[\mathbb{P}\Big(m + \frac{x_1^b(t)}{2},t\Big) + \mathbb{P}\Big(m - \frac{x_1^a(t)}{2},t\Big)\Bigg] + \\
        &-\mathbb{P}(m ,t)\Big[\frac{2\lambda \big(s(t) - 1\big) + 2\mu + 2\delta}{\Gamma(t)}\Big]
    \end{split}
\end{equation}
and
\begin{equation*}
    \langle m(t) \rangle = \sum_{m'=1}^{m_{MAX}}m'\mathbb{P}(m',t)
\end{equation*}
where $\mathbb{P}(m,t)$ is the probability of having a mid-price equal to $m$ at time $t$ and $m_{MAX}$ is the maximum mid-price we can have in the grid which represents the LOB i.e. $((K - 2) + (K-1))/2$.

Contrary to the master equation for the spread, the evolution of the mid-price depends on the probability $p_1$ that a LO is a sell. Therefore, Eq. \eqref{master_eq_midprice} holds for the standard ZI model only if $p_1 = 0.5$.

\begin{figure}
\includegraphics[width=0.5\linewidth]{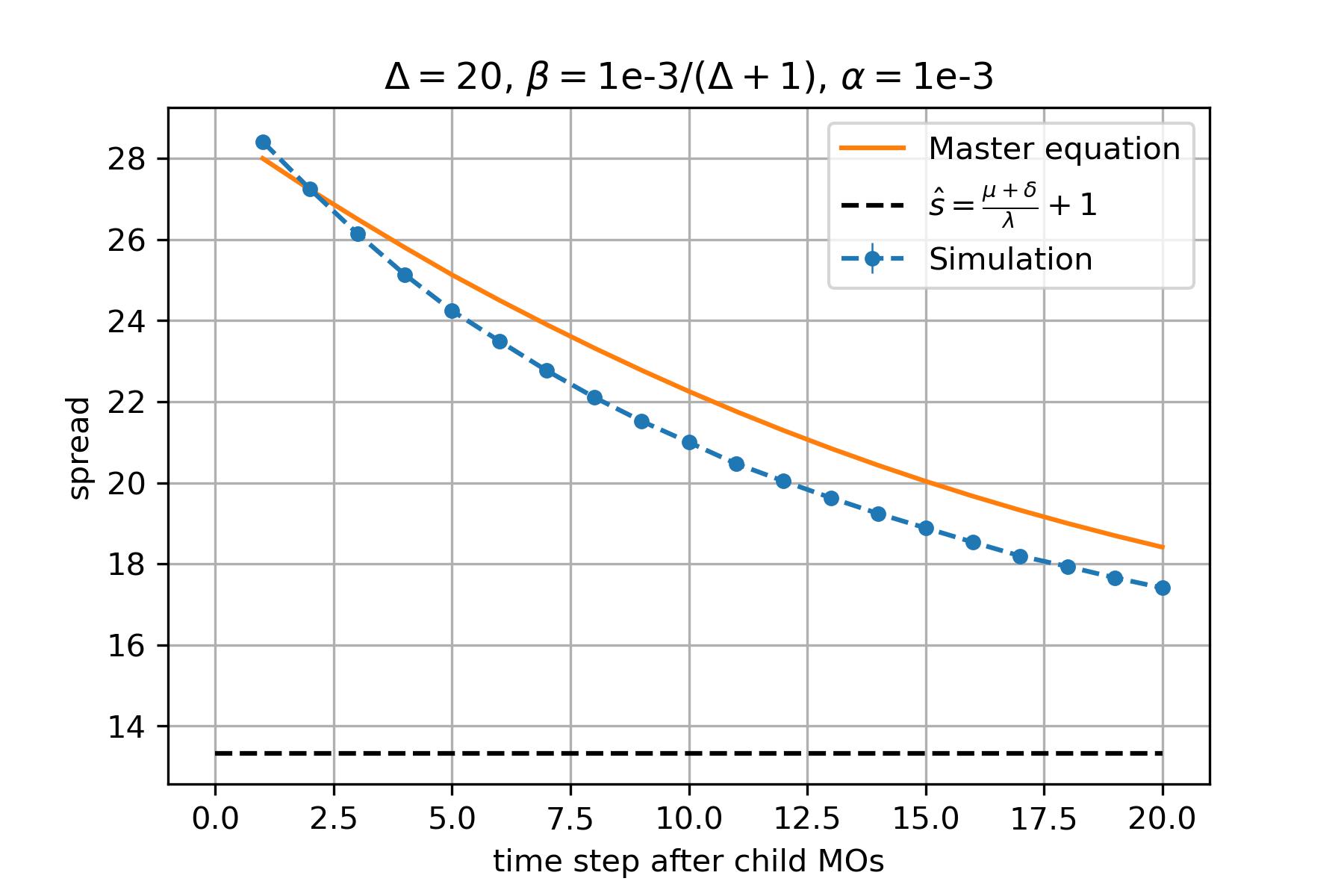}
\includegraphics[width=0.5\linewidth]{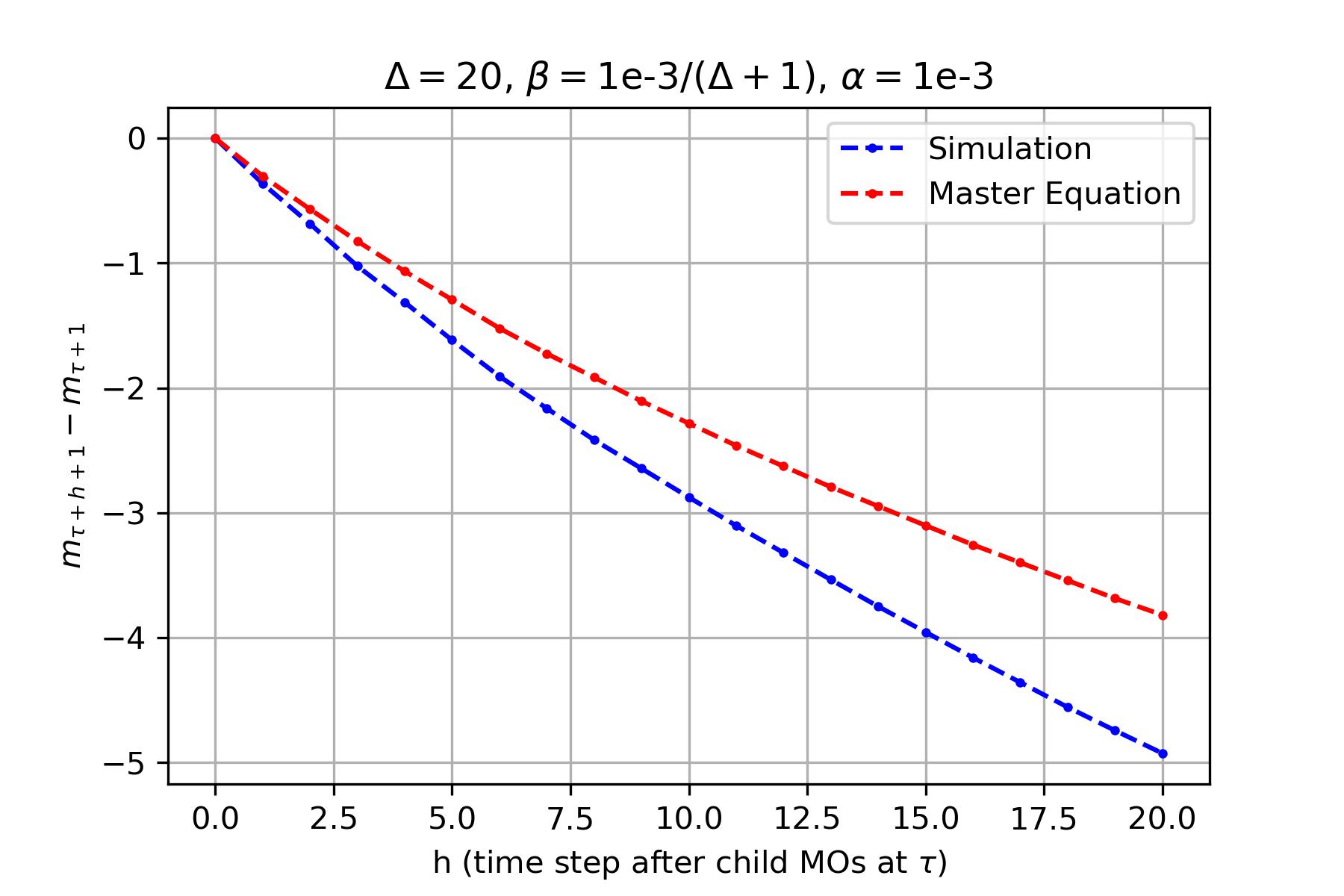}
\includegraphics[width=0.5\linewidth]{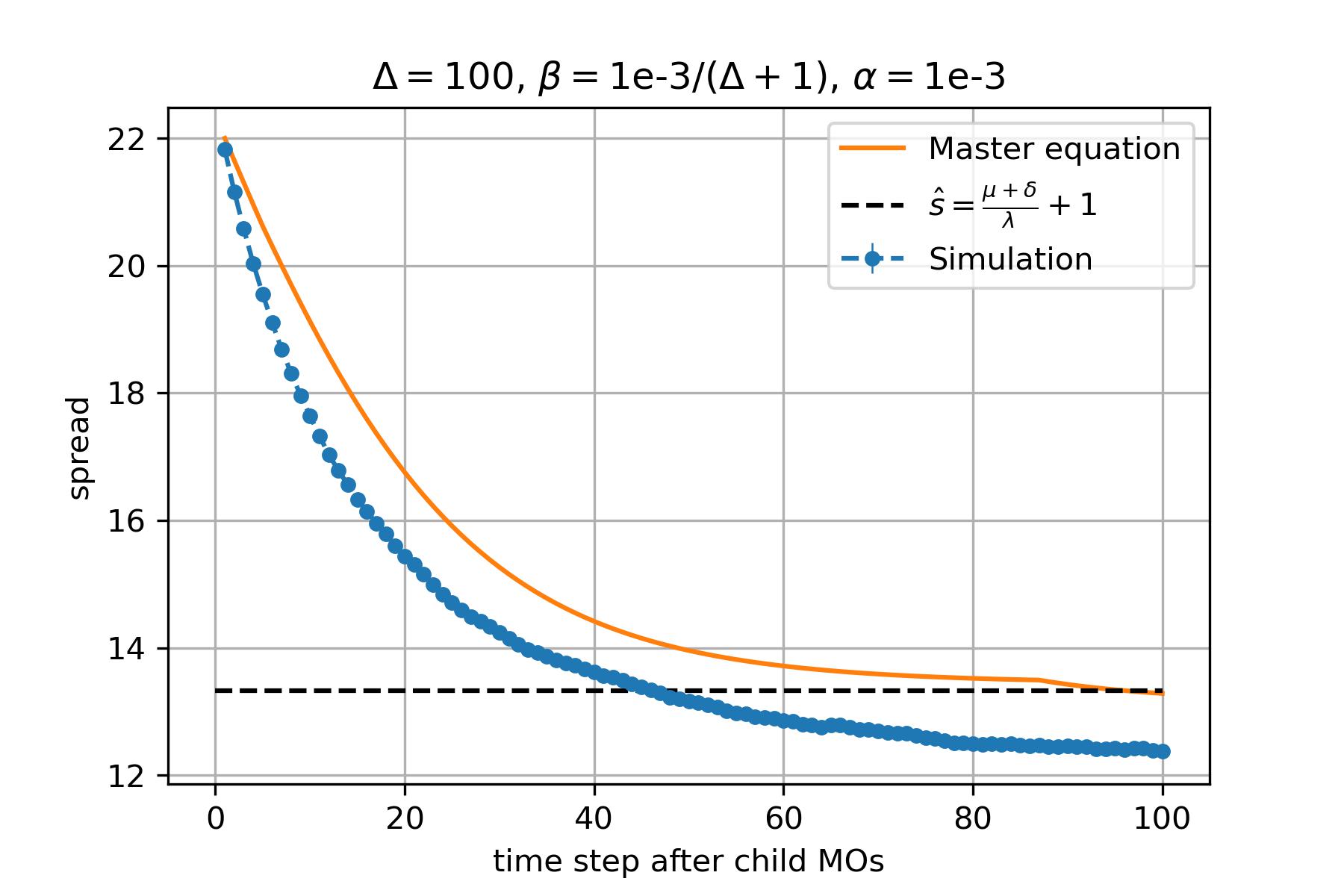}
\includegraphics[width=0.5\linewidth]{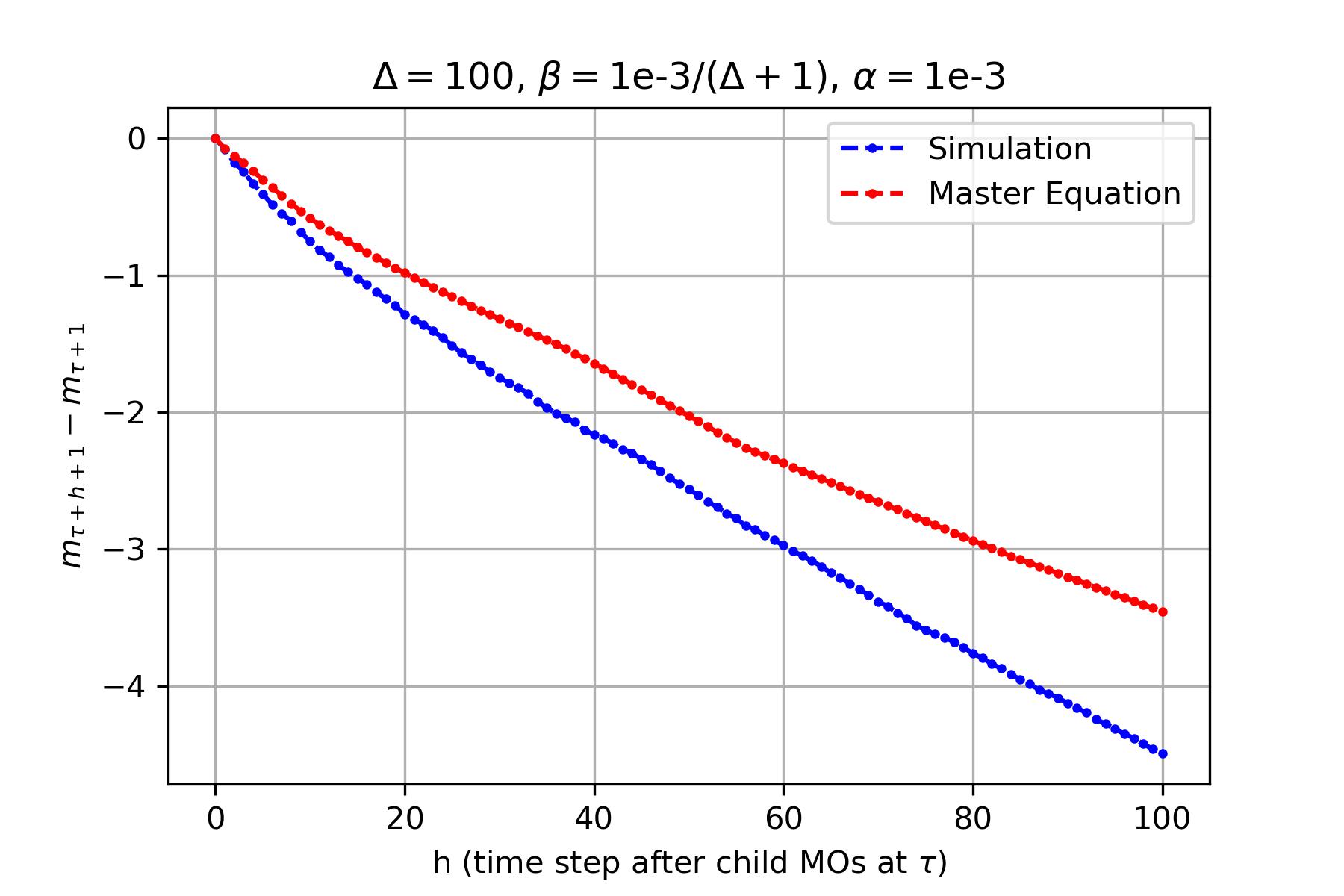}
    \caption{NMZI model, stationary regime. Spread (in ticks) as a function of the time step after child MOs. 
    Mid-price difference $m_{\tau + h + 1} - m_{\tau + 1} $ (in ticks) as a function of $h$ where $h$ is the time step after child MOs at time $\tau$.}
    \label{fig_mZI_stationary}
\end{figure}

\subsubsection{Numerical results}
We verify whether the outcomes predicted by the master equations we derived - are consistent with simulations' results. In order to avoid edge effects due to the vicinity to the concave regime, we focus on the \textit{deep stationary regime} that we define as the set of times greater than $t^* + 1/\beta$ and in such a way that $500$ child MOs are considered. We compute: the spread, the probability of a sell LO (given a LO) and the number of orders in the book as functions of the time step after child MOs; the mid-price difference $m_{\tau + h + 1} - m_{\tau + 1} $ as a function of $h$ where $h$ is the time step after child MOs at times $\tau$; the frequency of the different types of events in each time window between child MOs. They are shown in Fig. \ref{fig_mZI_stationary} and \ref{fig_mZI_stationary_2}-\ref{fig_mZI_stationary_3} in Appendix \ref{appendix_ModifiedZI} for $\beta =10^{-3}/(\Delta + 1), \ \alpha = 10^{-3}$ and two choices of $\Delta$. In particular, in Fig. \ref{fig_mZI_stationary}, we compare the evolution of the spread and the mid-price differences that we obtain from simulations, to the corresponding evolution resulting from the master equations \eqref{master_eq_spread} and \eqref{master_eq_midprice}. When we determine the evolution of the spread, we set the initial condition $\mathbb{P}(s, t=0) = \delta(s = s_0)$ where $s_0$ is the integer part of the mean spread after the child MOs in the stationary regime according to simulations. Concerning the mid-price, Eq. \eqref{master_eq_midprice} depends on the spread and we employ the values obtained with the corresponding master equation. We observe that the probability of a sell LO (given a LO), the number of orders and the first gap in the bid/ask side (displayed in Figures \ref{fig_mZI_stationary_2}-\ref{fig_mZI_stationary_3}) have a modest range of variability and we verified that considering them constant in the master equations negligibly impacts our results.

Focusing on the spread as a function of the time step after child MOs, we see that it reaches a maximum at step $1$ and then, it decreases. The maximum represents the spread after a child MO occurs and when $\Delta = 20$, it turns out to be $28.42$ ticks while when $\Delta = 100$, it is equal to $21.82$ ticks. As expected, they are both larger than the mean spread in the stationary regime which is $21.29$ and $14.08$ ticks respectively. We observe that the master equation well describes the evolution of the spread between two child MOs. Moreover, if we focus on the larger trading interval $\Delta = 100$, the spread reverts to the equilibrium value according to Eq. \eqref{spread_equi} i.e. $\hat{s} \simeq 13.33$ (dashed black line in Fig. \ref{fig_mZI_stationary}). This is also confirmed by computing the spread evolution with the master equation for times much larger than $\Delta$.

Similarly to the spread, the master equation of the mid-price is able to reproduce its behavior. It also emerges that employing a linear function in order to describe the mid-price evolution between two child MOs is a well-grounded choice. This is actually the behavior we expect, as introduced in the beginning of Subsection \ref{sec_master_equations}. 

\subsection{The impact components}\label{subsec_impact_components}
In the intervals between two child MOs, we observe a reversion of the mid-price. This is in contrast to the ZI, where there is not any reversion: the only component of the impact is the immediate price impact following each child MO and equal to half the first gap size (for small-tick assets). In the NMZI model, the interplay between the immediate impact and the reversion is at the root of the concavity in the mid-price path and for this reason we now analyze in depth this issue. We see that in the concave regime the main contribution which makes the overall impact decrease, is due to the reversion component, that increases in time. In the stationary regime, the overall child MOs' impact stabilizes to a stationary value.

Let us define the times at which the child MOs are executed as $t_j = \Delta j + j - 1$ with $j=1, \ldots, Q$ and the starting time of the simulation equal to $0$. We can describe the evolution of the mid-price during the execution as
\begin{equation}\label{eq_kernel_during_execution}
    m_t = G_{j_t}(t - t_{j_t} - 1), \ \ j_t = \Bigg\lfloor\frac{t}{\Delta + 1}\Bigg\rfloor, \ \ t \in (t_1, t_Q]
\end{equation}
where each $G_j(x)$ with $x \in [0, \Delta]$ and $j=1, \ldots, Q - 1$ is a non-increasing function describing the mid-price path in the time interval between the child MOs $j$ and $j+1$. Therefore, we have that:
\begin{itemize}
    \item the immediate price impact of the child MO $j$ is: $I^I_j = G_{j}(0) - G_{j-1}(\Delta)$;
    \item the reversion of the mid-price following the child MO $j$ is: $I^R_j = G_{j}(0) - G_{j}(\Delta)$;
    \item the overall price impact of the child MO $j$ is: $I^I_j - I^R_j$. 
\end{itemize}
Concerning the functional form of the functions $G_j(\cdot)$, we employ linear kernels given the findings of the preceding subsection. However, by studying the evolution after the execution ends with the same approach, in Subsection \ref{sec_after_execution} we see that the linear kernel is actually an approximation of the exponential case for trading intervals which are not too large.

\begin{figure}
\includegraphics[width=0.5\linewidth]{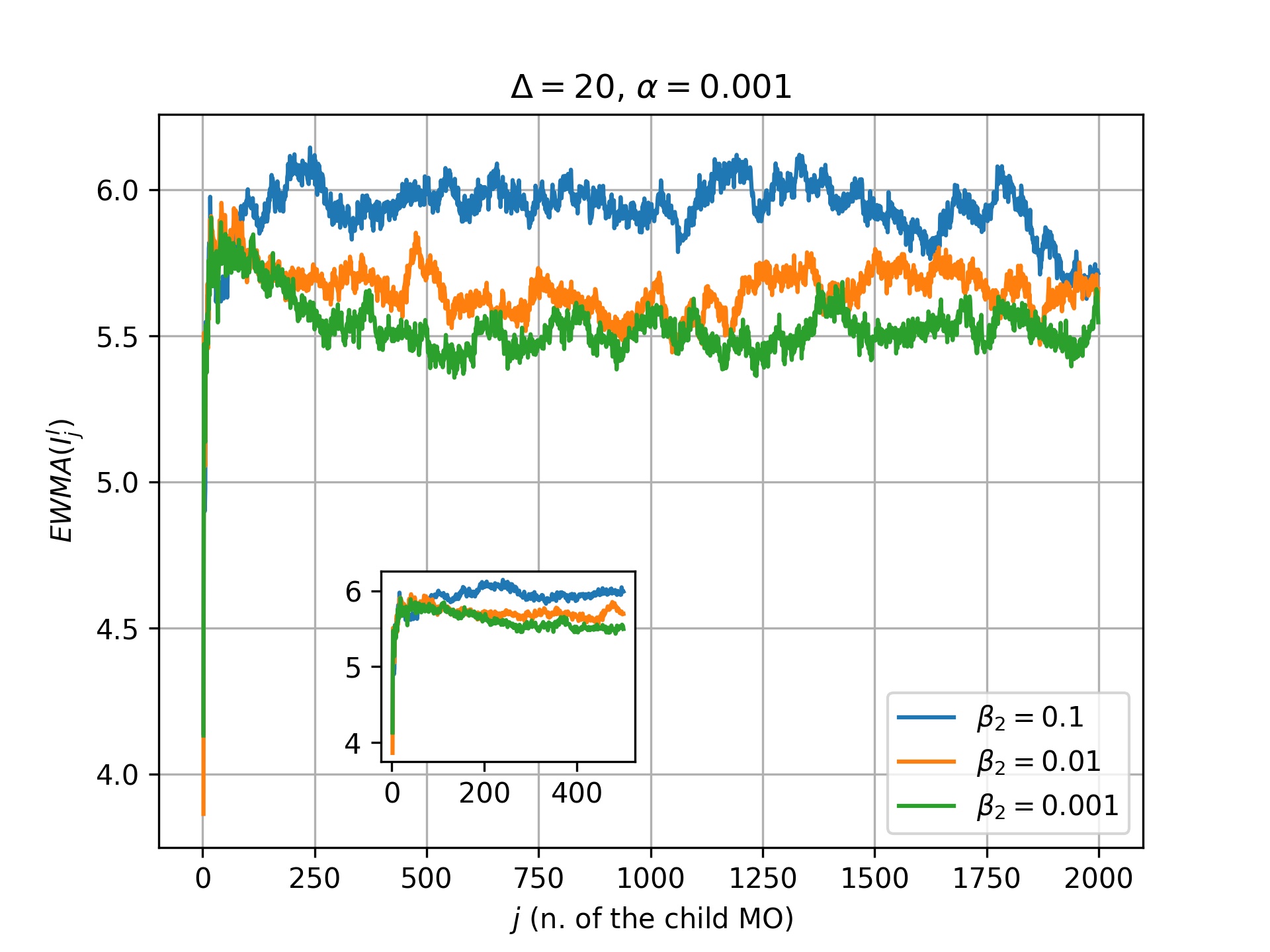}
\includegraphics[width=0.5\linewidth]{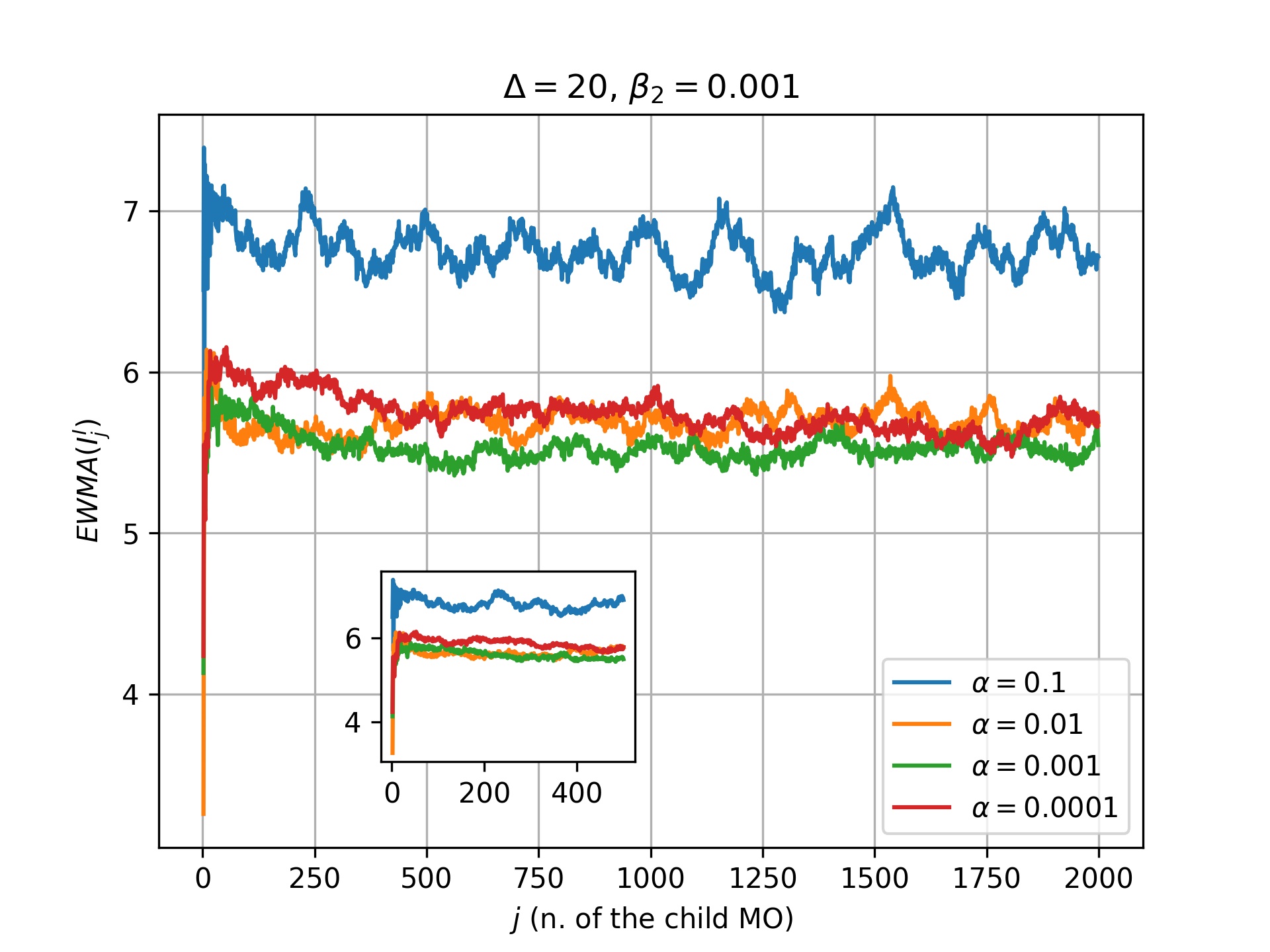}
    
    \caption{NMZI model. Evolution of the exponentially weighted moving averages (EWMAs) of the immediate component $I^I$ of the price impact. The inset plots focus on the first $500$ child MOs.}
    \label{impact_components_immediate_text}
\end{figure}

\begin{figure}

\includegraphics[width=0.5\linewidth]{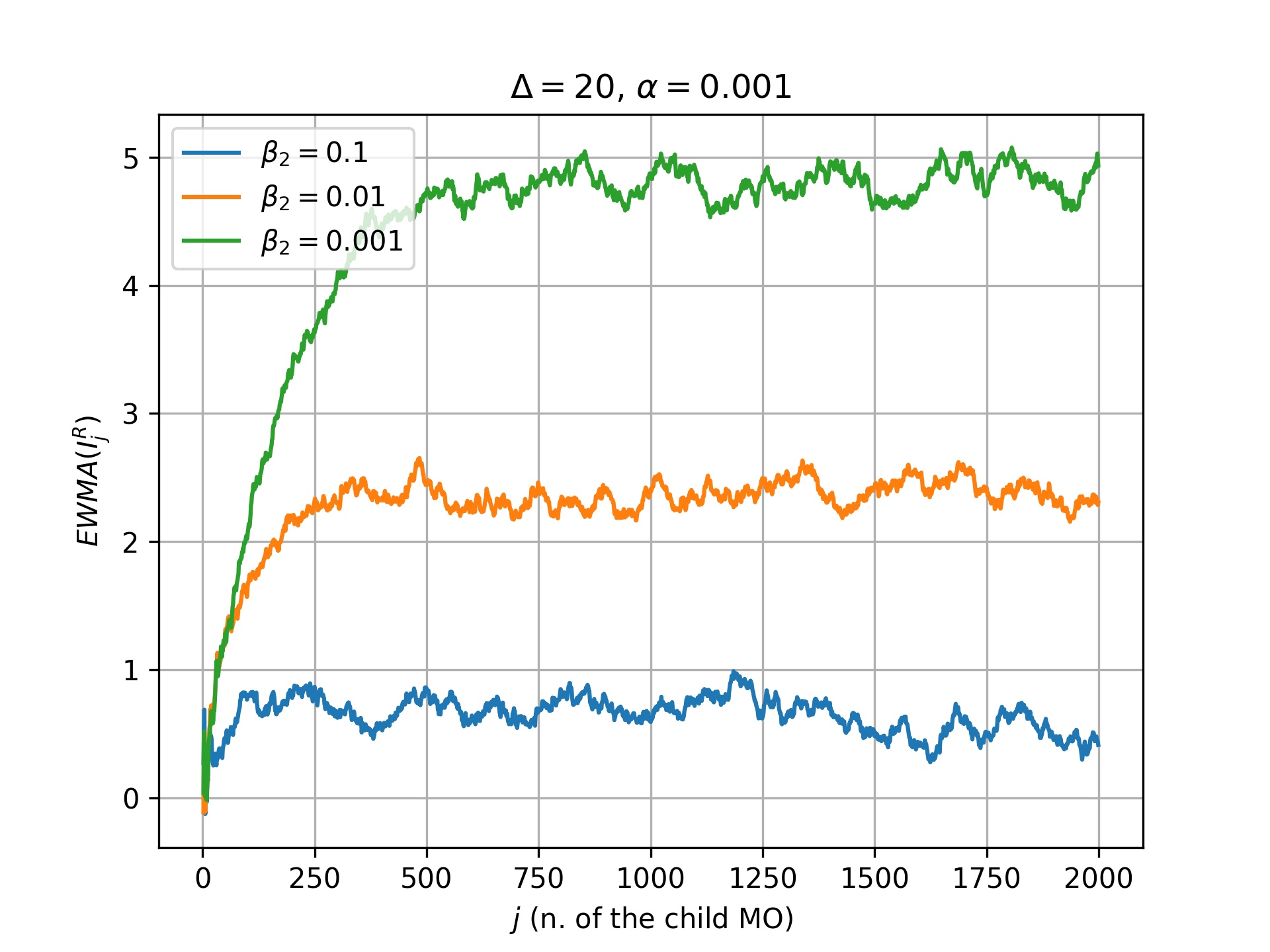}
\includegraphics[width=0.5\linewidth]{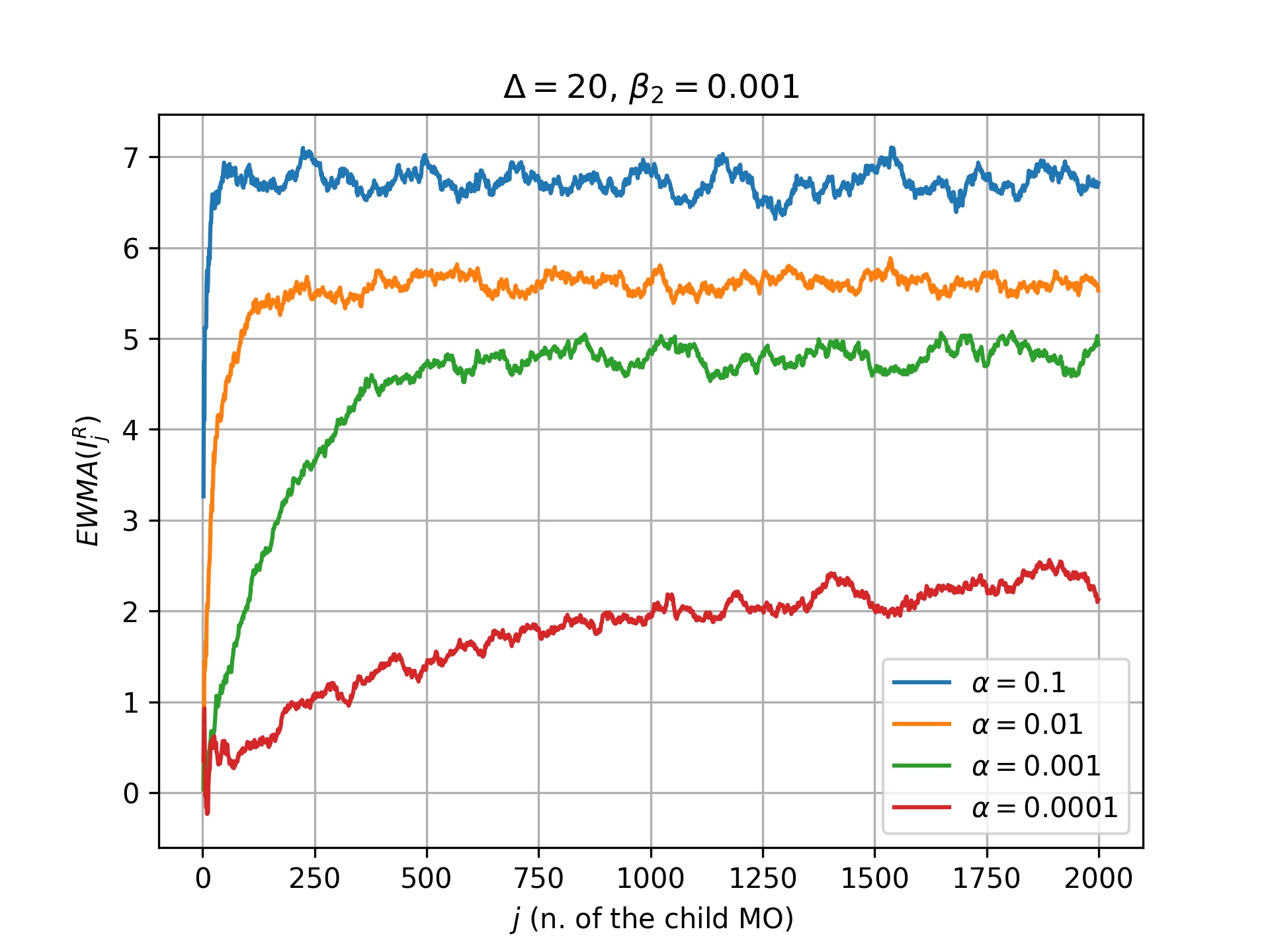}
    
    \caption{NMZI model. Evolution of the exponentially weighted moving averages (EWMAs) of the reversion component $I^R$ of the price impact. }
    \label{impact_components_reversion_text}
\end{figure}

\begin{figure}
\includegraphics[width=0.5\linewidth]{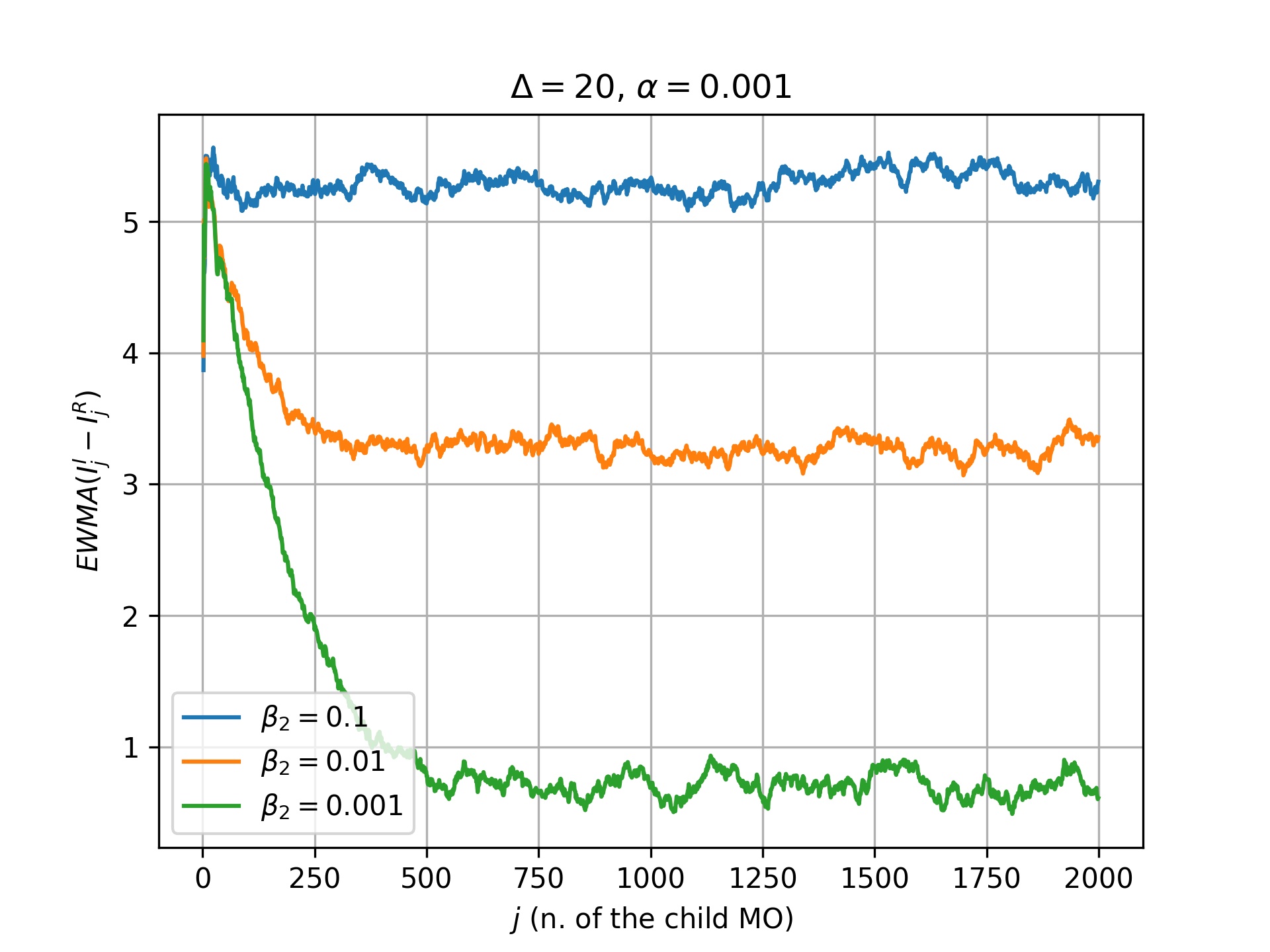}
\includegraphics[width=0.5\linewidth]{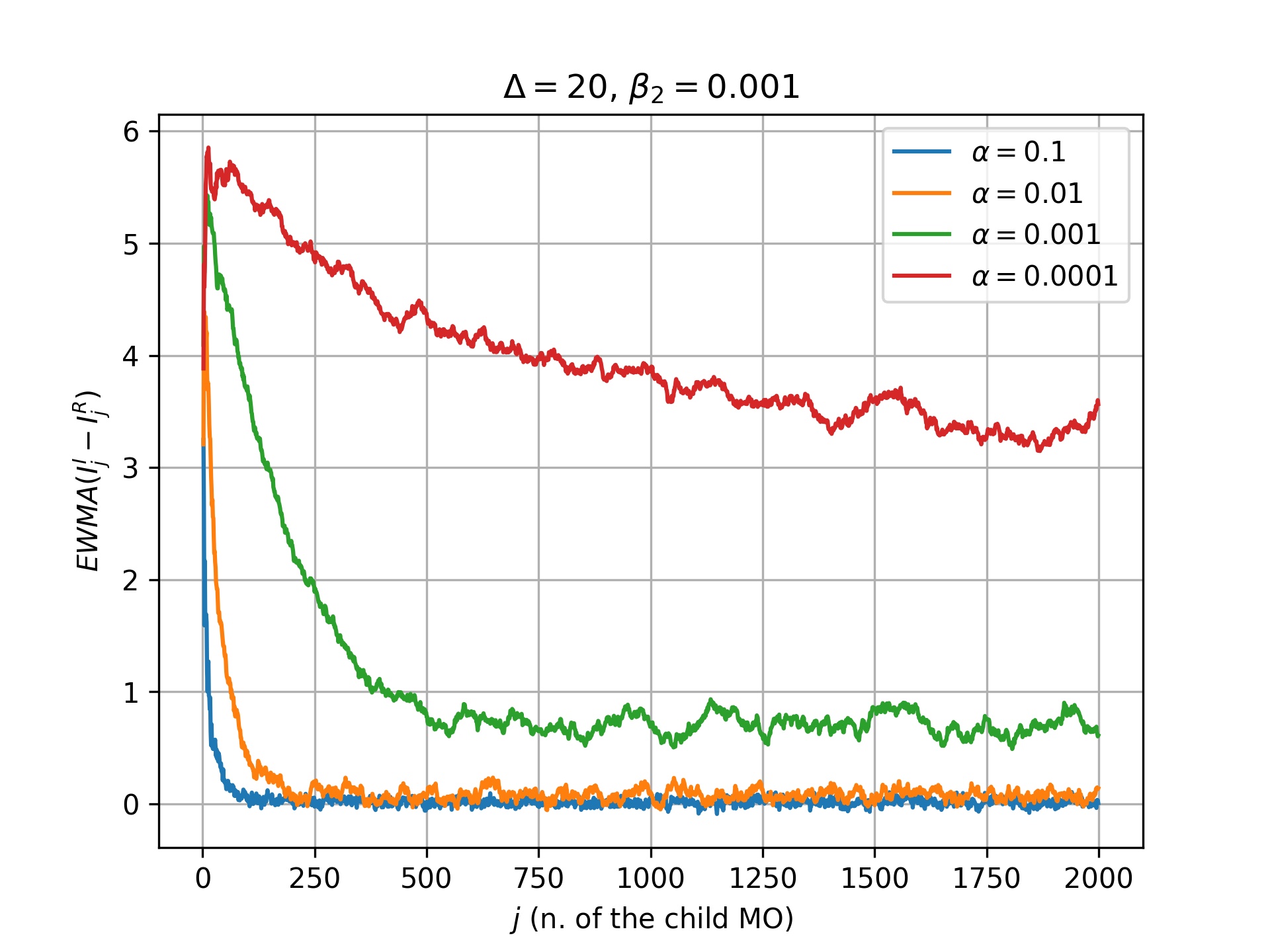}
    \caption{NMZI model. Evolution of the exponentially weighted moving averages (EWMAs) of the overall impact $I^I - I^R$. }
    \label{impact_components_total_text}
\end{figure}

By referring to Eq. \eqref{eq_kernel_during_execution}, we set: $$G_j(t - t_j - 1) = \eta_j\big(1 - \rho_j (t - t_j - 1)\big),$$ 
for
$$\ t \in [t_j + 1, t_{j+1}], \ j=1, \ldots, Q -1.$$
The parameter $\eta_j$ represents the value of the mid-price after the child MO $j$ while $\rho_j$ is the slope of the straight line which describes the reversion in the interval between the child MOs $j$ and $j+1$. By performing linear fits on the mid-price paths, we estimate the parameters of these kernels and the corresponding impact' components\footnote{Instead of performing linear fits in the intervals between child MOs, we also estimated the impact components directly from the mid-price path: the immediate impact of child MO $j$ as $m_{t_{j} + 1} - m_{t_j}$ and the reversion impact of child MO $j$ as $m_{t_{j} + 1} - m_{t_{j+1}}$. The results we obtain are analogous to the outputs of the linear fits, thus further corroborating the choice of linear kernels to describe the mid-price path between two child MOs. Because of that, plots are not displayed in the paper but they are available upon request.}, which are: 
$$I^I_j = \eta_j - \eta_{j-1} + \eta_{j-1}\rho_{j-1}\Delta, \ I^R_j = \eta_{j}\rho_{j}\Delta, \ I_j^I - I_j^R, \ j = 2, \ldots, Q - 1.$$
In Fig. \ref{impact_components_immediate_text}-\ref{impact_components_reversion_text}-\ref{impact_components_total_text}, we plot their exponentially weighted moving averages (EWMAs) for $\Delta = 20$, and several choices of $\beta_2$ and $\alpha$ (the results related to additional cases are plotted in Fig. \ref{impact_components_immediate}-\ref{impact_components_reversion}-\ref{impact_components_total} in Appendix \ref{appendix_ModifiedZI}). We consider the EWMAs since the impacts' evolutions are noisy. As expected, two regimes are identified. In the stationary regime, the impact stabilizes to a constant value\footnote{In some cases e.g. $\beta_2 = 10^{-3}$ and $\alpha = 10^{-4}$ (red curve in the right panel of Fig. \ref{impact_components_total_text}), the stationary regime is not reached in the plots since the transition to the stationary regime is slow and an higher number of child MOs would be needed. However, in order to better compare the results for different choices of parameters, in the plots we consider a metaorder of volume $Q=2,000$.}. When $\bar{R}_t$ is constant, the probability of sell LOs is constant and so is the reversion of the mid-price in the intervals between two child MOs. In parallel, in the stationary regime, the first gap size is approximately constant and so does the immediate impact component due to the execution of child MOs. Indeed, in Fig. \ref{fig_mZI_stationary_2} in Appendix \ref{appendix_ModifiedZI} we plot the evolution of the first gap size in the intervals between two child MOs in the stationary regime and we observe that the standard errors are negligible with respect to the values of the first gap size (averaged over $500$ child MOs). It follows that in the stationary regime the overall child MOs' impact is approximately constant. On the other hand, in the concave regime of the mid-price trajectory, we observe several facts which draw attention to the mechanism at the root of the concavity's emergence in our model.

{\bf Immediate impact component in the concave regime.} First, concerning the immediate component (see Fig. \ref{impact_components_immediate_text}), it initially increases for the first child MOs. Then, we can have two different behaviors. In the first case e.g. $\beta_2 = 10^{-3}$ and $\alpha = 10^{-3}$ (green curve in the left panel of Fig. \ref{impact_components_immediate_text}), following the initial increase, the immediate impact mildly decreases in time and then, stabilizes to its stationary value. The other possible behavior is that after the initial increase, the immediate impact stabilizes to its stationary value e.g. $\beta_2 = 10^{-1}$ and $\alpha = 10^{-3}$ (blue curve in the left panel of Fig. \ref{impact_components_immediate_text}). Both behaviors are explained by the evolution of the imbalance between the orders in the bid and ask side in the LOB, which is shown in Fig. \ref{imbalance_evolution} in Appendix \ref{appendix_ModifiedZI} for $\Delta = 20$, $\alpha=10^{-3}$ and $\beta_2 = 10^{-3}, \ 10^{-2}, \ 10^{-1}$. In all three cases, for the first approximately $10$ market events after the beginning of the execution, the imbalance is negative i.e. the number of orders in the bid side is greater than the number of orders in the ask side. For $\beta_2 = 10^{-3}, \ 10^{-2}$, this is highlighted in the inset plots. Therefore, at the beginning of the execution, the immediate impact increases since the executions of the child MOs deplete the ask side and the time-varying probability that a LO is a sell is still close to $1/2$.

If $\beta_2 = 10^{-1}$, after this increase the imbalance stabilizes to a negative stationary value. In this case, as we see in Subsection \ref{subsec_stationary_regime}, the stationary value of $\bar{R}^*$ is small compared to the other cases investigated (e.g. $\beta_2 = 10^{-3}$) and it is such that the stationary value of the probability that a LO is a sell is close to $1/2$. This implies that, as we observe from the blue curve in the left panel of Fig. \ref{impact_components_reversion_text}, the reversion is small compared to the other cases and the depletion of the LOB due to the metaorder execution is not compensated by the submission of sell LOs which replenish the ask side. 
    
    On the other hand, if $\beta_2 = 10^{-3}$ or $10^{-2}$, before the stabilization of the immediate impact to a constant value, there is a decrease which corresponds to an increase of the imbalance. If $\beta_2 = 10^{-3}$, the transition to the stationary regime is slower compared to $\beta_2 = 10^{-2}$ and the stationary value of the imbalance is higher. This is mirrored by the path of the immediate impact, which has a  more accentuated decrease before the stationary regime. Indeed, a longer memory (smaller $\beta_2$) corresponds to a longer concave regime. Anyway, for both $\beta_2 = 10^{-3}$ and $\beta_2 = 10^{-2}$, the decrease of the immediate impact before the stationary regime is mild.

{\bf Reversion impact component in the concave regime.} The reversion (see Fig. \ref{impact_components_reversion_text}) increases and this is attributed to the evolution of our indicator of the past price dynamics $\bar{R}_t$ and its role in the time varying probability that a LO is a sell. 

{\bf Overall impact in the concave regime.} The overall impact (see Fig. \ref{impact_components_total_text}) decreases, making the price path concave, and the major contribution in the decrease is ascribable to the behavior of the reversion component. Therefore, introducing a model that mimics the behavior of traders who have reservation prices and post LOs in response to the price trend, proves to be pivotal in reproducing the well-known concavity of the price impact during the execution of metaorders.
\\

So far, we focused on the behavior of the impact components which determine the concavity in our model. However, understanding the role of the parameters $\beta_2$ and $\alpha$ in our results, is crucial since it allows to shed light on the financial interpretation of our model. We observe that lower values of $\beta_2$ i.e. $\bar{R}_t$ with longer memory, lead to a more pronounced reversion (see left panel of Fig. \ref{impact_components_reversion_text}) and indeed, as previously shown in Subsection \ref{subsec_stationary_regime}, they correspond to longer concave regimes in the mid-price paths. In the framework of the interpretation of our model in terms of traders with reservation prices, this means that, if traders use longer values of the memory in the evaluation of the price trend, then the reversion mechanism and their response to the upward price trend is stronger. 

Concerning $\alpha$, lower values of this parameter make the transition to the stationary regime slower (see right panel of Fig. \ref{impact_components_reversion_text}). As we explain in Subsection \ref{subsec_stationary_regime}, the derivative of the probability that a LO is a sell with respect to $\bar{R}_t$ can be seen as a measure of its transition speed to $1$. We also observe that low values of $\alpha$ as $\alpha = 10^{-4}$ lead to lower variability of the overall impact (see right panel of Fig. \ref{impact_components_total_text} and Fig. \ref{impact_components_total} in Appendix \ref{appendix_ModifiedZI}). This is due to the fact that if $\alpha$ is small such that even though $\bar{R}_t$ varies, $\mathbb{P}(\text{sell LO}{| \text{LO})}$ does not deviate consistently from $1/2$, then this probability, that is at the core of our model, is not enough informative about the price dynamics we aim to consider in the sampling. The reaction of traders to the price trend is extremely mild and so, our model does not consistently differ from the ZI.

Additionally, for $\alpha = 10^{-1}$ and $\beta_2 = 10^{-3}$, the overall impact is about 0 in the stationary regime (see the blue curve in the right panel of Fig. \ref{impact_components_total_text}), leading to a flat trajectory of the mid-price. This is due to the fact that, as we say in Subsection \ref{subsec_stationary_regime}, with high values of $\alpha$, the probability that a LO is a sell tends fast to $1$. Therefore, the reaction of traders to the upward price trend totally compensates the immediate impact due to the child MOs. We also observe that if $\alpha = 10^{-1}$ and $\beta_2 = 10^{-2}$ or $\beta_2 = 10^{-1}$, the overall impact is not null in the stationary regime (see Fig. \ref{impact_components_total} in Appendix \ref{appendix_ModifiedZI}). In these cases, the values of $\bar{R}^*$ are even smaller than what we have for $\beta_2 = 10^{-3}$ and the effect of the high value of the reaction intensity in the time-varying probability of the LOs' sign is weakened. 

To sum up these observations related to $\alpha$, if traders react strongly, the upward trend of the price becomes rapidly compensated by the submission of sell LOs, making the mid-price become approximately constant. If instead, traders mildly react to arising price trends, the reversion entity is moderate and slowly converges to a constant value.

\subsection{Price decay after the execution}\label{sec_after_execution}
Empirically, it is well known that a buy metaorder pushes the price up during its execution, with a concave path. When the execution ends, so does the buying pressure and this leads to a price reversion (see \cite{moro_2009, brokman2015, zarinelli_2015, bucci2018}). This behavior is represented in Fig. \ref{fig_meta_order_impact_th}.

As we have seen in Section \ref{section_sf}, the ZI model is a linear model of price impact. During the execution of a buy metaorder, the price increases linearly, with a slope equal to the response function obtained for the model without any execution. During the intervals between two child MOs, the price remains approximately constant and so does after the execution ends. The NMZI model we propose is able to overcome these limitations and as shown in previous subsections, during the execution of a metaorder, the concavity of the price path emerges. 
In the following, we focus on the behavior of the mid-price after the end of the execution and we show that our model is able to mimic the well-known price reversion.

Similarly to Eq. \eqref{eq_kernel_during_execution}, we can describe the price evolution after the execution of the metaorder ends as
\begin{equation}\label{eq_kernel_after_execution}
    m_t = G_{after}(t - t_{Q} - 1) , \ \ t > t_{Q}
\end{equation}
where $G_{after}(x)$ with $x \in [0, +\infty)$ is the corresponding function. In order to infer the functional form of this kernel, we argue the following. The functional form which describes the reversion of $\bar{R}_t$ after the execution ends, is linked to the functional form that describes the mid-price path after the execution. Given the simulations' results, our hypothesis is that $\bar{R}_t$ decays exponentially i.e. $\bar{R}_{t'} = ae^{-bt'}$ where $t' = (t - t_Q - 1)$ and $t > t_{Q}$. This relation and Eq. \eqref{eq_recursive_EWMA_returns} imply that:
$$ae^{-bt'} - \gamma a e^{-b(t'-1)} = r_{t'} \iff r_{t'} = \bar{a}e^{-bt'}, $$ 
where $$\bar{a} = a(1 - \gamma e^b), \ \gamma = e^{-\beta} = e^{-\beta_2/(\Delta + 1)}. $$
Since $r_{t'}$ represent mid-price differences, we end up with
\begin{equation}\label{eq_midprice_after_ex}
\begin{split}
     m_{t} &= m_{t_Q + 1} 
 + \int_{t_{Q} + 1}^t ds \ \bar{a}e^{-b(s - t_Q - 1)} = c - \frac{\bar{a}}{b}e^{-b(t - t_Q - 1)} \\
 c &= m_{t_Q + 1} + \frac{\bar{a}}{b} \\
 t &\in [t_{Q} + 2, +\infty)
\end{split}
\end{equation}
with $\bar{a} < 0, \ b > 0$, i.e. we expect that the mid-price path decays exponentially with parameters that are linked to the ones which enter the fit of $\bar{R}_t$. We notice that at the end of the decay of $\bar{R}_t$, the model reverts to the standard ZI since the probability that a LO is a sell becomes approximately equal to $1/2$. However, a permanent component of the impact can still persist and it corresponds to the difference between the parameter $c = \lim_{t \rightarrow +\infty} m_{t}$ and the mid-price just before the beginning of the execution. 

\begin{figure}
\includegraphics[width=0.5\linewidth]{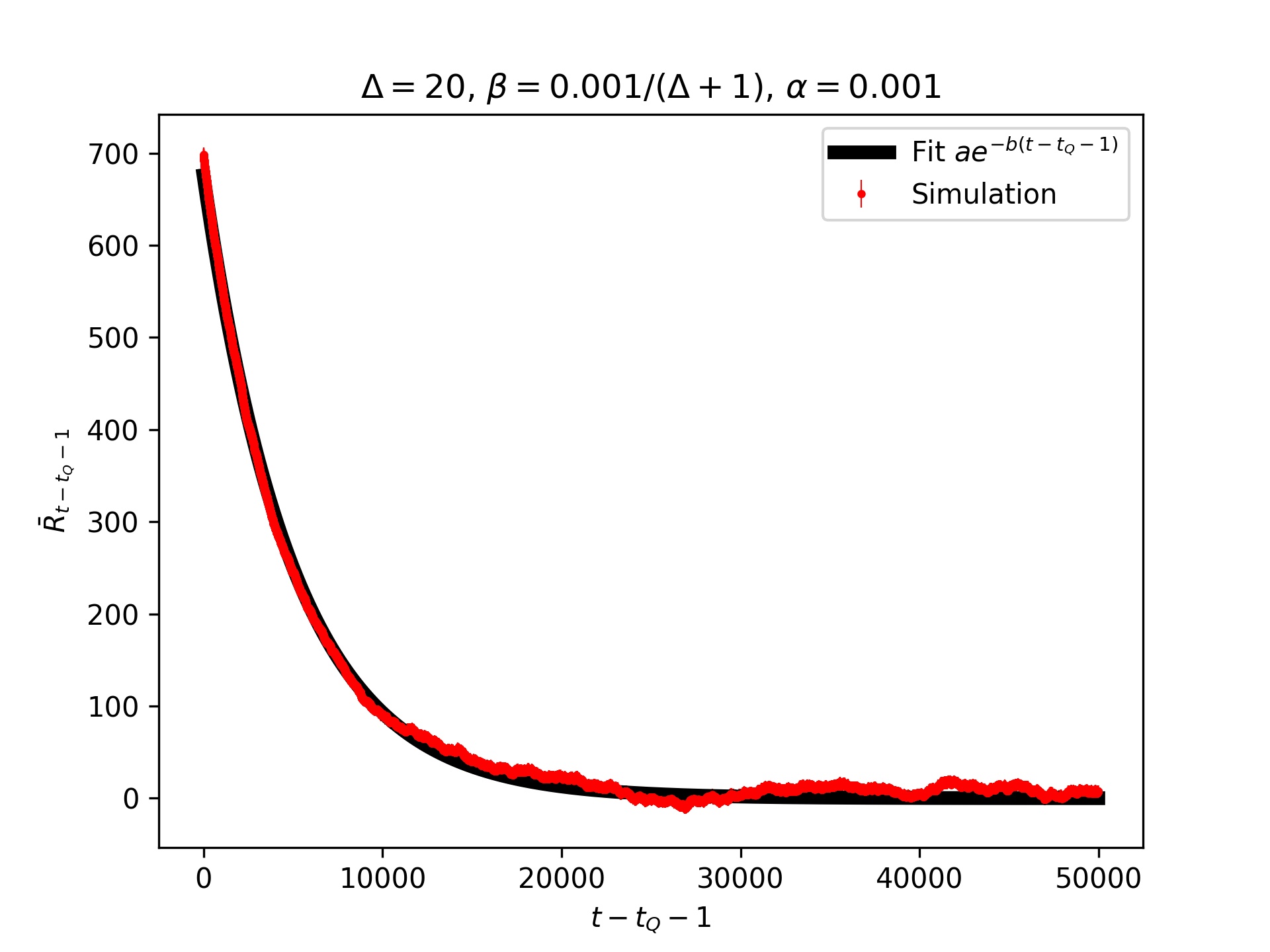}
\includegraphics[width=0.5\linewidth]{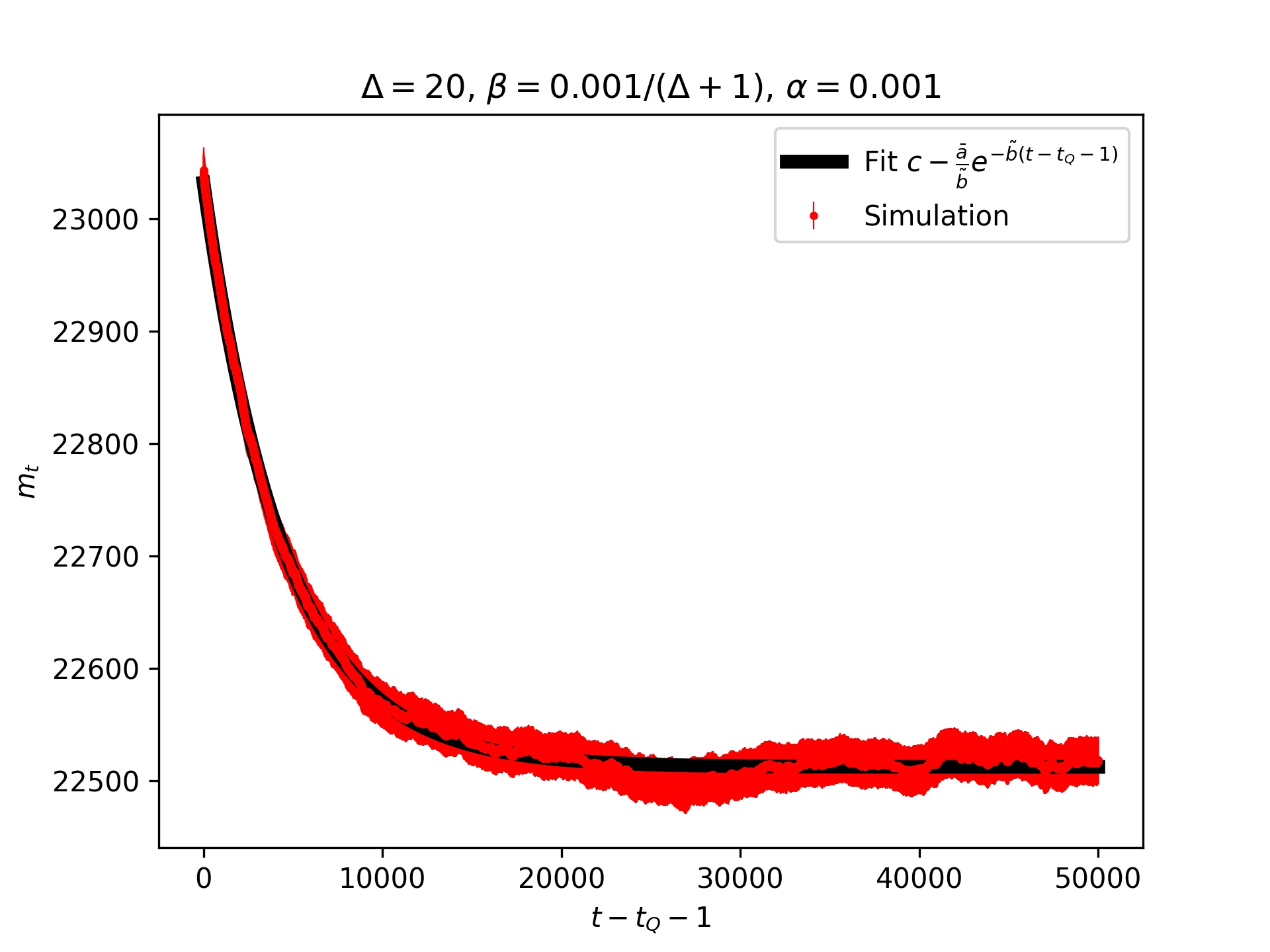}
\includegraphics[width=0.5\linewidth]{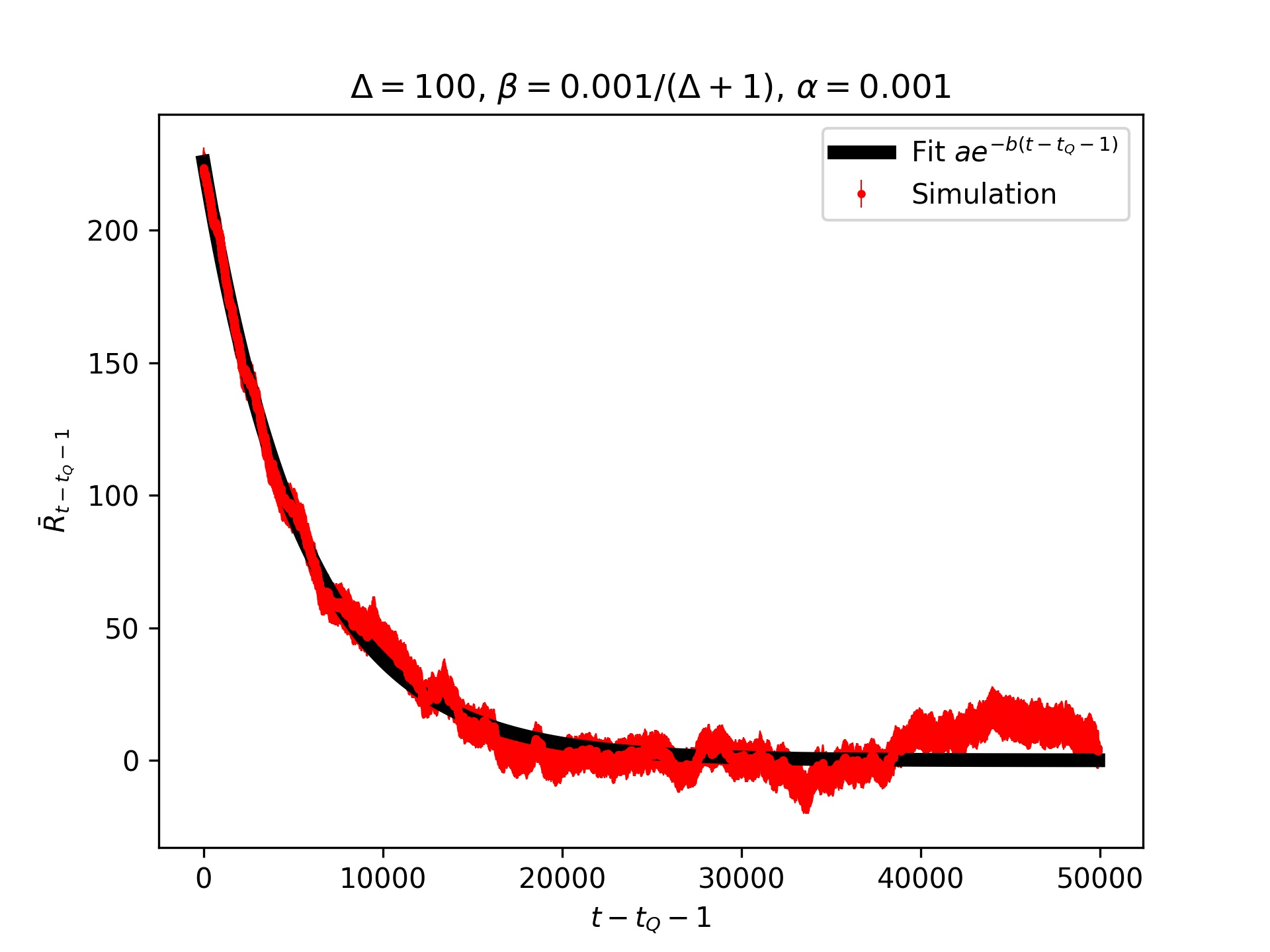}
\includegraphics[width=0.5\linewidth]{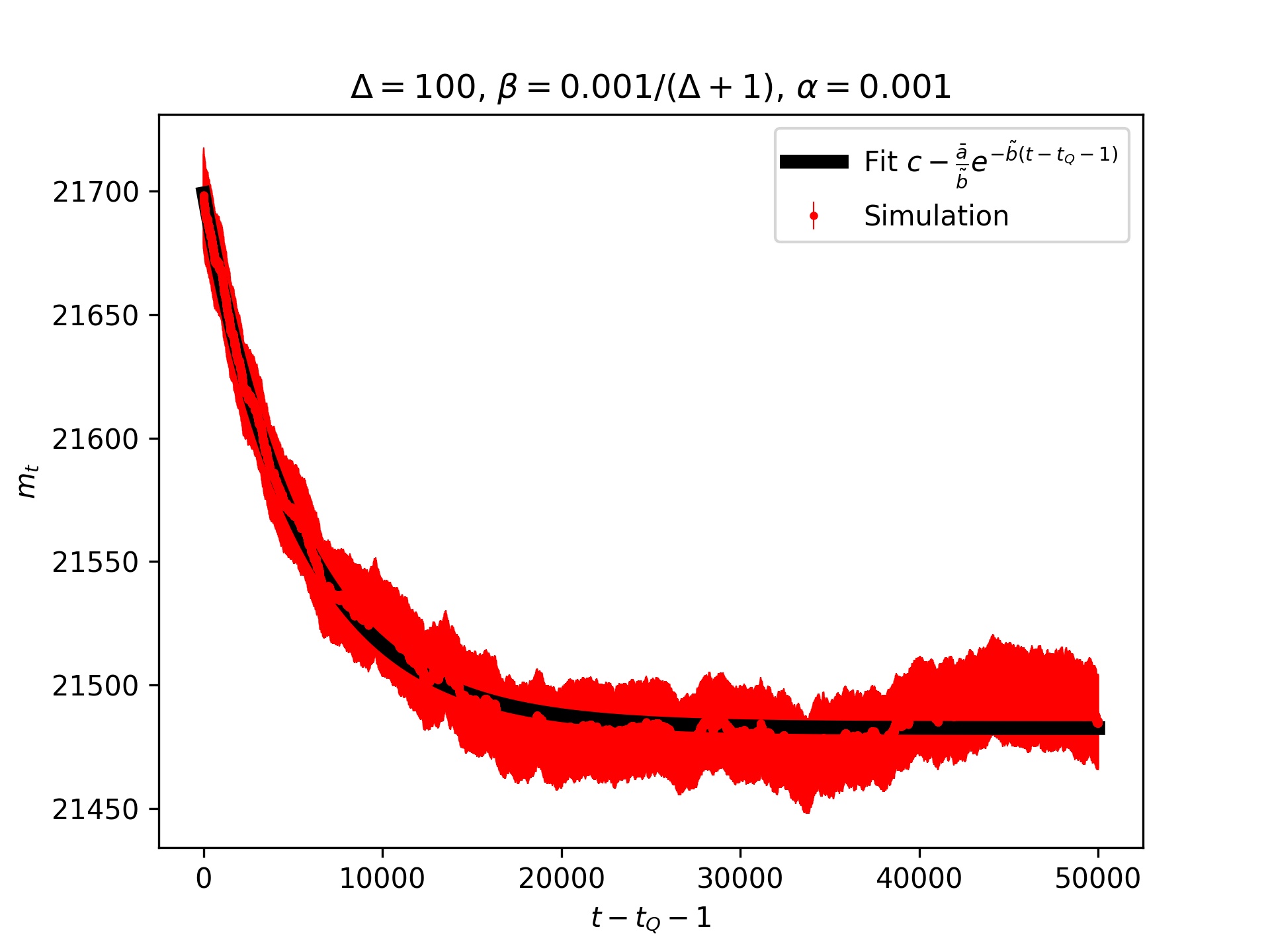}
    \caption{NMZI model. Evolution in time of $\bar{R}_t$ and the mid-price after the execution of the metaorder ends. $\Delta = 20$: $a  = (675.95 \pm 0.27)$ ticks,
$b = 0.000199 \pm 0.00001$, $\bar{a} = (-0.1141\pm 0.001)$ ticks, $\tilde{b} = 0.000219 \pm 0.00001$, $c = ( 22512.12 \pm 0.04)$ ticks, $a(1 - \gamma e^{b}) = -0.1004$ ticks, $m_{t_Q + 1} + \bar{a}/b_2 = 22523.30$ ticks. $\Delta = 100$: $a  = (225.83 \pm 0.20)$ ticks,
$b = 0.000179 \pm 0.00001$, $\bar{a} = (-0.0402\pm 0.001)$ ticks, $\tilde{b} = 0.000186 \pm 0.00001$, $c = (21482.59 \pm 0.04)$ ticks, $a(1 - \gamma e^{b}) = -0.0381$ ticks, $m_{t_Q + 1} + \bar{a}/\tilde{b} = 21481.83$ ticks.}
    \label{fit_after_execution}
\end{figure}

In Fig. \ref{fit_after_execution}, we focus on the case $\beta_2 = \alpha = 0.001$ and two choices of trading intervals i.e. $\Delta = 20, \ 100$. The evolution in time of $\bar{R}_t$ and the mid-price after the execution of the metaorder ends, are shown together with the exponential fits $$\bar{R}_{t - t_Q - 1} = ae^{-b(t - t_Q - 1)} \ \text{and} \ m_{t}= c - \frac{\bar{a}}{\tilde{b}}e^{-\tilde{b}(t - t_Q - 1)}, \ t \geq t_Q + 1.$$ We observe that the exponential fits are suitable to describe the reversion of $\bar{R}_t$ and of the mid-price. Moreover, by considering the estimated parameters that are reported in the figure's caption, we note that the agreement with the theoretical findings related to the equality between the parameters of the fits of the two quantities, qualitatively holds i.e. $b\simeq \tilde{b}$, $\bar{a} \simeq a(1 - \gamma e^b)$ and $c\simeq m_{t_Q + 1} + \bar{a}/\tilde{b}$. 

The decay after the execution of the metaorder is analogous to the reversion we observe between two child MOs, with the difference that the next child MO is posted with an infinite trading interval and the exponential mid-price reversion cannot be anymore approximated by a straight line. Therefore, the role played by the parameters $\beta_2$ and $\alpha$ in determining the post-execution decay  is equivalent to what happens for the reversion between two child MOs. In the following, we investigate how the permanent impact, the peak impact and the magnitude and the speed of the reversion, vary as functions of $\beta_2$ and $\alpha$. 

As we observe in Subsection \ref{subsec_impact_components}, when traders use a longer memory in the evaluation of the price trend i.e. smaller $\beta_2$, the entity of the reversion increases and the concavity in the mid-price trajectory is more pronounced. Similarly, after the end of the execution of the metaorder, the mid-price slowly reverts to its initial value if the price trend indicator $\bar{R}_t$ is computed with a memory that is extremely large. This can be seen by comparing Fig. \ref{fig_mid_price_paths_trading_interval_mZI} and Fig. \ref{fig_mid_price_paths_trading_interval_mZI_beta1e-6} in Appendix \ref{appendix_ModifiedZI}. Both represent the mid-price trajectory for the execution of a metaorder with trading interval $\Delta = 20$, volume $Q=2,000$ (left) and $Q = 100$ (right) however, in the former figure it is $\beta_2 = 10^{-3}$, while in the latter it is $\beta_2 = 10^{-6}$. As expected, when the memory of $\bar{R}_t$ increases, the limiting value of the price $c$ approaches the mid-price before the execution starts and therefore, the permanent impact tends to $0$. Analogously, given that the reversion is stronger and the overall impact of each child MO is weaker when $\beta_2$ is smaller, the peak impact is smaller if the memory of the price trend indicator $\bar{R}_t$ is larger. This is further highlighted in the left panel of Fig. \ref{compare_midprice_beta20.001} in Appendix \ref{appendix_ModifiedZI}, where we plot the evolution of the mid-price before, during and after the execution of a metaorder of volume $Q = 100$, with $\Delta = 20$, $\alpha = 10^{-3}$ and several values of $\beta_2$.

Also the behavior of the permanent impact and the peak impact as a function of $\alpha$ can be explained by the behavior of the reversion. When the intensity of reaction of the traders to the price trends, i.e. $\alpha$, is weaker, so does the reversion mechanism. This implies that both the permanent impact and the maximum impact increase, as represented in the right panel of Fig. \ref{compare_midprice_beta20.001} in Appendix \ref{appendix_ModifiedZI} where we plot the evolution of the mid-price before, after and during the execution of a meta-order of volume $Q = 100$, with $\Delta = 20$, $\beta_2 = 10^{-3}$ and several values of $\alpha$. 

Moreover Fig. \ref{fig_heatmaps_after_execution} in Appendix \ref{appendix_ModifiedZI} sheds light on the role played by both $\beta_2$ and $\alpha$ in the decay after the ending of the execution. This figure presents heat-maps representing  $-\bar{a}/\tilde{b}$ and $\log(2)/\tilde{b}$ as functions of $\beta_2$ and $\alpha$ for $\Delta = 20, 100$. 
The ratio $\bar{a}/\tilde{b}$ is the amount of the price reversion, whereas $\log(2)/\tilde{b}$ is a measure of the decay speed of the mid-price. As expected, when $\beta_2$ decreases with $\alpha$ fixed, $-\bar{a}/\tilde{b}$ increases, i.e. the magnitude of the reversion after the execution ends, is larger when the memory of $\bar{R}_t$ is longer.  
Moreover, when $\beta_2$ ($\alpha$) decreases with $\alpha$ ($\beta_2$) fixed, $\log(2)/\tilde{b}$ increases: the time to revert to a constant value is larger for $\alpha$ and $\beta_2$ smaller. This is consistent with what we have previously noted concerning the role of the two parameters in the reversion between two child MOs: as $\alpha$ or $\beta_2$ decreases, the transition to the stationary regime is slower.

Additionally, we consider the role of the metaorder size $Q$ in determining the speed of impact decay. By fitting the mid-price trajectories  after the execution in Fig. \ref{fig_mid_price_paths_trading_interval_mZI} and \ref{fig_mid_price_paths_trading_interval_mZI_beta1e-6}, we observe that the speed of decay $\tilde{b} (\simeq b)$  is approximately independent of $Q$ and of whether the stationary regime is reached during the metaorder execution:
\begin{itemize}
    \item  $\beta_2 = 10^{-3}$ (Fig. \ref{fig_mid_price_paths_trading_interval_mZI}): if $Q = 100$, $\tilde{b} = 0.000228 \pm 0.000001$ and if $Q = 2,000$, $\tilde{b} = 0.000219 \pm 0.00001$; 
    \item  $\beta_2 = 10^{-6}$ (Fig. \ref{fig_mid_price_paths_trading_interval_mZI_beta1e-6}): if $Q = 100$, $\tilde{b} = 0.000157 \pm 0.000001$ and if $Q = 2,000$, $\tilde{b} = 0.000168 \pm 0.000001 $.
\end{itemize}

We also note that the entity of the reversion after the execution, with respect to the peak impact, decreases as $Q$ increases for given $\beta_2$ and $\alpha$. Indeed, if we consider $\beta_2 = \alpha = 10^{-3}$, $\Delta = 20$ (as in Fig. \ref{fig_mid_price_paths_trading_interval_mZI}), and several values of the metaorder size i.e. $Q = [10, 100, 10^3, 10^4]$, the reversion amounts at $79.13\%$, $73.56\%$, $40.24\%$, $7.28\%$ of the peak impact, where both the peak impact and the price level at the end of the reversion ($c$ in the exponential fit of the mid-price decay) are normalized with respect to the price before the beginning of the execution. Therefore, if the memory that is considered in the evaluation of the past price trend, manages to cover only part of the dynamics during the execution i.e. $Q \gg 1/\beta_2$, the reaction mechanism to the price trend is weaker. This is coherent with our previous finding that, for a given number of total shares, considering a longer memory leads to a stronger reversion.

\begin{table}[]
    \centering
    \begin{tabular}{c||c|c|c||c|c|c||c|c|c}
    & $\bar{R}^*$ &$r^*$ & $\tau^{*}$ & $I^I$ & $I^R$ & $I^I - I^R$ & peak & speed of & permanent \\
    & & & & & & & impact & reversion & impact \\
    \hline
        $\beta_2 \downarrow$ & $\uparrow$ & $\downarrow$ & $\uparrow$ & $\simeq$ & $\uparrow$  & $\downarrow$ & $\downarrow$& $\downarrow$ & $\downarrow$\\
         $\alpha \downarrow $ & $\uparrow$ & $\uparrow$ & $\uparrow$ & $\simeq$ & $\downarrow$  & $\uparrow$ & $\uparrow$ & $\downarrow$ & $\uparrow$\\
    \end{tabular}
    \caption{NMZI model. Summary of the role played by $\beta_2$ and $\alpha$. The meaning of the symbols $\uparrow$, $\downarrow$ and $\simeq$ is \textit{increases}, \textit{decreases} and \textit{is approximately constant}. The double vertical lines separate three blocks: the first refers to the stationary regime, the second to the impact components and the third to the reversion after the execution ends.}
    \label{tab_role_params}
\end{table}

To summarize our findings, Table \ref{tab_role_params} shows the considerations we made throughout the paper about the role played by the parameters $\beta_2$ and $\alpha$ in our model. The analogy between the behavior of the reversion between two child MOs and of the decay after the end of the execution is highlighted.

\section{Conclusion}\label{section_conclusion}
In this work we aim to model and reproduce the behavior of the price impact when a metaorder is executed, by introducing a novel statistical model of the limit order book, the Non-Markovian Zero Intelligence. As its name suggests, it is the non-Markovian counterpart of the pre-existing Zero Intelligence model \cite{daniels_2003, smith_2003, farmer_2005}. In the original model, limit orders, market orders and cancellations in the LOB are modeled as three independent Poisson processes. Despite its simplicity, this model reproduces well several stylized facts, but it misses to capture the well-known concavity of the market impact, which is a desirable feature for a simulator, since it allows to test trading strategies in a realistic framework. In order to overcome this limitation, we propose our non-Markovian variant: contrary to the ZI model which considers equally probable sell and buy LOs, we define the probability that a LO is a sell, as price path dependent. Financially, this can be interpreted as traders who have reservation prices and submit LOs in response to the price trend. This apparently small modification leads to substantial accomplishments in terms of the price impact. During the execution of a metaorder, the concavity of the market impact emerges. Indeed, two regimes can be distinguished: initially, the price path is concave then, after a large number of child MOs, it stabilizes to a stationary regime which is characterized by a linear impact. In the first regime, the concavity originates because of the interplay between two mechanisms. For a given child MO of the metaorder e.g. buy, there is an immediate impact which makes the mid-price increase and more sell LOs are triggered. Consequently, more sell LOs in the spread occur and so, the mid-price decreases. This causes the price reversion in the intervals between two child MOs and this reversion results to be increasing with time. In parallel, also the immediate impact component varies in time since so does the imbalance between sell and buy LOs, making the first gap size in the ask side change. However, we observe that this variation plays a marginal role: the pivotal contribution is due to the reversion mechanism and overall, the impact proves to be decreasing in time, causing the concavity. Finally, after the execution of the metaorder, the price path exponentially decreases. 

In addition to analyze in depth the mechanisms at the root of the price impact in our model, we investigate how the results depend on the two main parameters. The former represents the inverse of the memory to evaluate the price trend which enters the probability that a LO is a sell. If this parameter is infinite, our model collapses to the ZI. We observe that if the memory is longer, the concave regime is longer, the convergence to the stationary regime is slower, the reversion mechanism is stronger and the permanent impact is smaller. The second parameter also enters the definition of the probability that a LO is a sell but it represents the intensity of the reaction to the price trend. The case with this parameter equal to $0$ corresponds to the ZI model. Our results point out that if the reaction intensity is stronger, then the concave regime is shorter, the convergence to the stationary regime is faster, the reversion mechanism is stronger and the permanent impact is smaller. To sum up, the shape of the market impact during and after the execution of a metaorder is well-captured when the decisions to submit LOs vary in time as a response to the price trend that is evaluated with a long memory.

The ability of our model to capture the market impact's concavity and the price reversion after the execution of a metaorder ends, is accompanied by three other bright sides, which make it an innovative and relevant contribution. This model works for small-tick assets: the modeling of their dynamics is under explored in the literature, given its challenging nature. Also, this model is completely explainable and even though it is not fully analytical, several relations can still be derived. For instance, we derive master equations which describe the evolution of the spread and the mid-price in the intervals between two child MOs. 

Several extensions could be interesting to study in future works. In this paper, we focus on the market impact associated with the executions of metaorders with constant speed but undoubtedly, more complex trading strategies could be investigated. Moreover, we consider a specific choice of our indicator of the past price trend $\bar{R}_t$, that is the exponentially weighted mid-price return. However, instead of the exponential damping factor, different kernels, as a power-law function, could be employed. Similarly, we define the time-varying probability that a LO is a sell as a sigmoid function of $\bar{R}_t$ but, of course, other functional forms could be considered. Finally, similarly to what we do for LOs, also the cancellation signs' sampling could be performed via a time-varying probability that depends on the past price trend.

\section*{Acknowledgements}
\noindent 
FL  acknowledges support from the grant PRIN2022 DD N. 104 of February 2, 2022 ”Liquidity and systemic risks in centralized and decentralized markets”, codice proposta 20227TCX5W - CUP J53D23004130006 funded by the European Union NextGenerationEU through the Piano Nazionale di Ripresa e Resilienza (PNRR).\\

\newpage
\appendix
\section{LOBSTER data }\label{appendix_lobster}
As mentioned in the main text, the parameters $\lambda, \mu, \delta, q_0$, which enter the Zero Intelligence model and the variant we propose, are estimated by relying on LOBSTER data. In this Appendix we provide information about the format of these data and the pre-processing steps we perform.

Given a stock, for each day, two files of equal length are available: the message and the order book files. Row $j$ of the former contains details about the event which leads to the new order book state that is reported in row $j$ of the order book file. In particular, the details about each event are: time (with nanosecond precision), type, ID, size, dollar price, direction; each order book state is represented by the first $10$ queues with their prices and volumes.

By referring to \cite{bouchaud_book_2018}, the pre-processing steps we perform are the following:
\begin{itemize}
    \item we clean the data from trading halts;
    \item we remove the opening and closing auctions;
    \item we check whether there are observations for which best ask prices are greater than their corresponding best bid prices. It is not our case otherwise we would have dropped them;
    \item we handle the split executions of limit orders by grouping them;
    \item we drop hidden orders.
\end{itemize}

Additionally, we drop the observations in the first and last hours of trading. This is common practice when working with high-frequency data. Indeed, these two moments of the day are characterized by higher volatility and more intense trading activity.

A \textit{Python} class which allows to load and clean a LOBSTER data set is provided in the Github repository \href{https://github.com/adeleravagnani/non-markovian-zero-intelligence-lob-model}{adeleravagnani/non-markovian-zero-intelligence-lob-model}.

\section{Zero Intelligence model: parameters' estimation and simulation algorithm}\label{sec_parameters_estimation}
In order to estimate the parameters of the ZI model, we follow the procedure outlined in \cite{bouchaud_book_2018}. Let us define $\Omega_{LO}$ as the set of LOs that are submitted at the best quotes or within the spread, $\Omega_{C}$ as the set of cancellations at the best quotes, $\Omega_{MO}$ as the set of MOs. We also define: $N_{LO}=|\Omega_{LO}|$, $N_{MO}=|\Omega_{MO}|$, $N_{C} = |\Omega_{C}|$, $N = N_{LO} + N_{MO} + N_{C}$. LOs and cancellations at deeper levels are not taken into consideration in the parameters' estimation since the events at the best quotes and within the spread are the most relevant for the price dynamics.

The equations for parameters' estimation are the following \cite{bouchaud_book_2018}:
\begin{itemize}
    \item order size $q_0$:
    \begin{equation}
    q_0 = \frac{1}{N_{LO}}\sum_{x \in \Omega_{LO}} q_x
\end{equation}
where $q_x$ is the size (in number of shares) of order $x$;
\item total MO arrival rate per event $2\mu$:
    \begin{equation}
    2\mu = \frac{1}{N}\sum_{x \in \Omega_{MO}} \frac{q_x}{q_0};
\end{equation}
\item total LO arrival rate per event $\lambda$:
    \begin{equation}
    \lambda = \frac{1}{n_{ls}}\Bigg(\frac{1}{N}\sum_{x \in \Omega_{LO}} \frac{q_x}{q_0}\Bigg) = \frac{1}{2\Bigg(1 + \langle \lfloor \frac{\hat{s}_{LO}}{2}\rfloor\rangle\Bigg)}\Bigg(\frac{1}{N}\sum_{x \in \Omega_{LO}} \frac{q_x}{q_0}\Bigg)
\end{equation}
where $n_{ls}$ is mean number of available price levels inside the spread and at the best quotes and $\hat{s}_{LO}$ is the average spread before LOs;
\item total cancellation rate per unit volume and event $2\delta$:
    \begin{equation}
    2\delta = \frac{1}{N}\sum_{x \in \Omega_{C}} \frac{q_x}{\bar{q}}, \  \ \bar{q} = \frac{\hat{q}_{b_1} + \hat{q}_{a_1}}{2}
\end{equation}
where $\hat{q}_{b_1/a_1}$ is the mean size at the best bid/ask quote.
\end{itemize}

After estimating these parameters, the LOB can be simulated according to Algorithm \ref{alg_ZI}. 

\begin{algorithm}
\caption{Standard/Non-Markovian Zero Intelligence model}
\small
\begin{algorithmic}\label{alg_ZI}
\REQUIRE $\lambda$, $\nu$, $\delta$: LO, MO, cancellation rates\\
\REQUIRE $K$: number of tick levels in the grid which represents the LOB \\
\REQUIRE $iterations_0$: iterations to reach equilibrium \\
\REQUIRE $iterations$: iterations for the simulation \\
\STATE Place a buy order for each price $p_i \leq K/2$ and a sell order for each $p_i > K/2$ ($p_i \in [0, K - 1]$) 
\FOR{$t=1, \ldots, iterations_0 + iterations$}
\STATE Define $n_{orders}(t) ={n_{orders}^{bid}}(t) + {n_{orders}^{ask}}(t)$ as the total number of orders in the LOB
\STATE Compute $(\Lambda, M, \Delta)/\Gamma$ where $\Lambda = \lambda K$, $M = 2\mu$, $\Delta = \delta n_{orders}(t)$, $\Gamma = \Lambda + M + \Delta$
\STATE Draw order type
\IF{LO}
\IF{Standard Zero Intelligence model}
    \STATE Pick sign randomly
\ELSIF{Non-Markovian Zero Intelligence model}
    \STATE Pick sign with probability defined in Eq. \eqref{eq_probability_sellLO}
\ENDIF
\STATE Place the order in a random price level
\ELSIF{MO}
\STATE Pick sign randomly
\STATE Execute the order at the best bid or best ask
\ELSIF{cancellation}
\STATE Pick sign $+1$ with prob ${n_{orders}^{bid}}(t)/{n_{orders}}(t)$ and $-1$ with prob ${n_{orders}^{ask}}(t)/{n_{orders}}(t)$
\STATE Cancel random order in the book
\ENDIF
\STATE Center the LOB around the mid-price
\IF{$t > iterations_0$}
\STATE Save updates
\ENDIF
\ENDFOR
\end{algorithmic}
\end{algorithm}

\section{Additional results related to the Non-Markovian Zero Intelligence model}\label{appendix_ModifiedZI}
\begin{figure}
\centering
    \includegraphics[scale = 0.5]{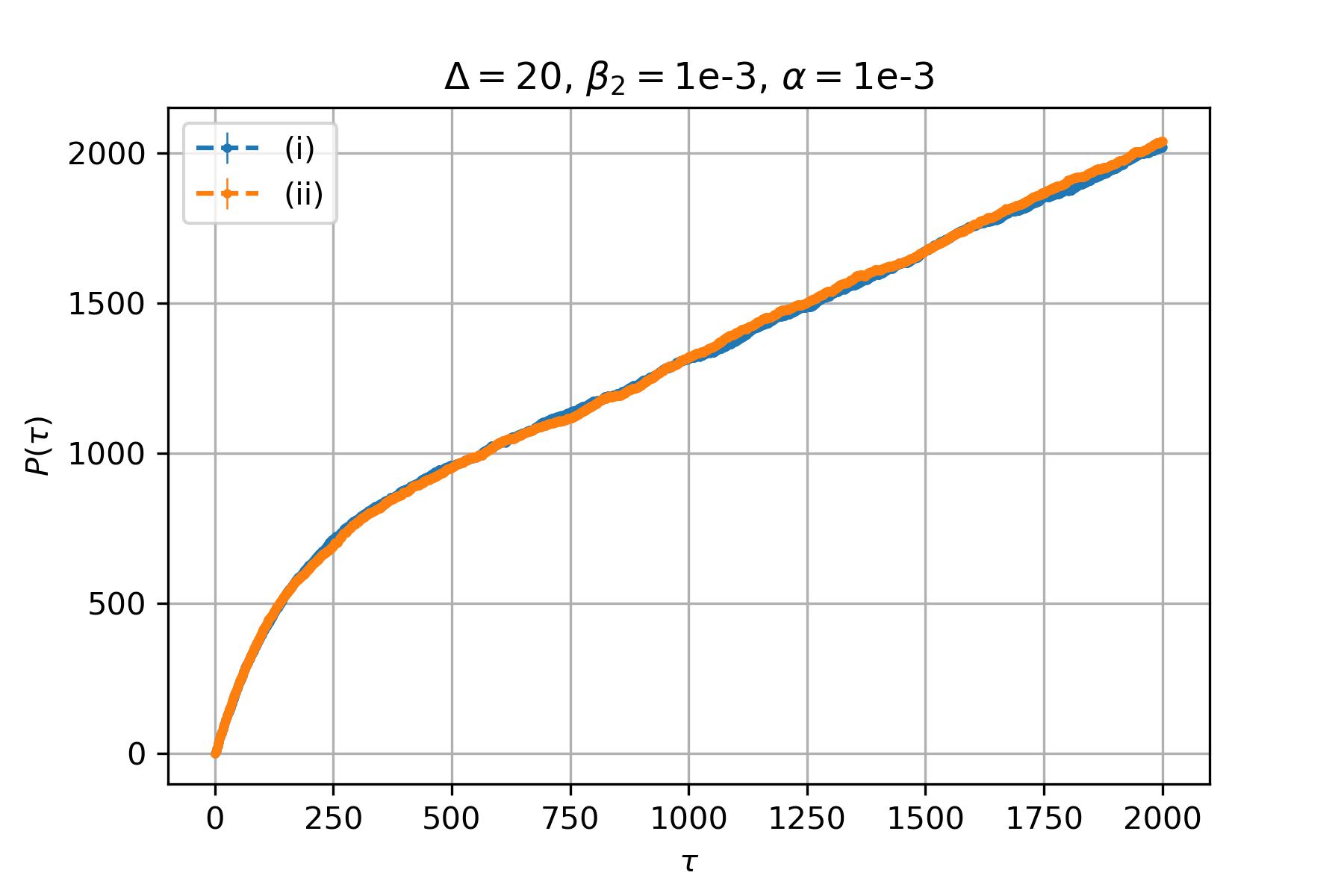}
    \includegraphics[scale = 0.5]{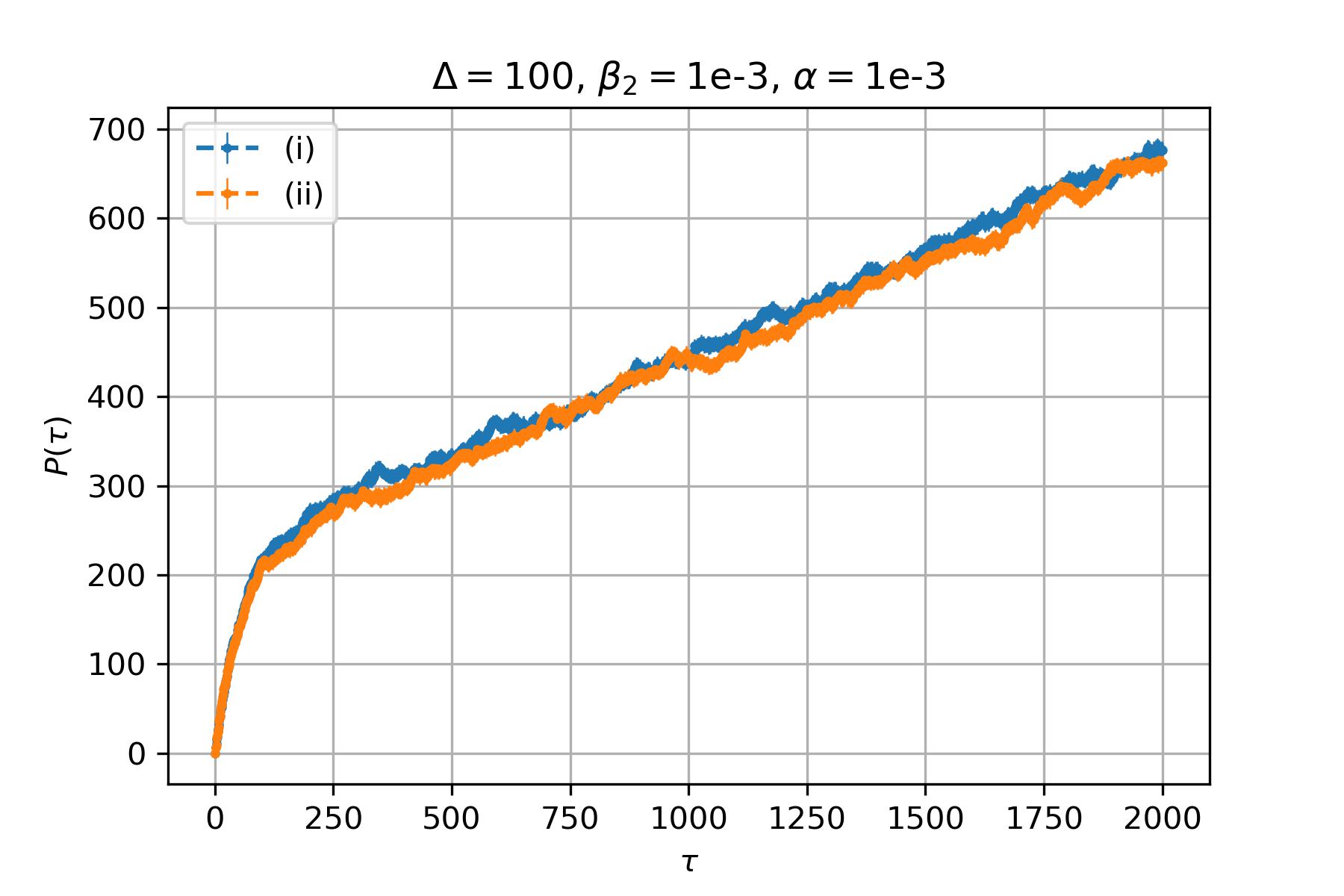}
    \includegraphics[scale = 0.5]{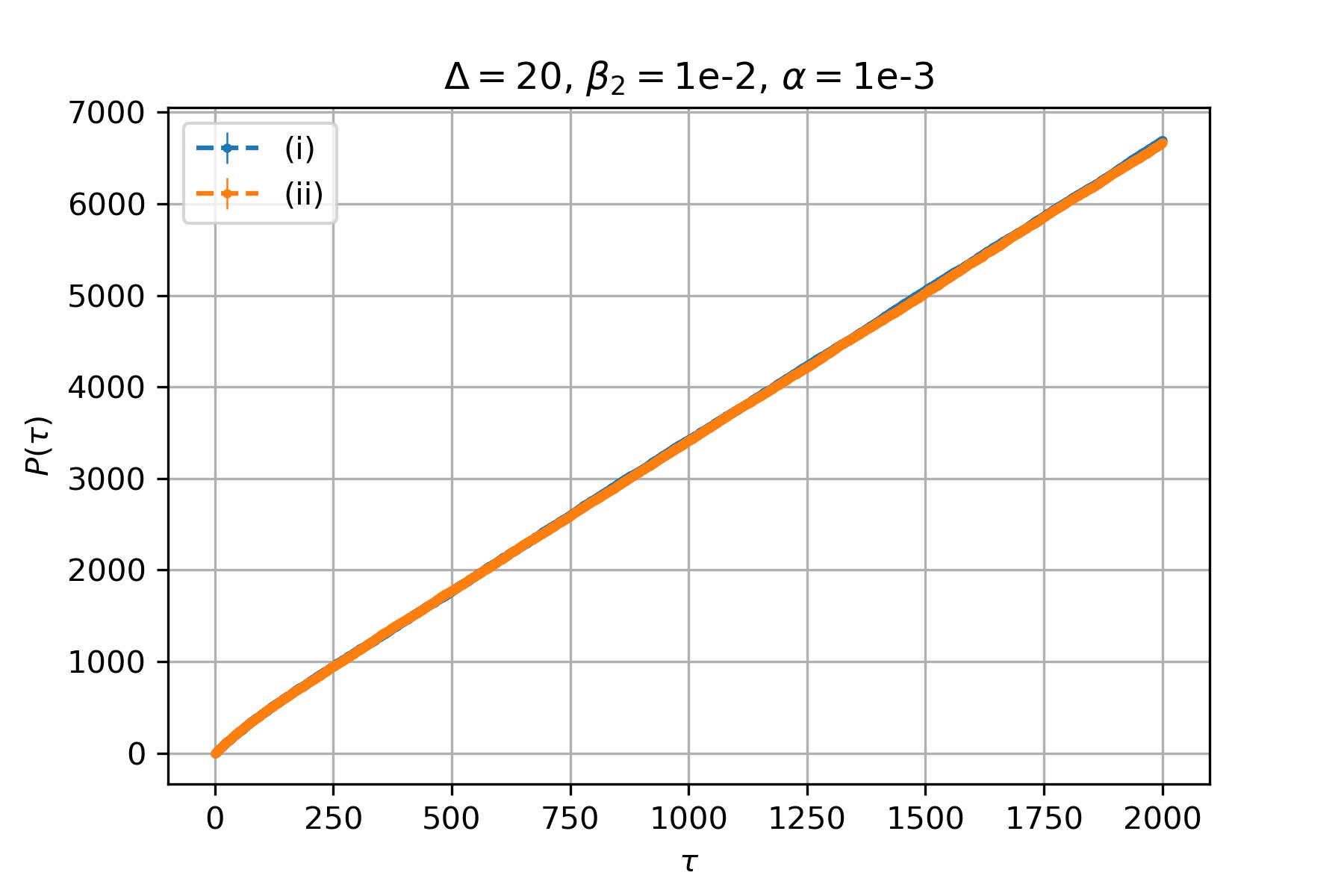}
    \includegraphics[scale = 0.5]{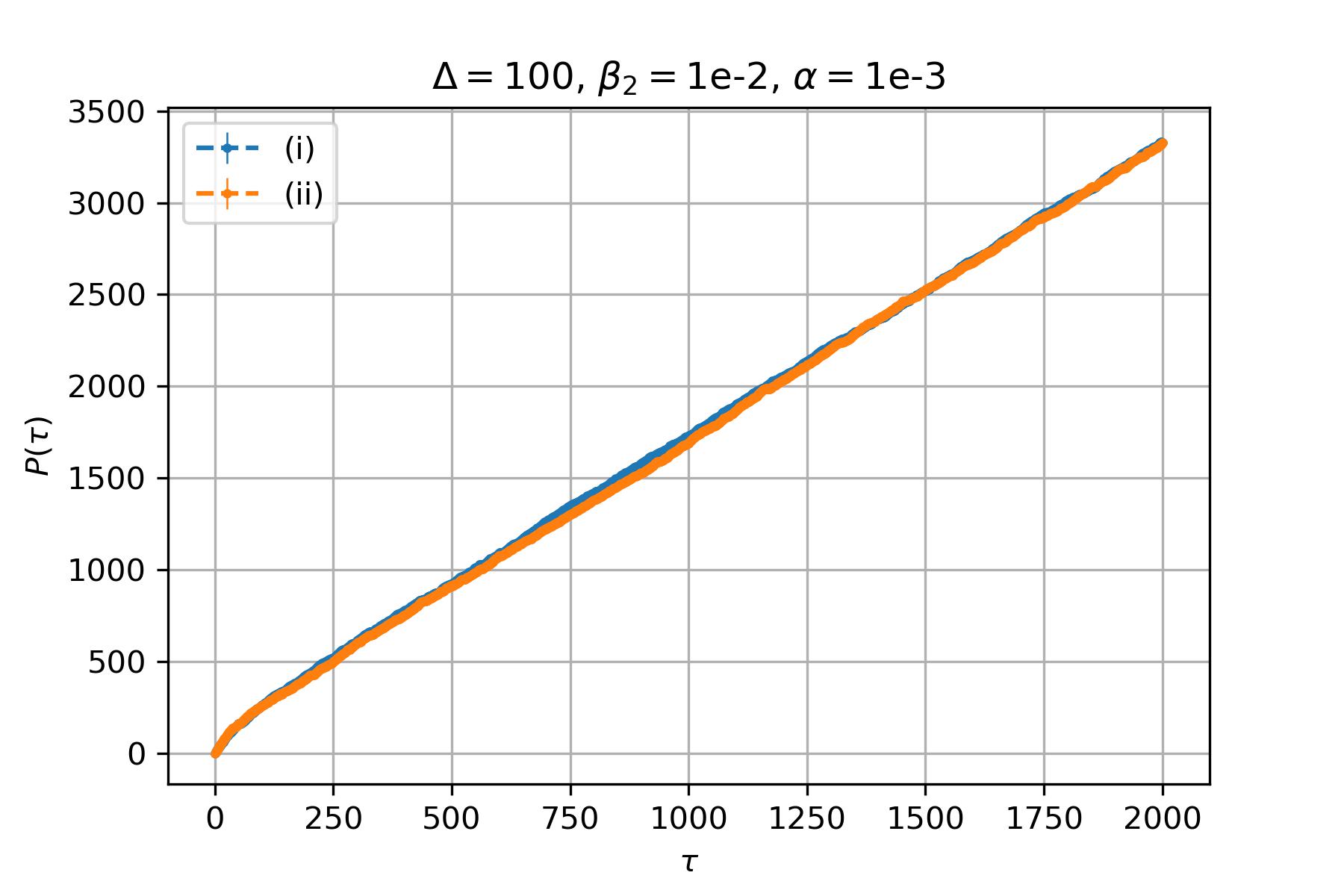}
    \caption{NMZI model. Path of the mid-price difference versus the number of the child MOs of the metaorder. It is defined as $P(\tau) = \langle (m_{\tau} - m_1)\epsilon \rangle_{sim}$ where $\tau=1, \ldots, Q$ and $m_{\tau}$ is the mid-price before the $\tau^{th}$ child MO. The metaorder has total volume $Q= 2,000$ and buy direction. In the legend, (i) and (ii) refer to the descriptions in Eq. \eqref{eq_EWMAreturns_descriptions}.}
    \label{fig_response_mZI}
\end{figure}

\begin{figure}
\includegraphics[scale = 0.5]{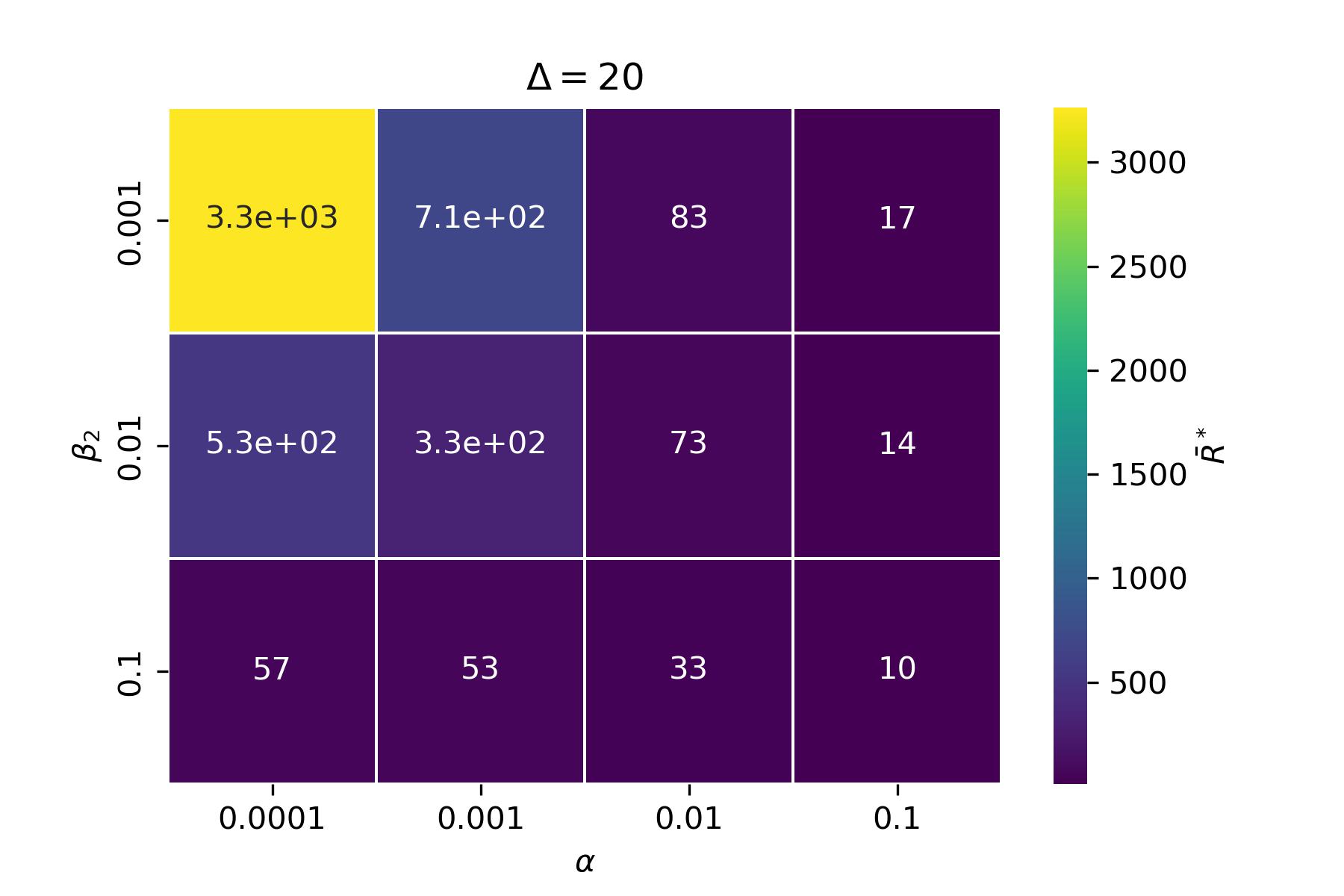}
    \includegraphics[scale = 0.5]{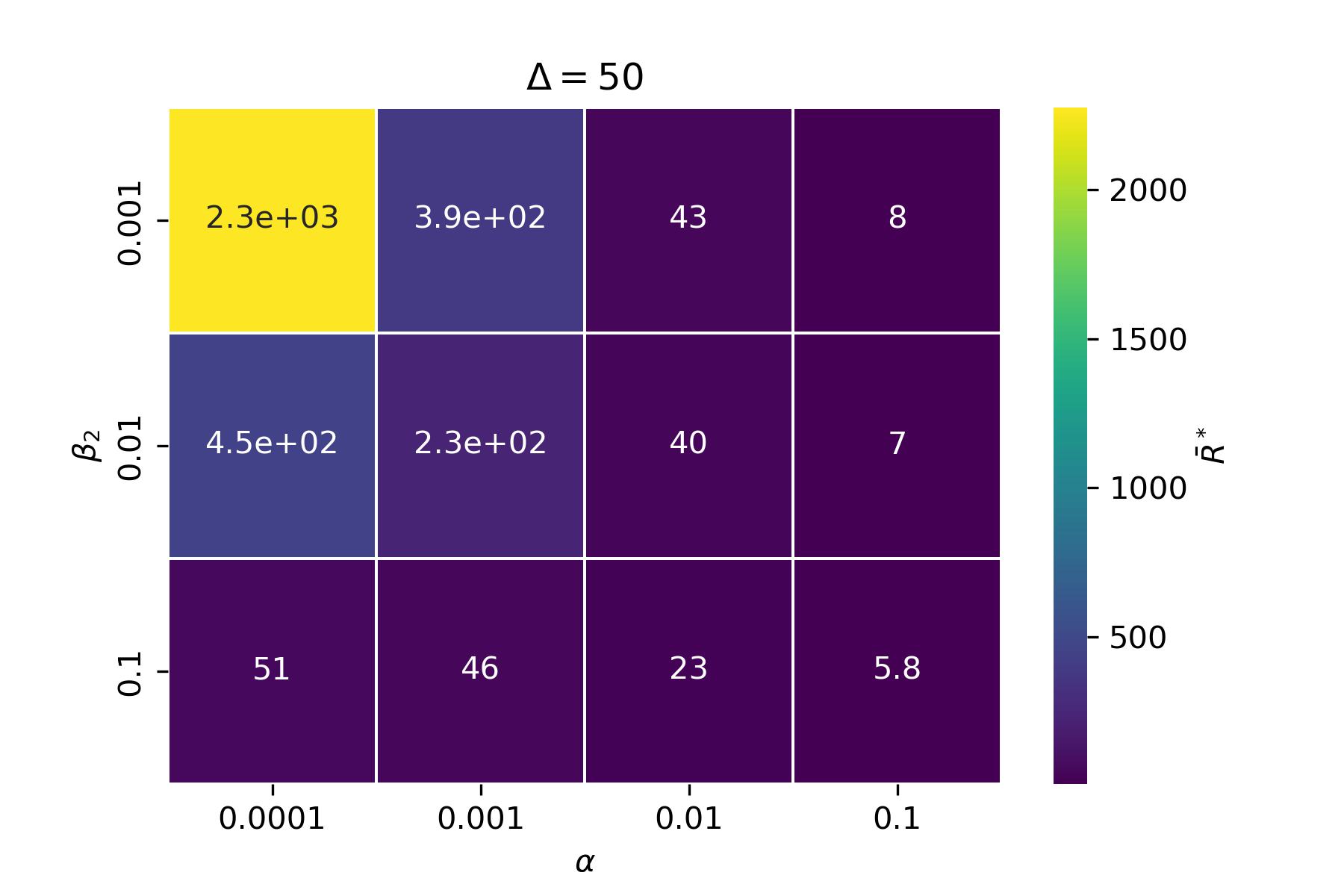}
    \includegraphics[scale = 0.5]{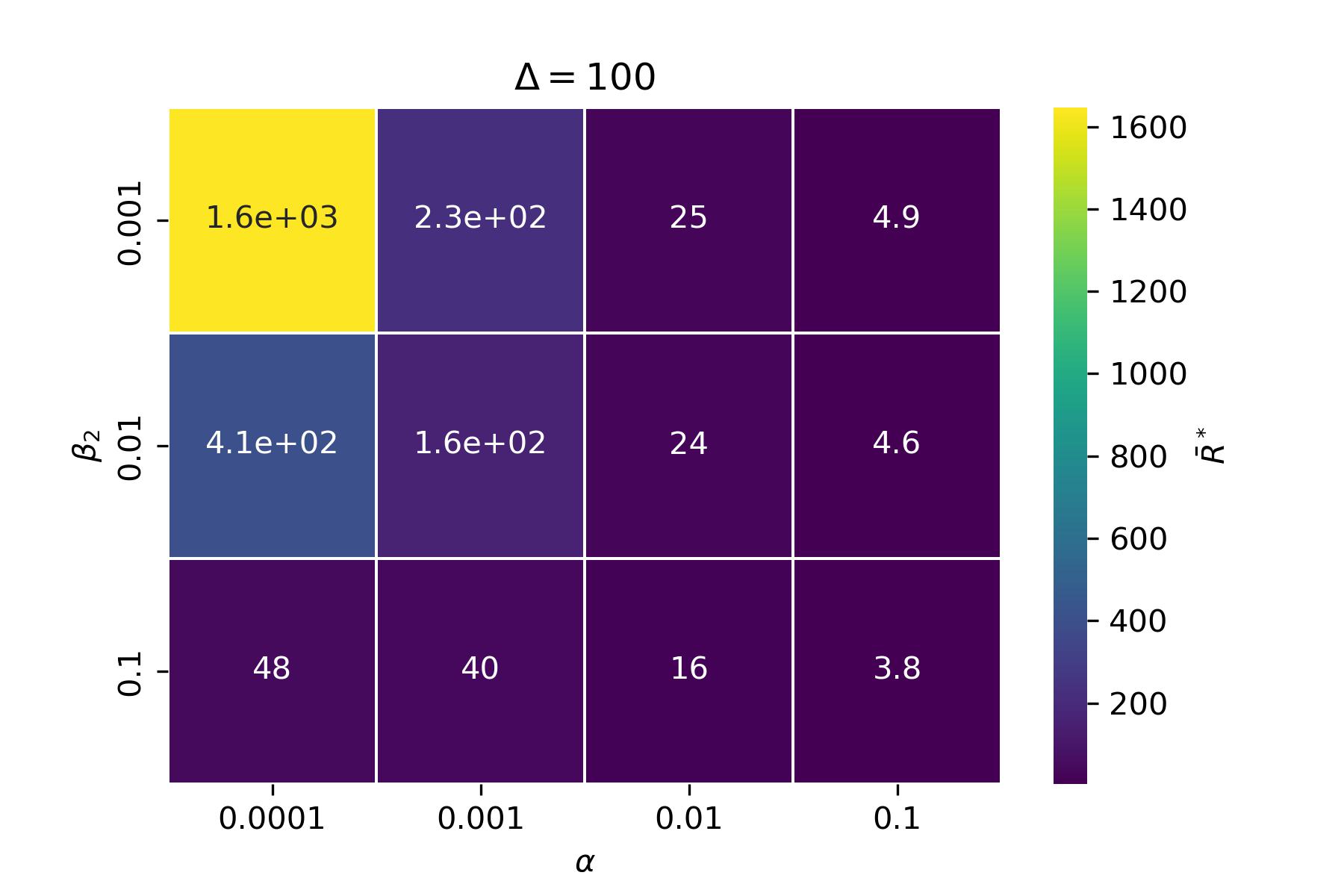}
    \caption{NMZI model. Heat maps representing the stationary value $\bar{R}^*$ when executing a buy metaorder.}
    \label{fig_heatmap_Rstar_mZI}
\end{figure}

\begin{figure}
    \includegraphics[scale = 0.5]{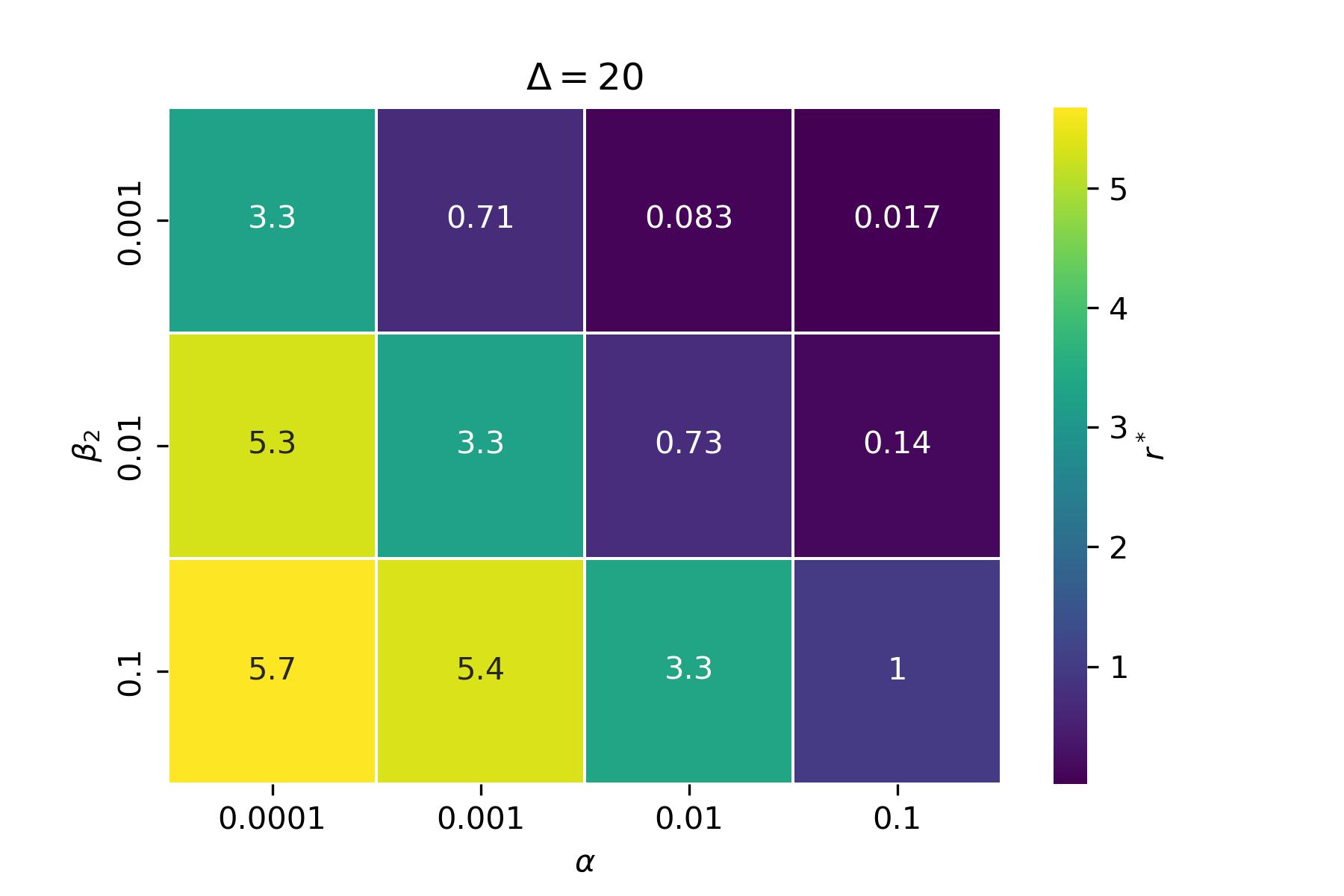}
    \includegraphics[scale = 0.5]{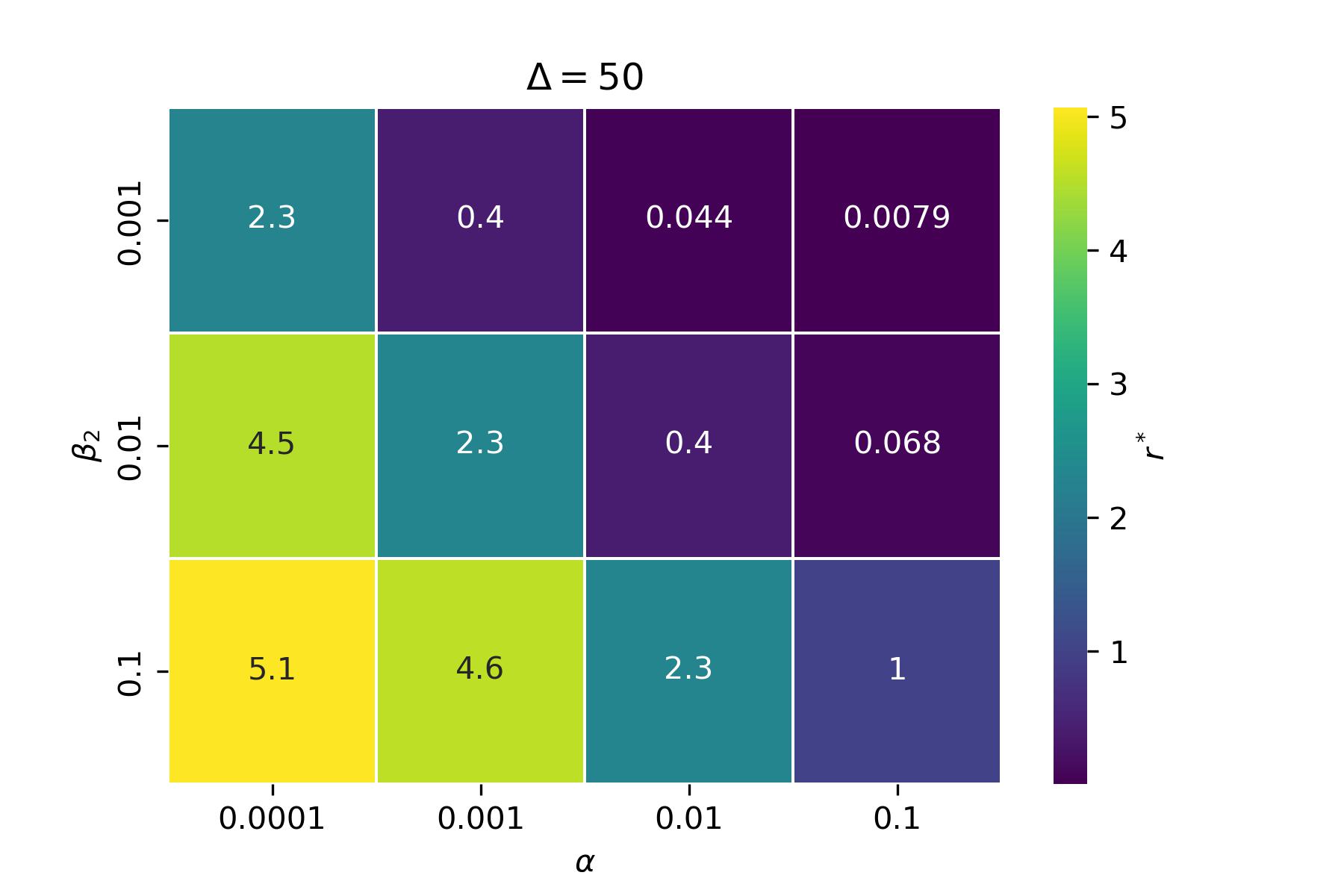}
    \includegraphics[scale = 0.5]{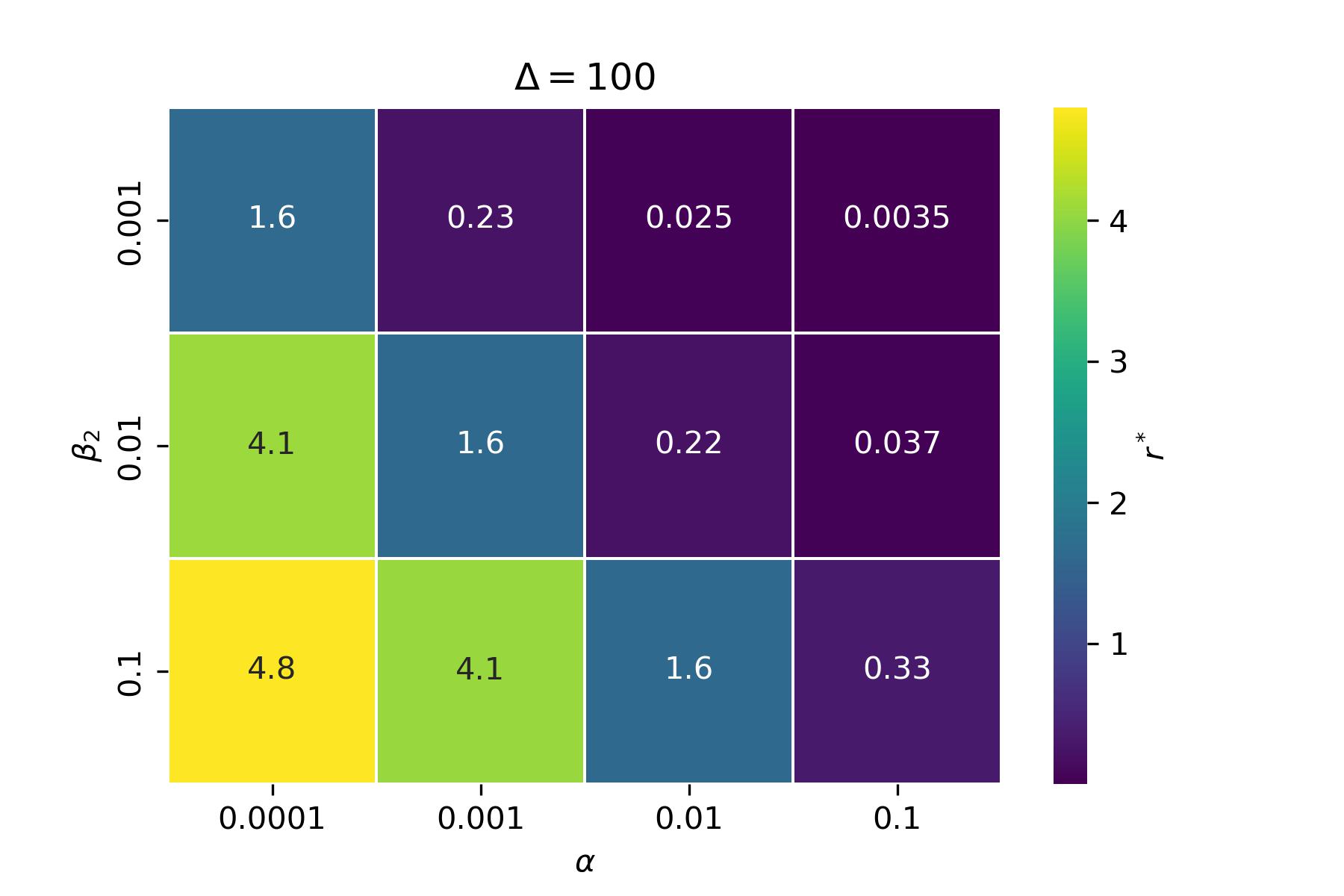}

    \caption{NMZI model. Heat maps representing the stationary value $r^*$ when executing a buy metaorder.}
    \label{fig_heatmap_mathcalRstar_mZI}
\end{figure}

\begin{figure}
    \includegraphics[scale = 0.5]{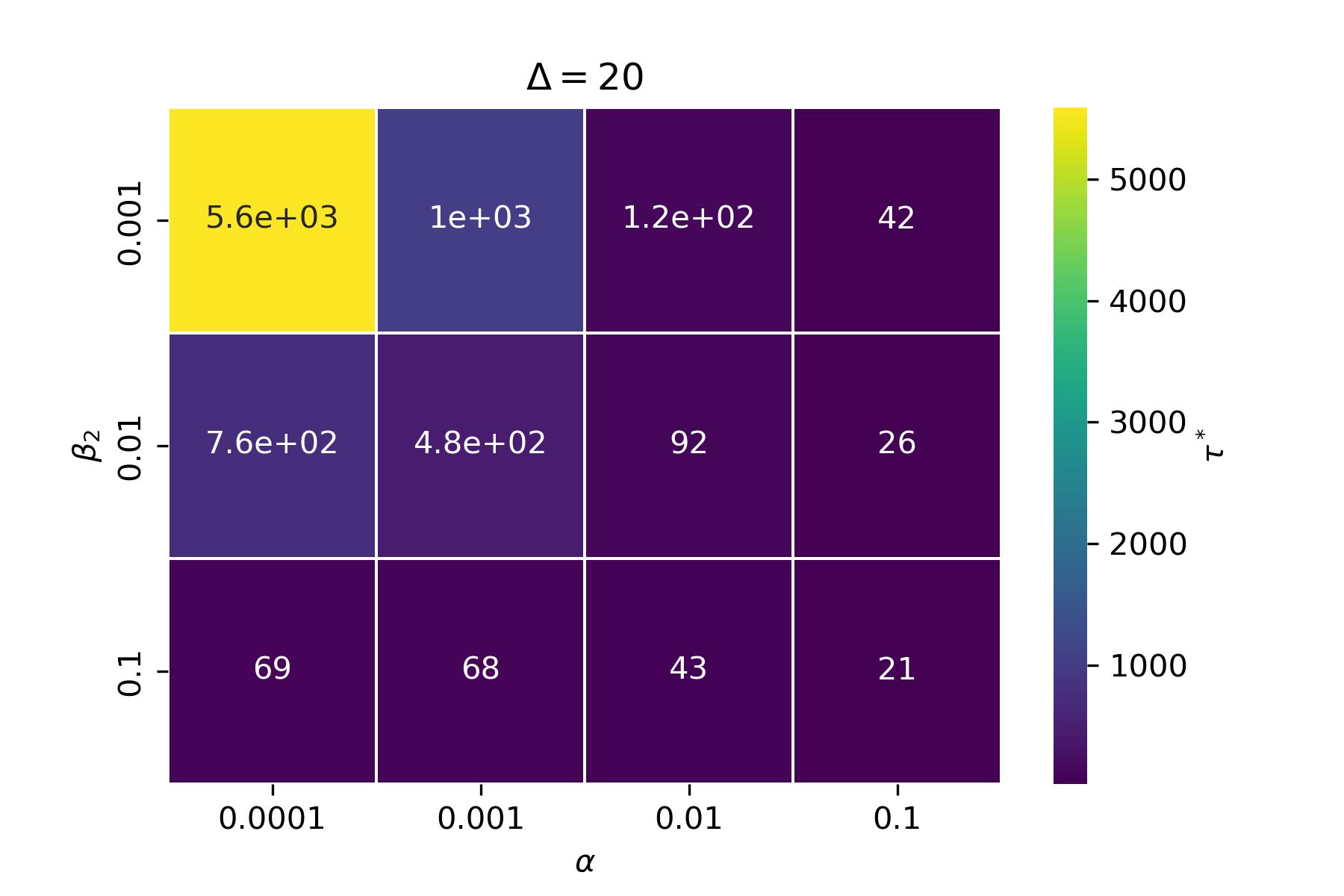}
    \includegraphics[scale = 0.5]{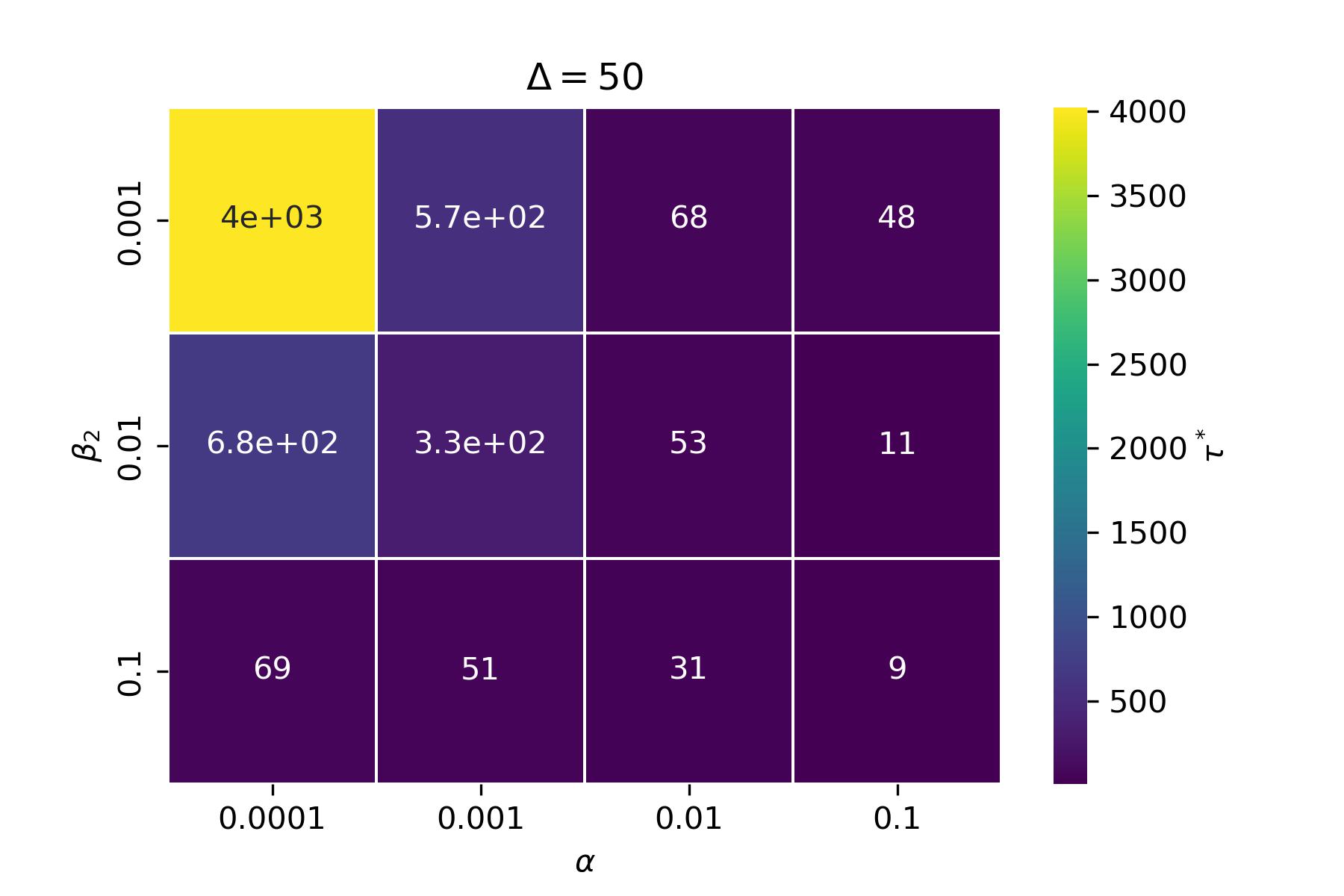}
    \includegraphics[scale = 0.5]{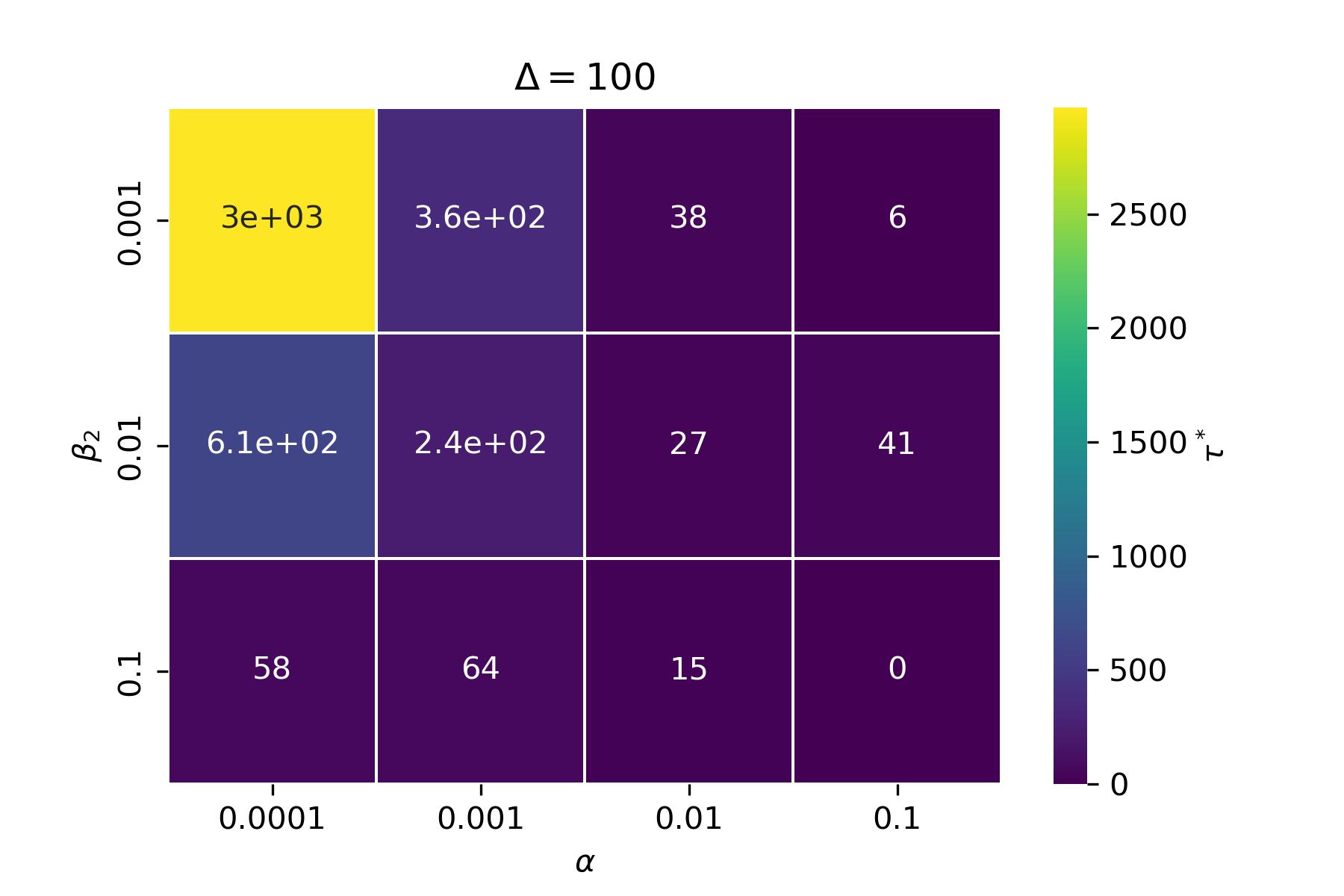}

    \caption{NMZI model. Heat maps representing $\tau^*$ when executing a buy metaorder. $\tau^*$ identifies the beginning of the stationary regime in terms of the number of the child MOs.}
    \label{fig_heatmap_tstar_mZI}
\end{figure}

\begin{figure}
\includegraphics[width=0.5\linewidth]{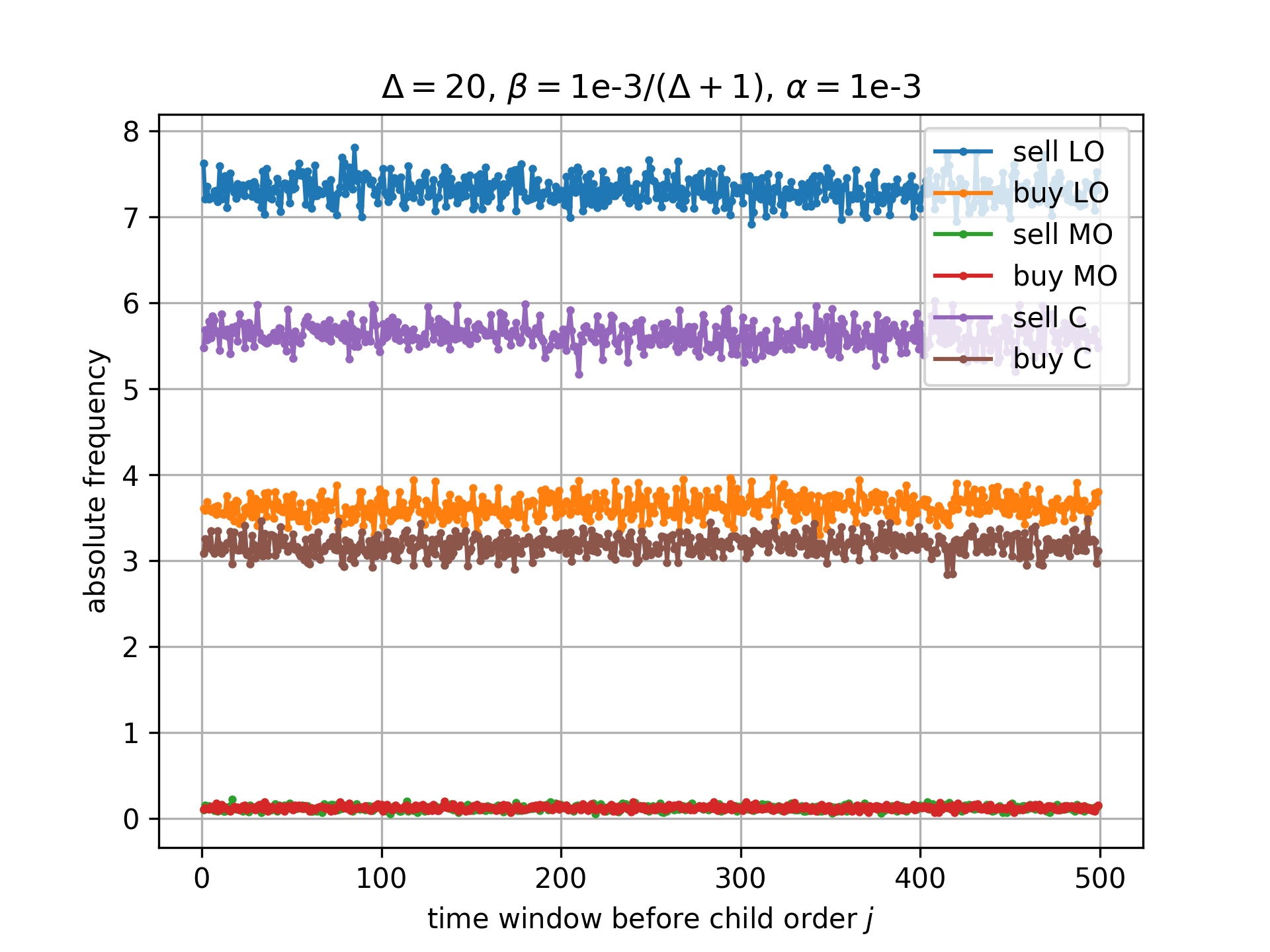}
\includegraphics[width=0.5\linewidth]{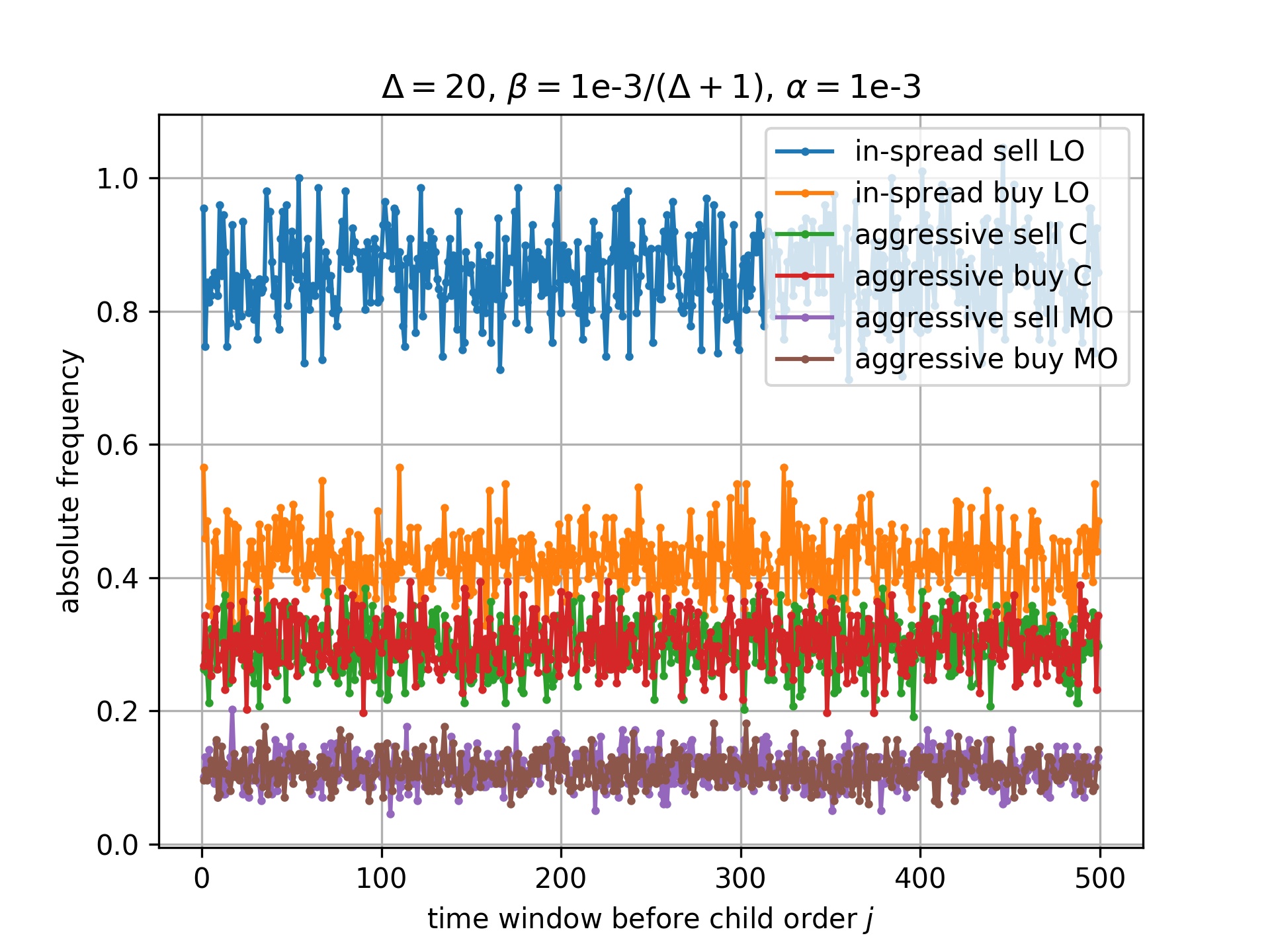}
\includegraphics[width=0.5\linewidth]{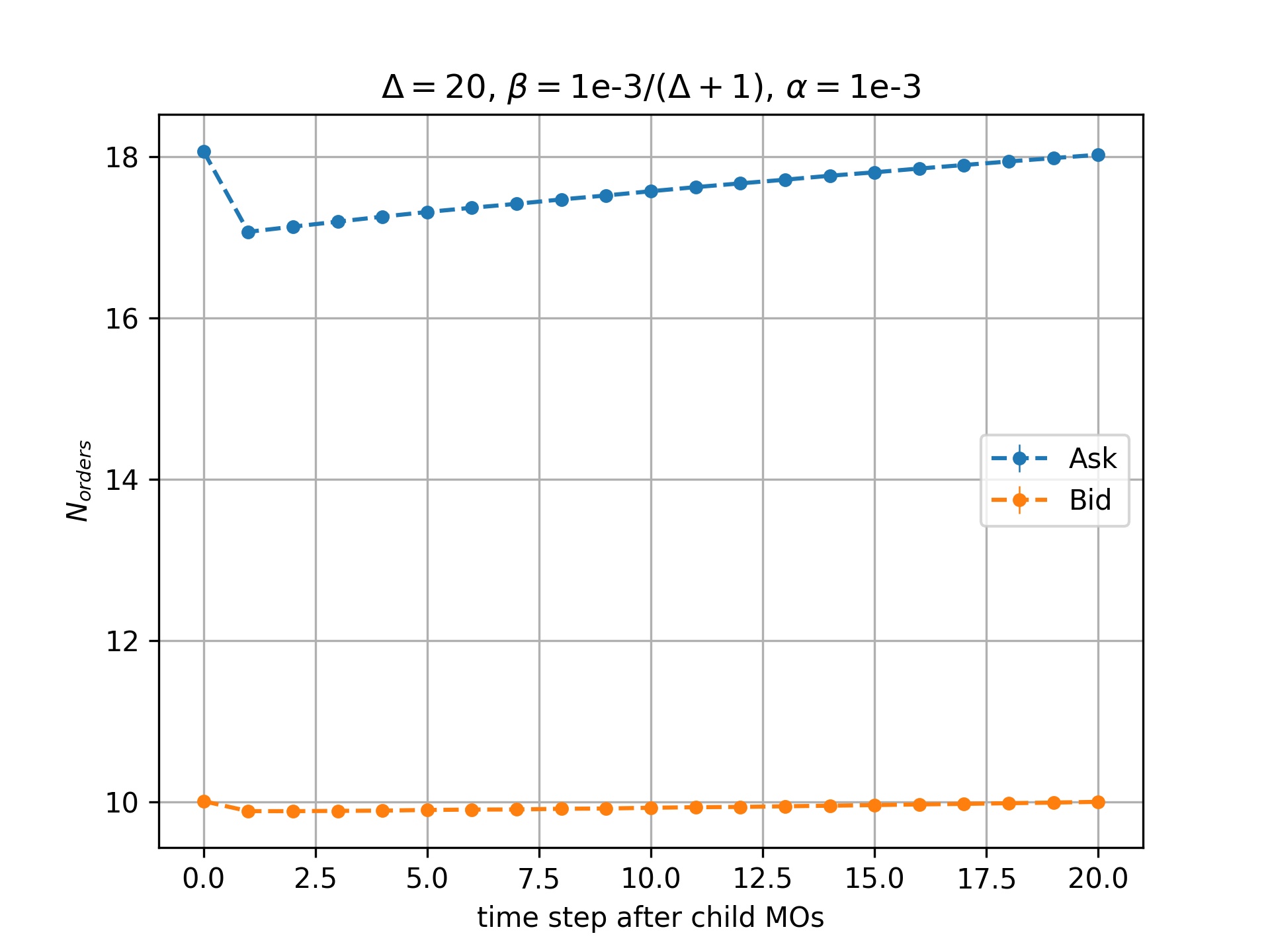}
\includegraphics[width=0.5\linewidth]{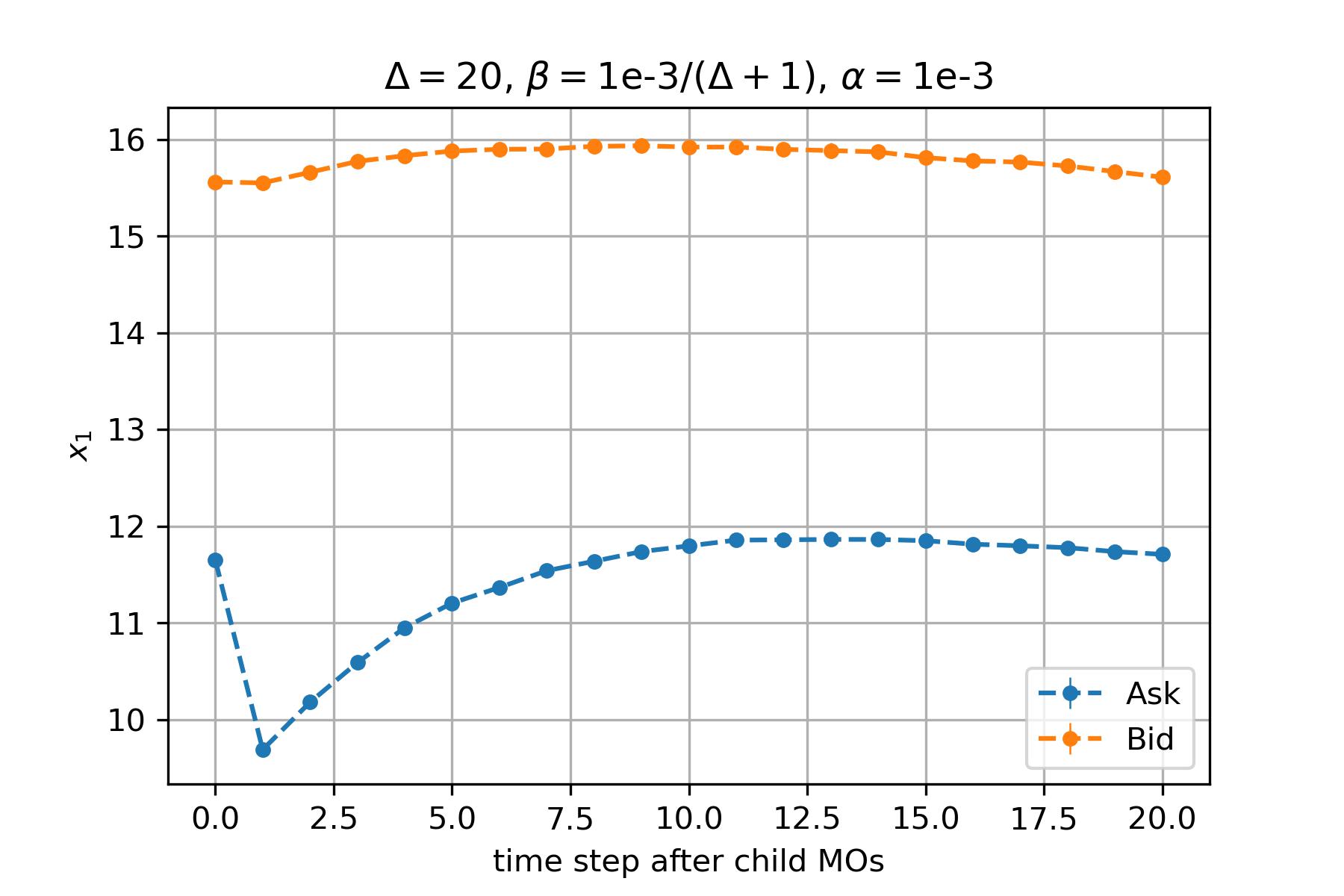}
\includegraphics[width=0.5\linewidth]{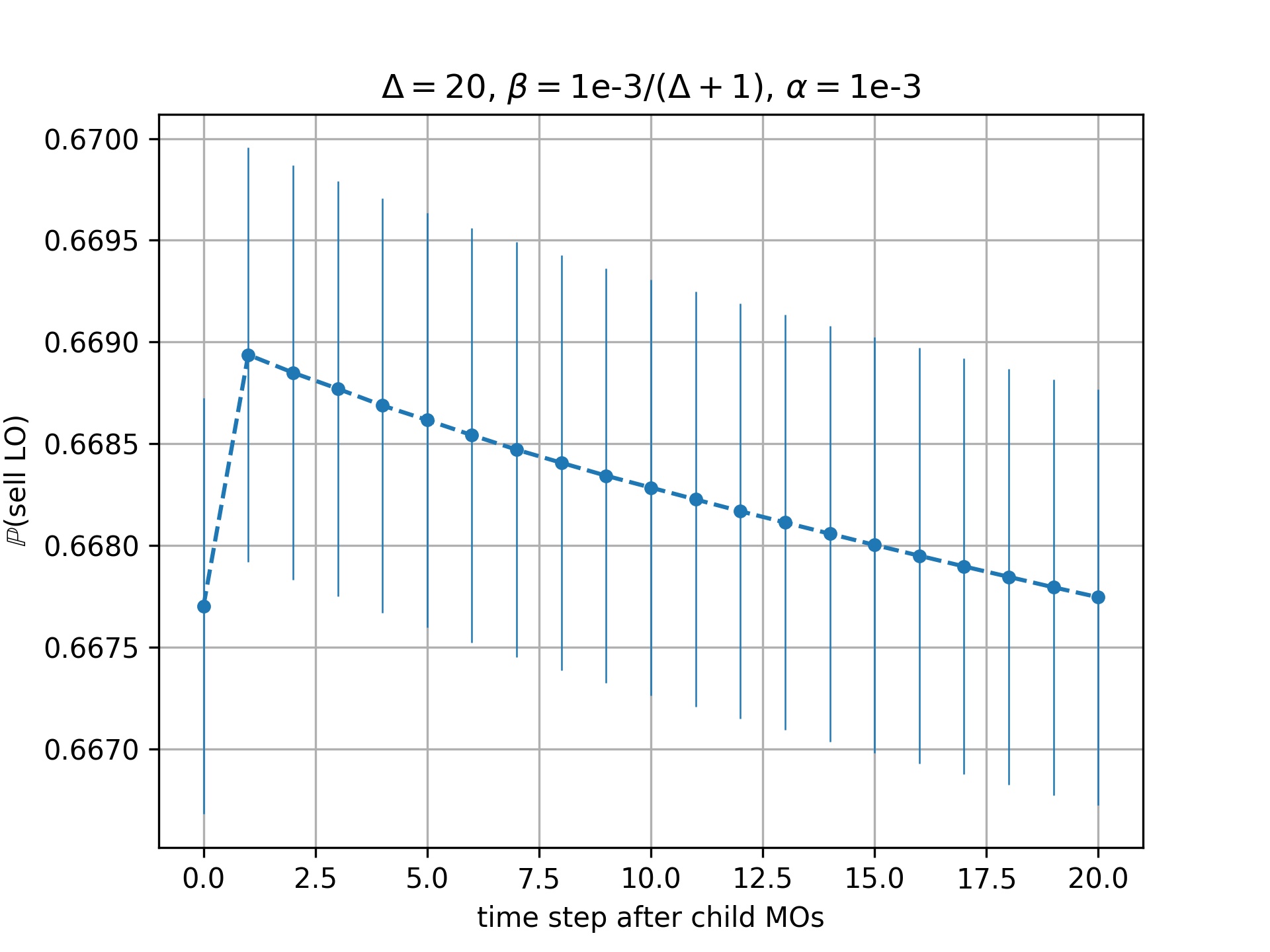}
    \caption{NMZI model, stationary regime. Frequencies of the different types of events in each time window between child MOs (MOs and cancellations are aggressive if they lead to a mid-price change). Number of orders, first gap sizes and probability that a LO is a sell as functions of the time step after child MOs.}
    \label{fig_mZI_stationary_2}
\end{figure}

\begin{figure}
\includegraphics[width=0.5\linewidth]{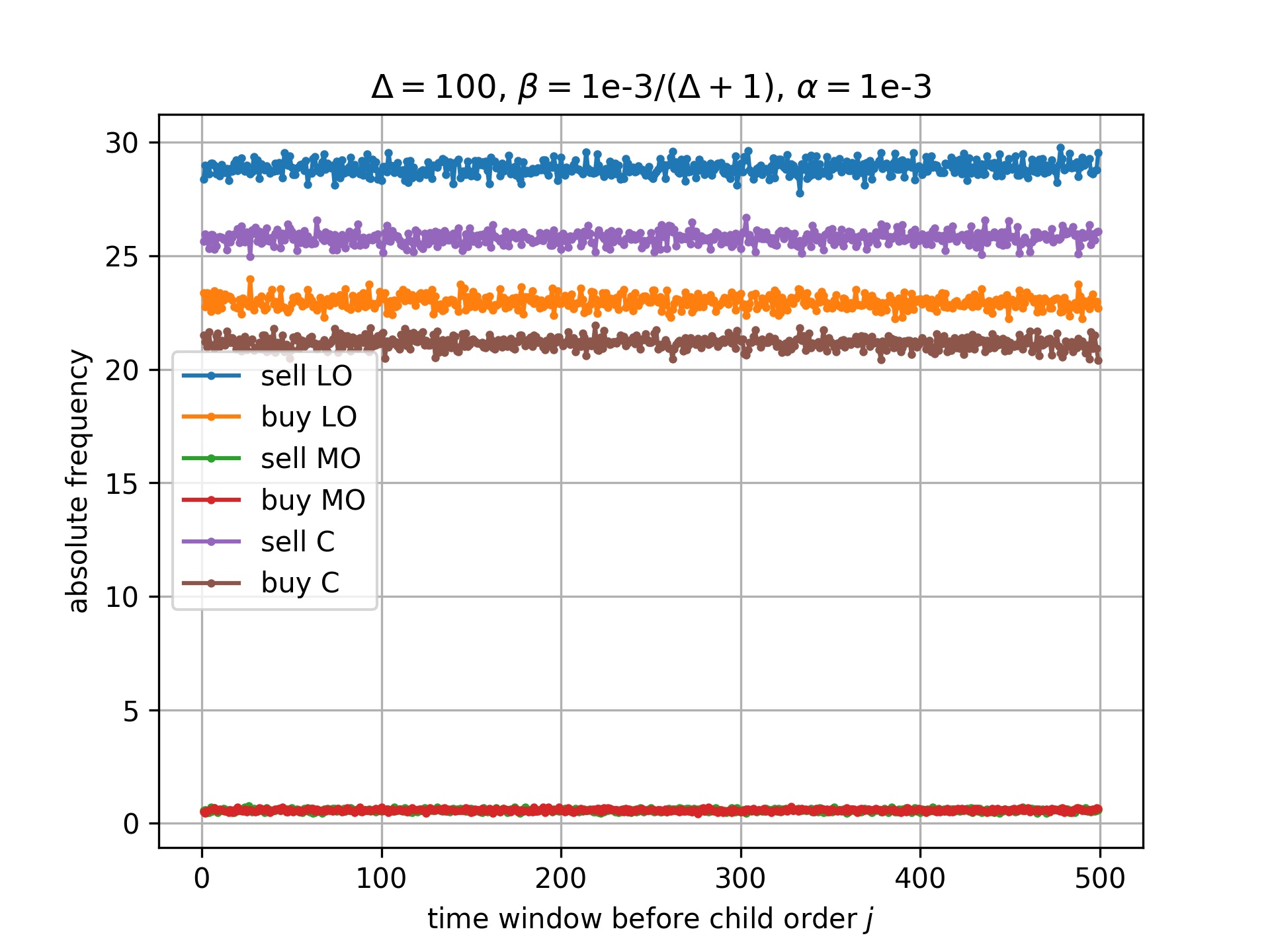}
\includegraphics[width=0.5\linewidth]{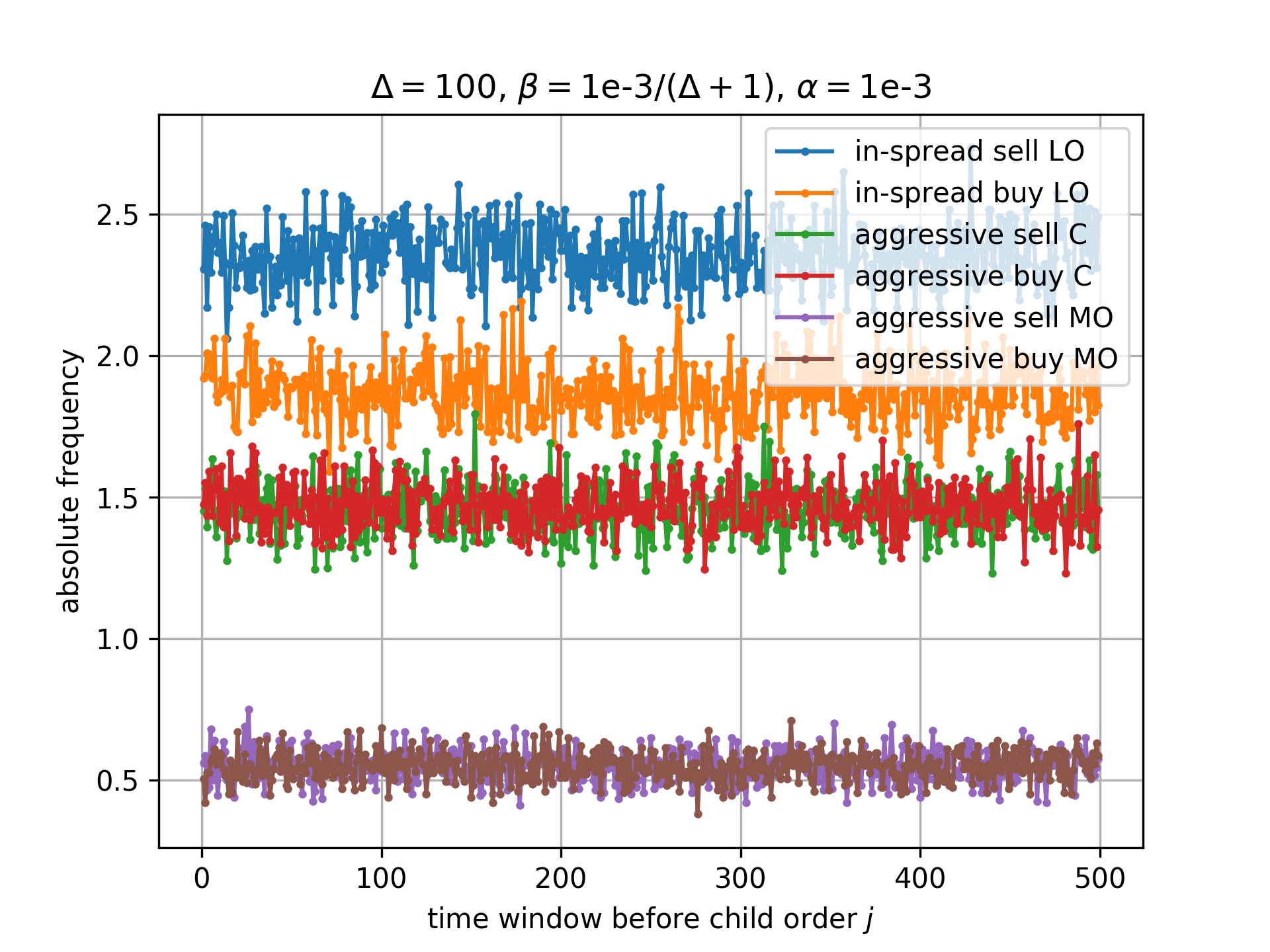}
\includegraphics[width=0.5\linewidth]{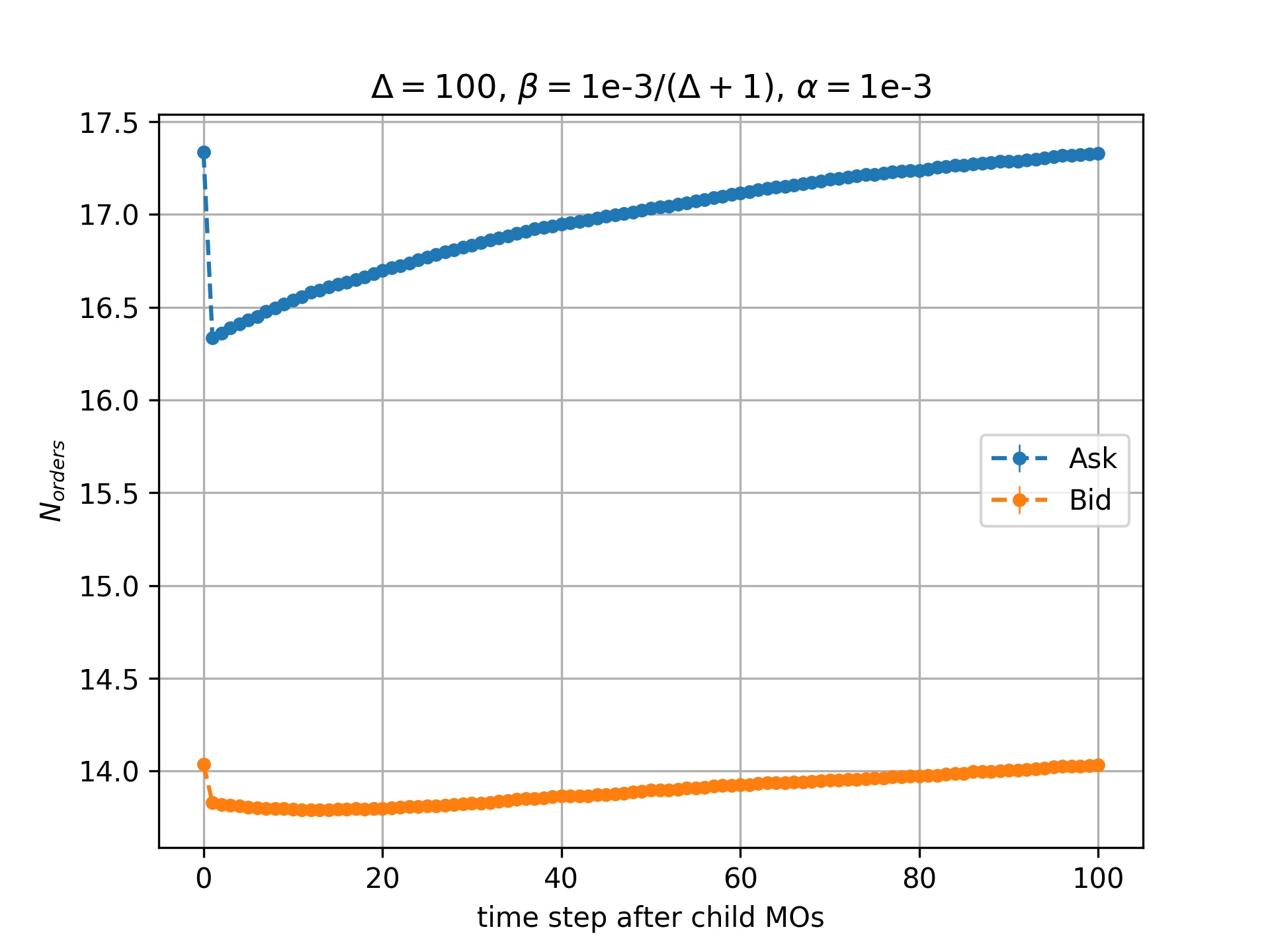}
\includegraphics[width=0.5\linewidth]{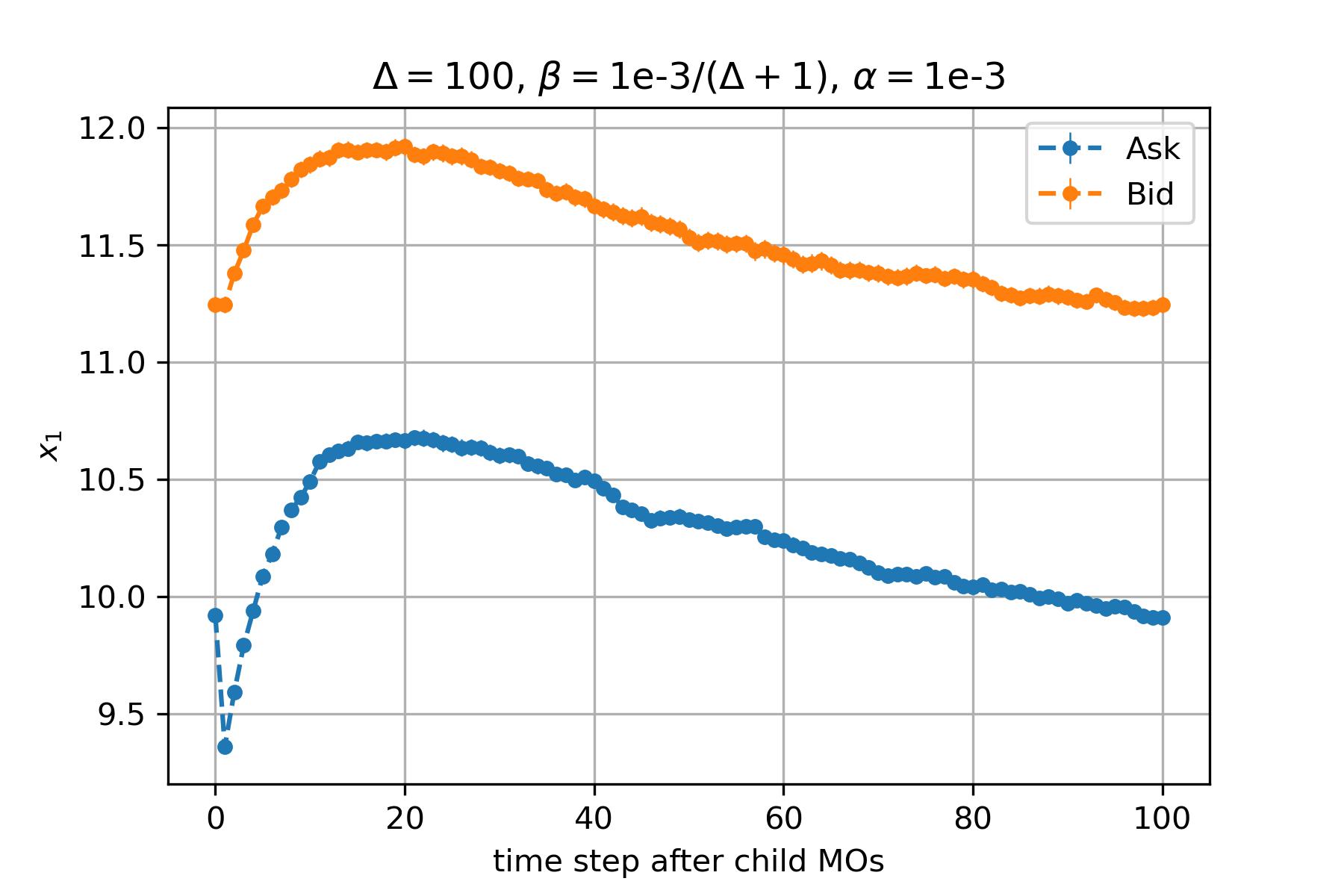}
\includegraphics[width=0.5\linewidth]{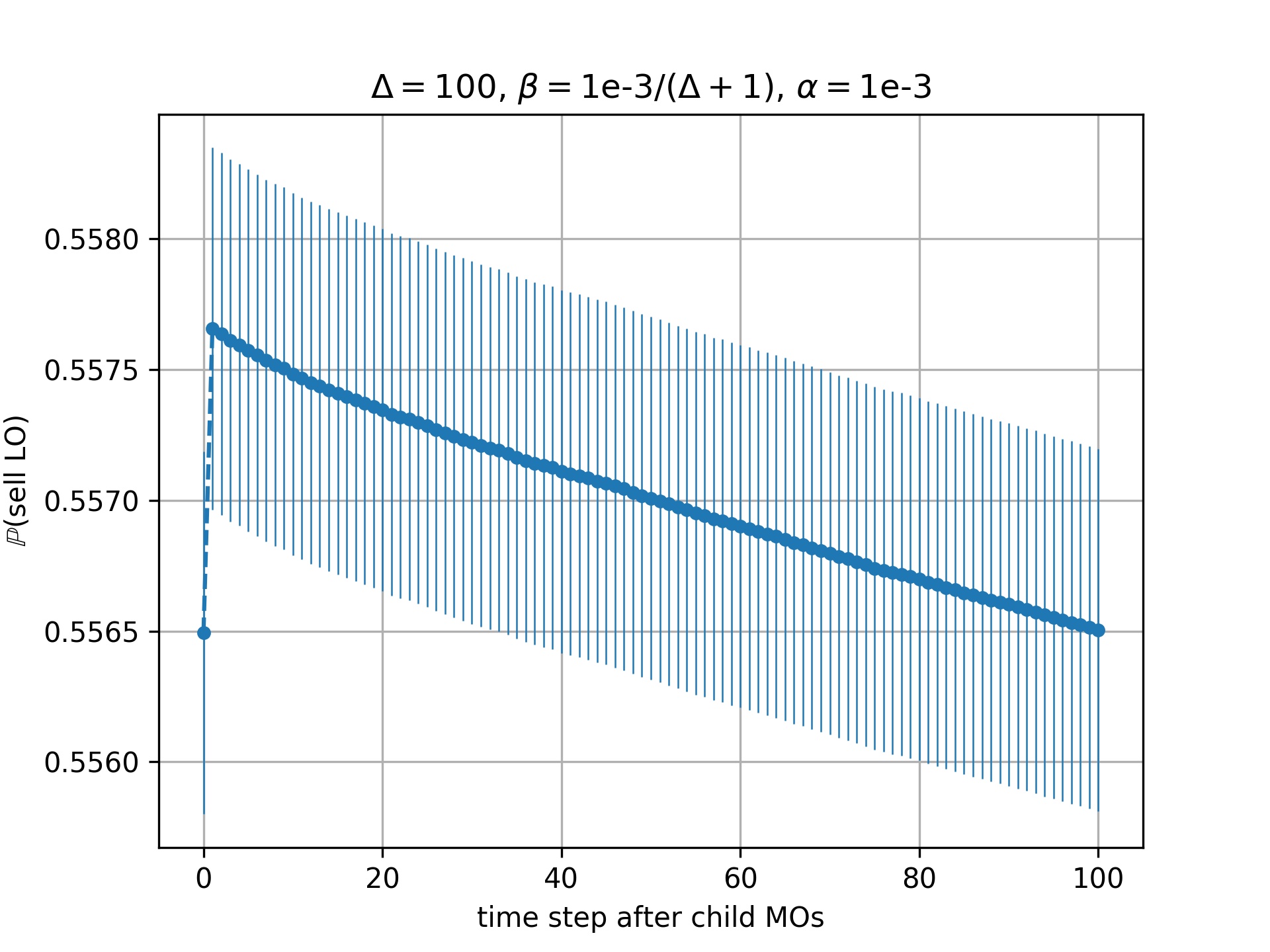}
    \caption{NMZI model, stationary regime. Frequencies of the different types of events in each time window between child MOs (MOs and cancellations are aggressive if they lead to a mid-price change). Number of orders, first gap sizes and probability that a LO is a sell as functions of the time step after child MOs.}
    \label{fig_mZI_stationary_3}
\end{figure}

\begin{figure}
\includegraphics[width=0.5\linewidth]{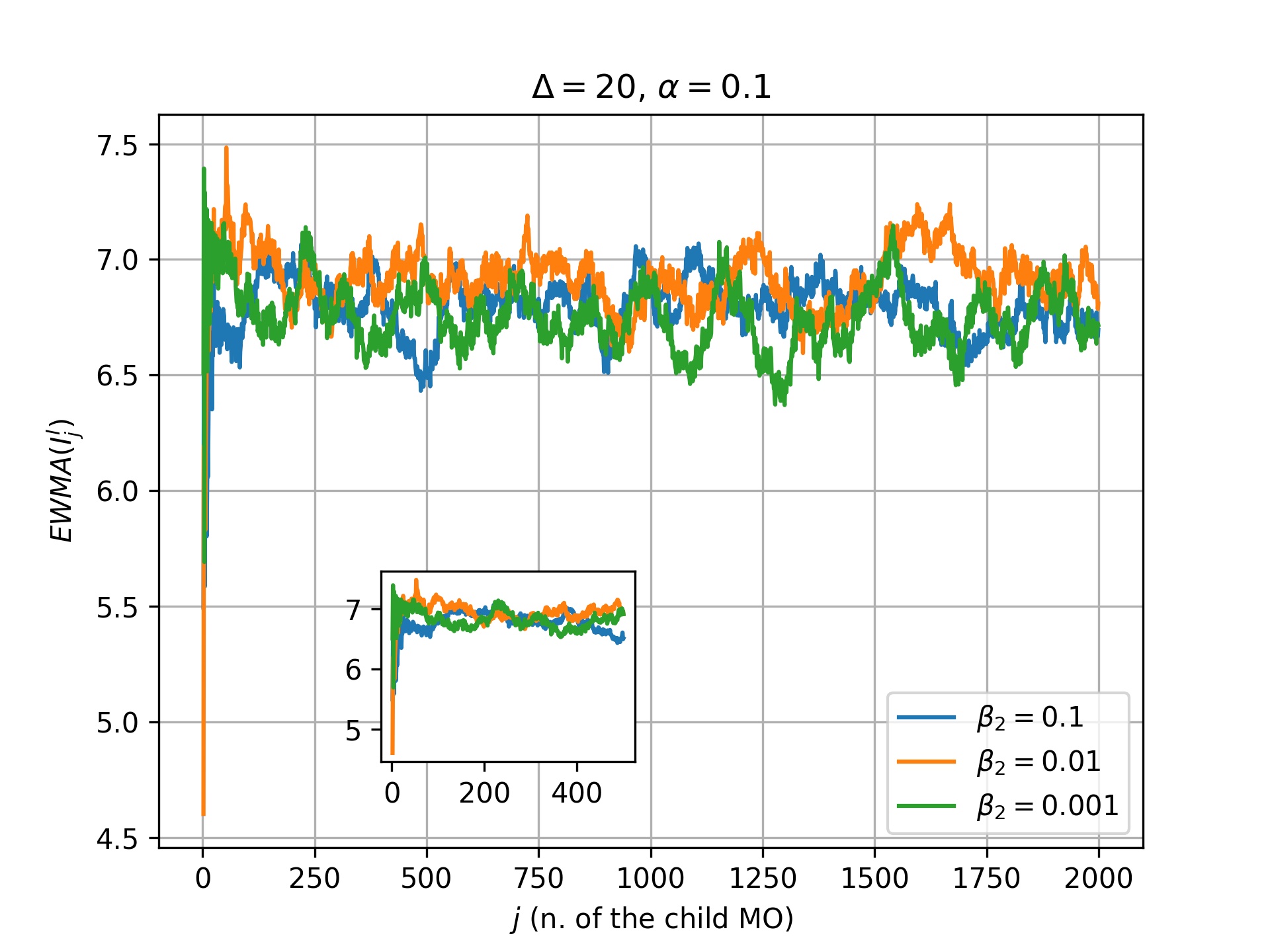}
\includegraphics[width=0.5\linewidth]{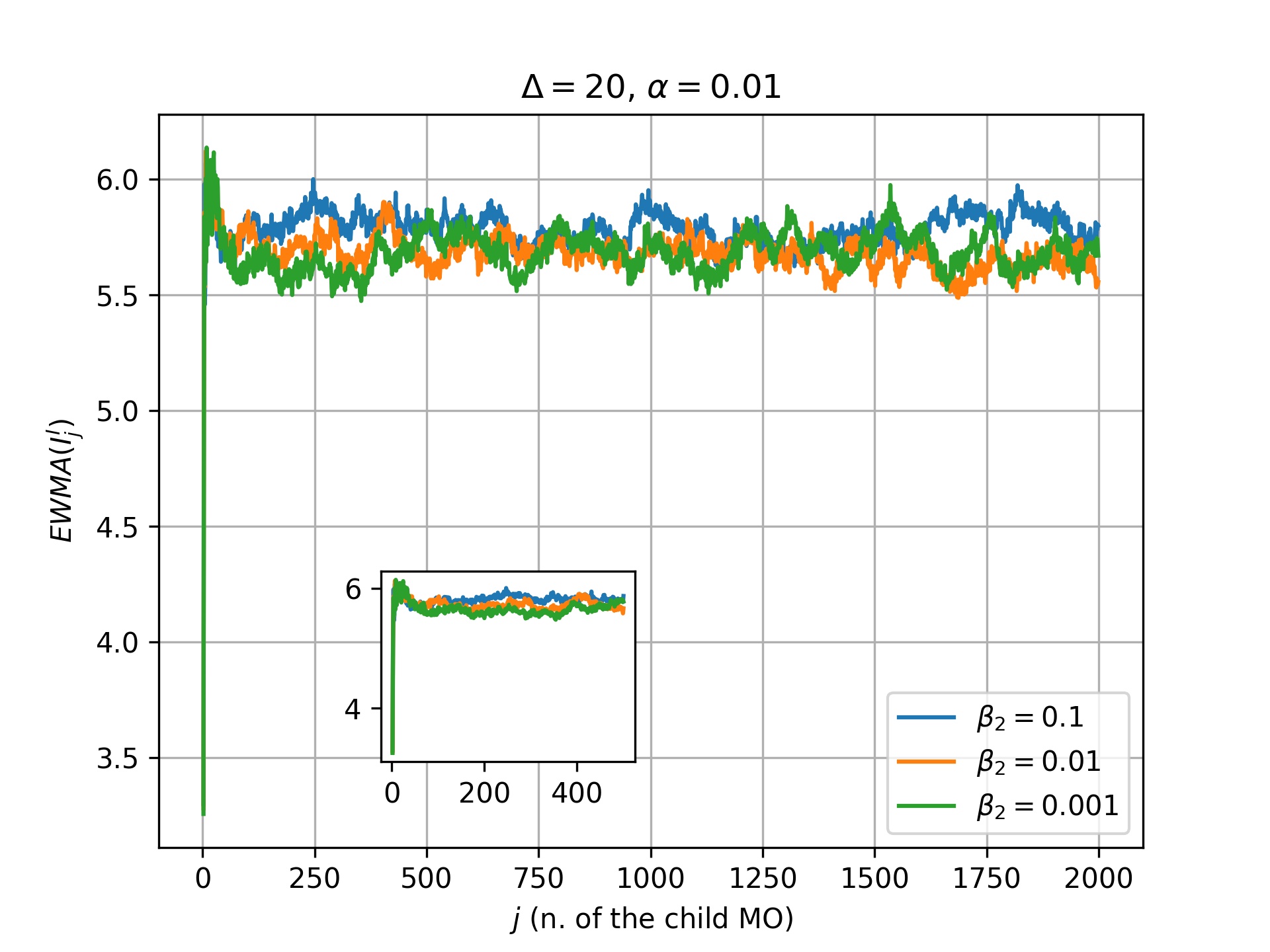}
\includegraphics[width=0.5\linewidth]{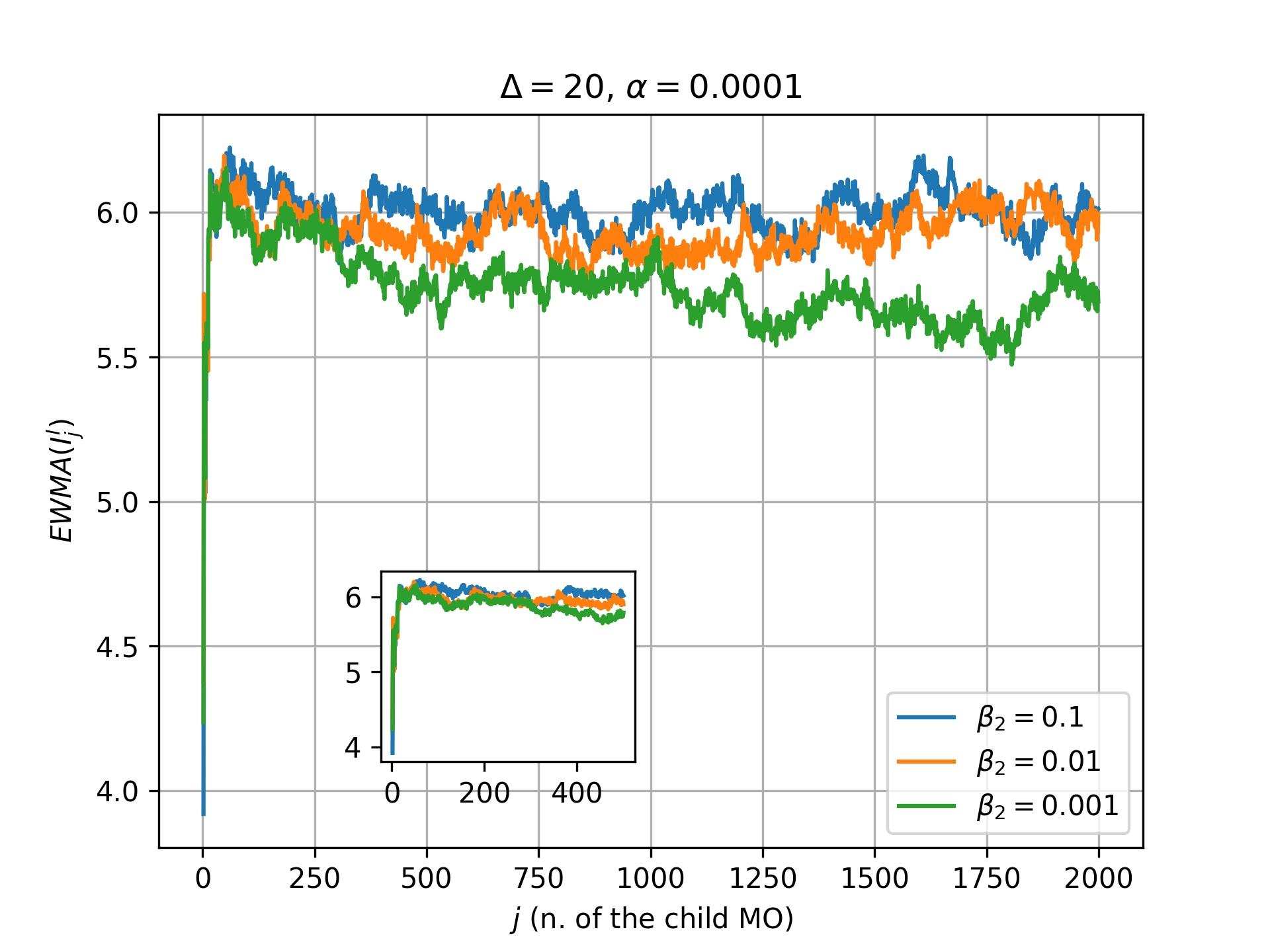}   
\includegraphics[width=0.5\linewidth]{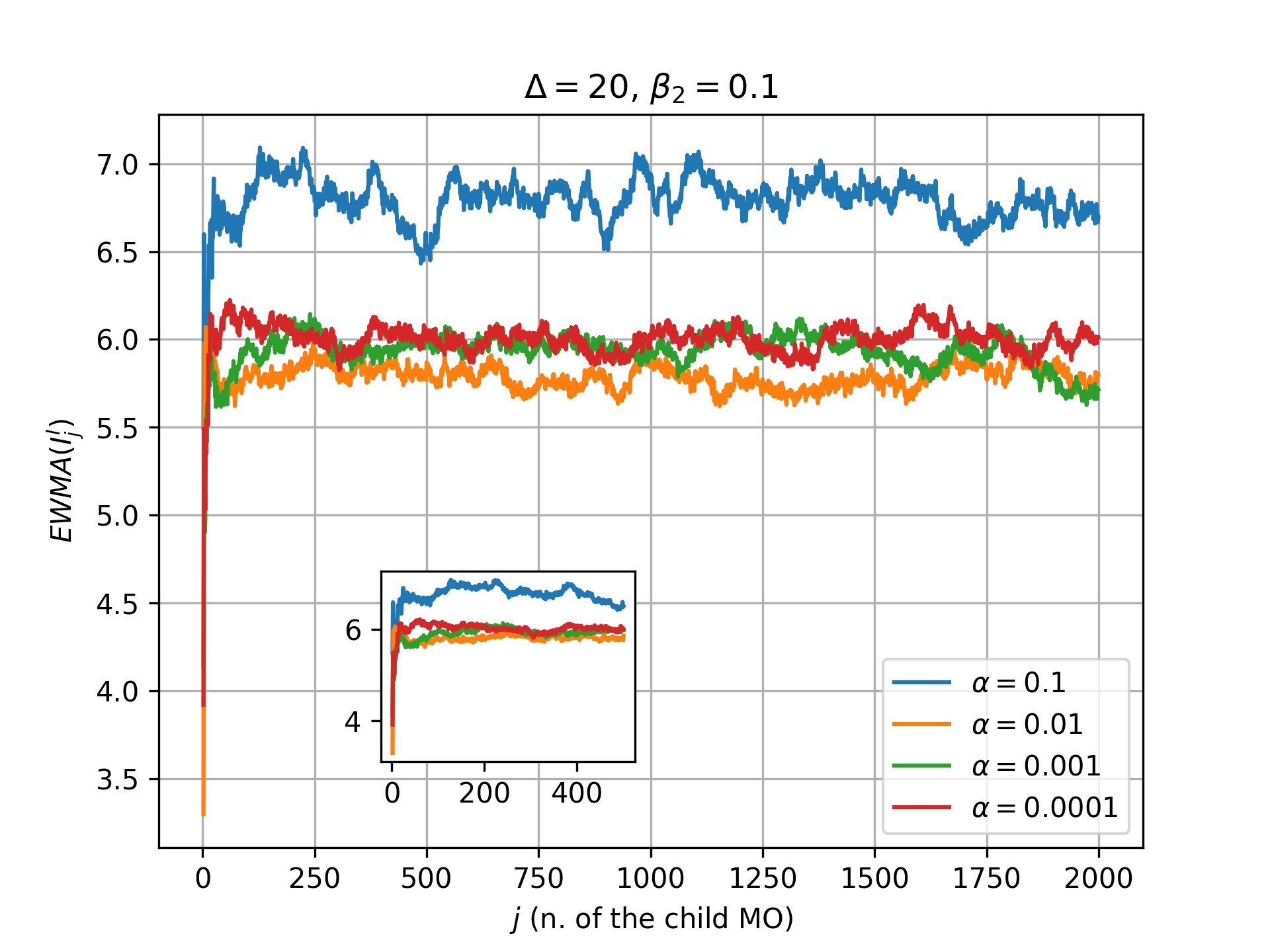}
\includegraphics[width=0.5\linewidth]{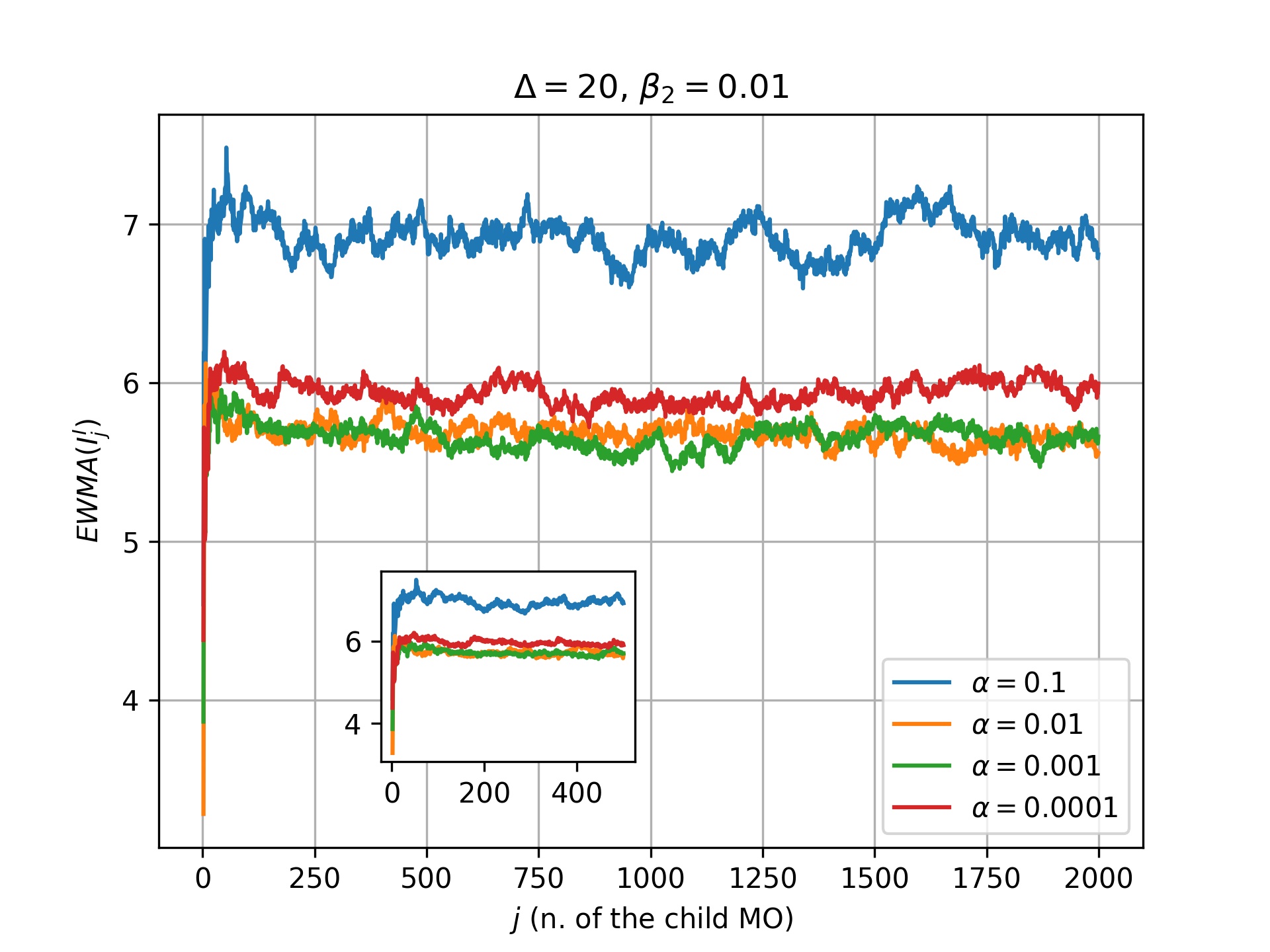}
    
    \caption{NMZI model. Evolution of the exponentially weighted moving averages (EWMAs) of the immediate component $I^I$ of the price impact. The inset plots focus on the first $500$ child MOs.}
    \label{impact_components_immediate}
\end{figure}

\begin{figure}
\includegraphics[width=0.5\linewidth]{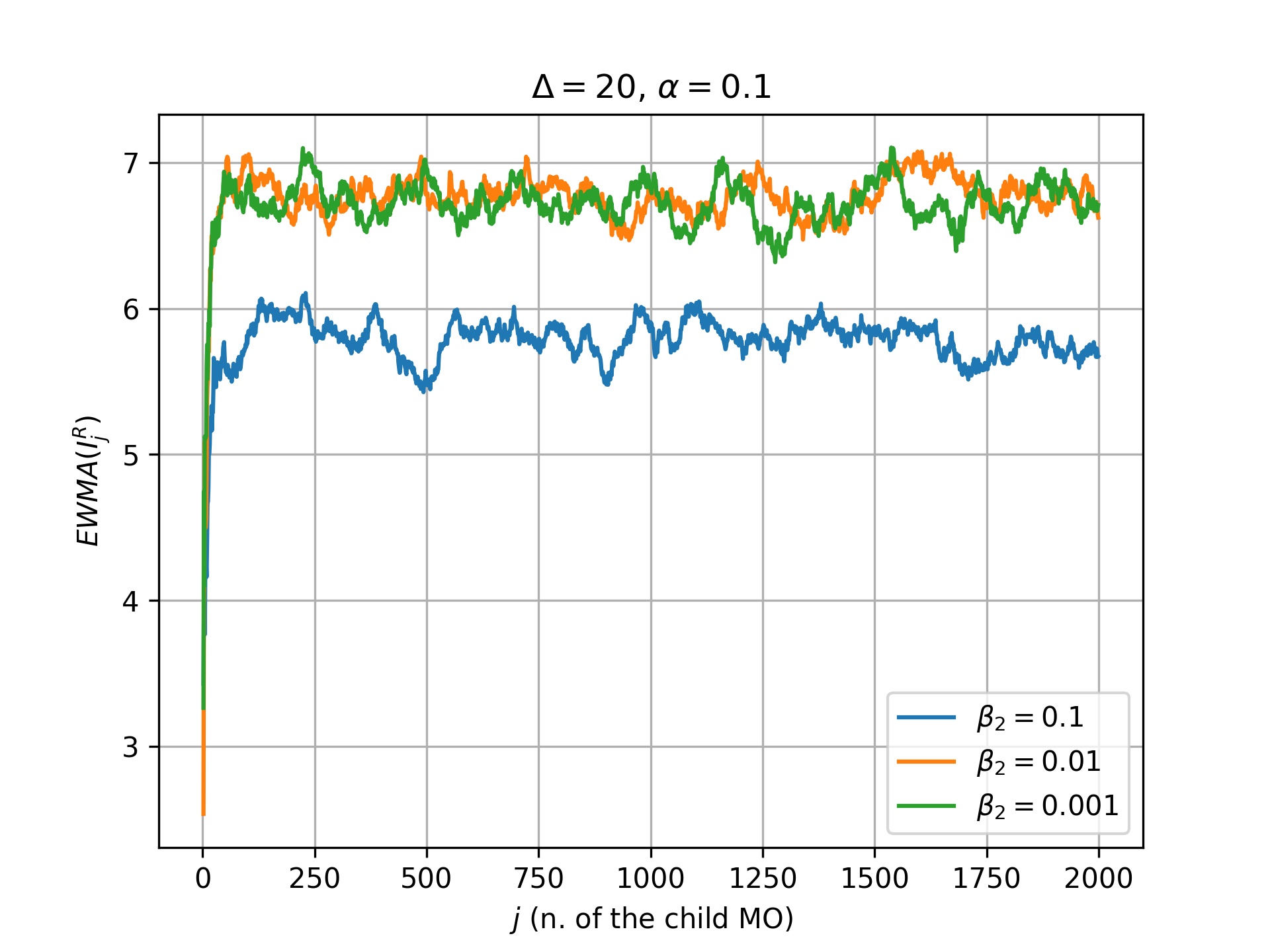}
\includegraphics[width=0.5\linewidth]{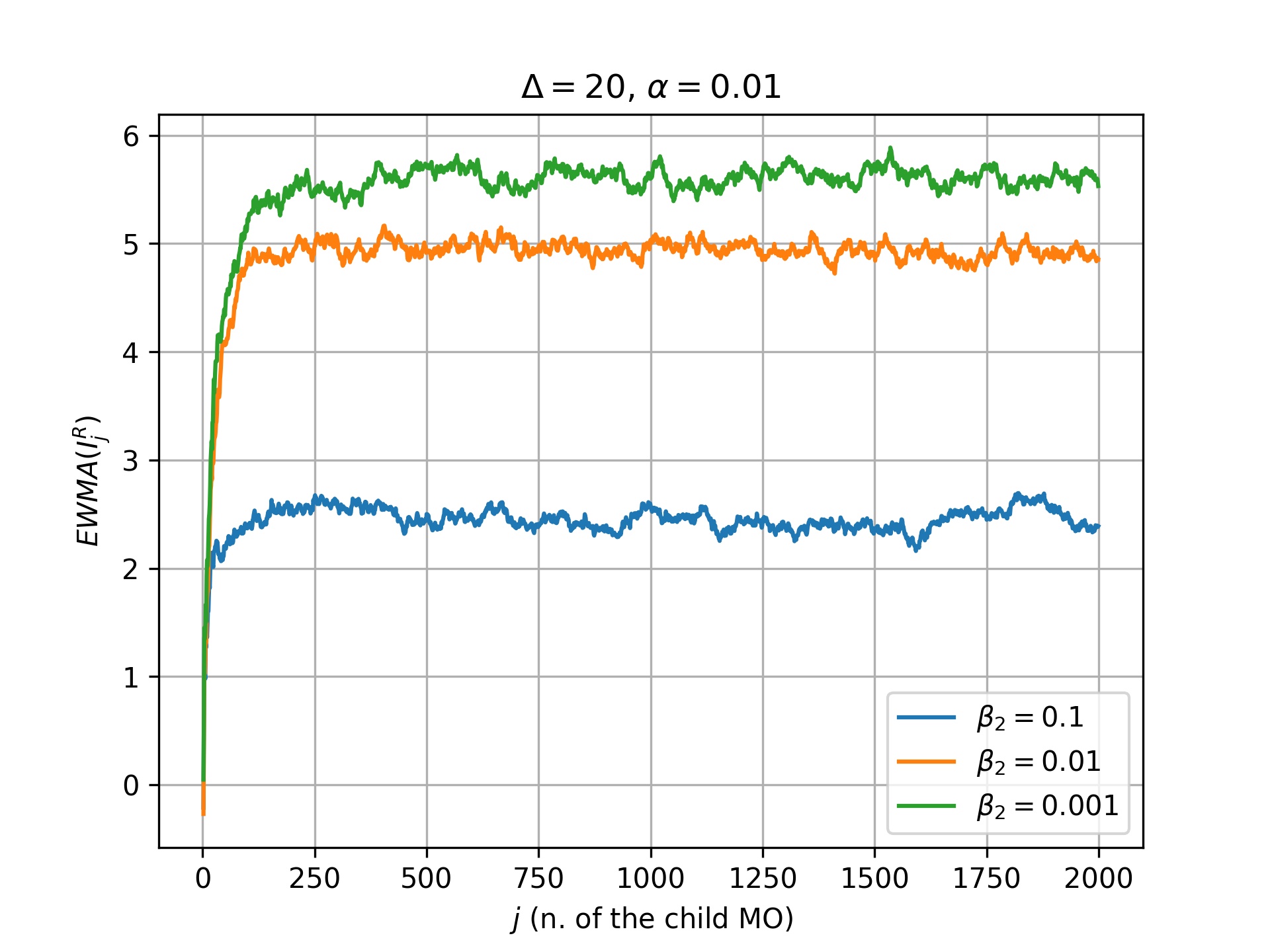}
\includegraphics[width=0.5\linewidth]{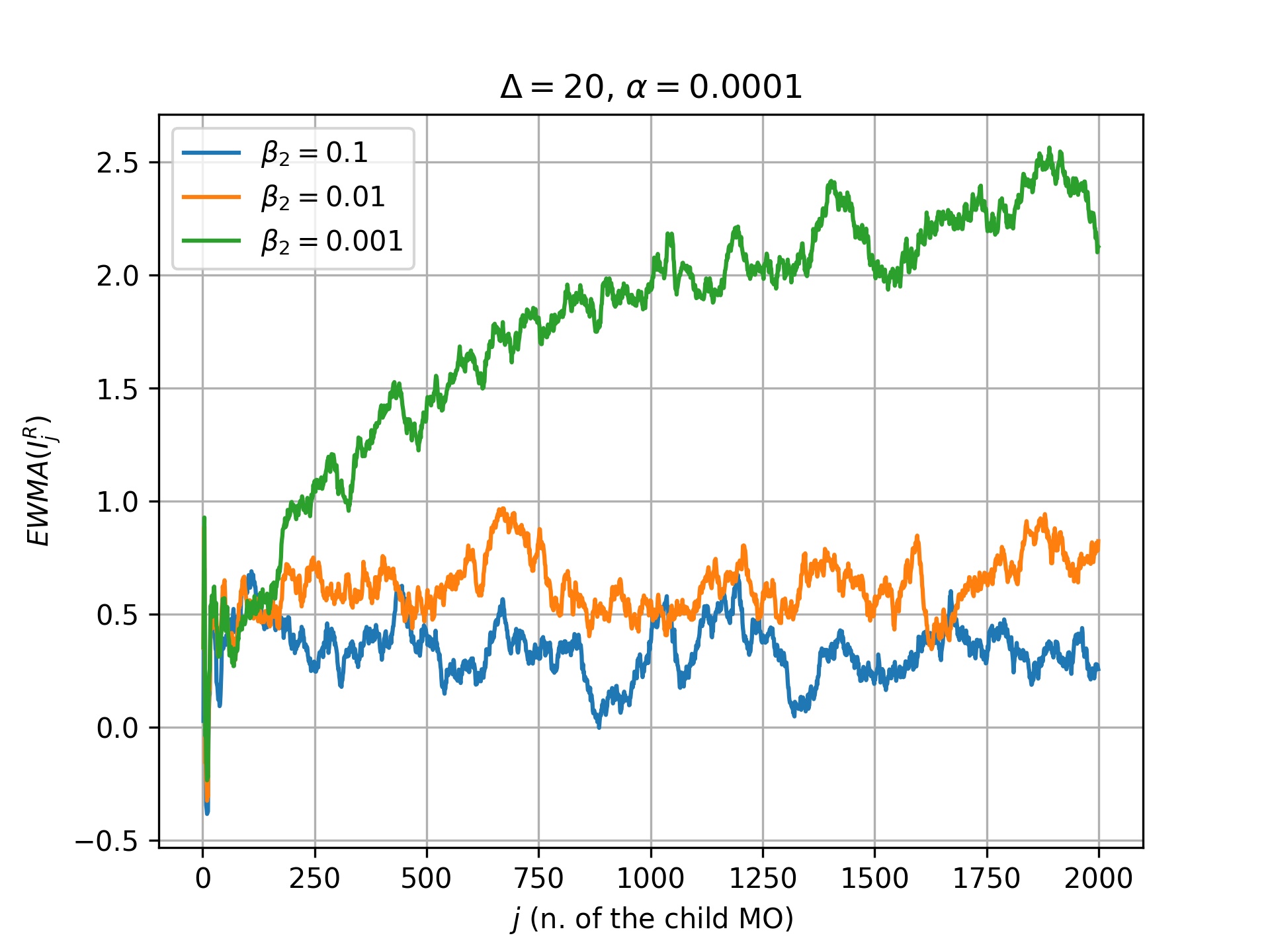}   
\includegraphics[width=0.5\linewidth]{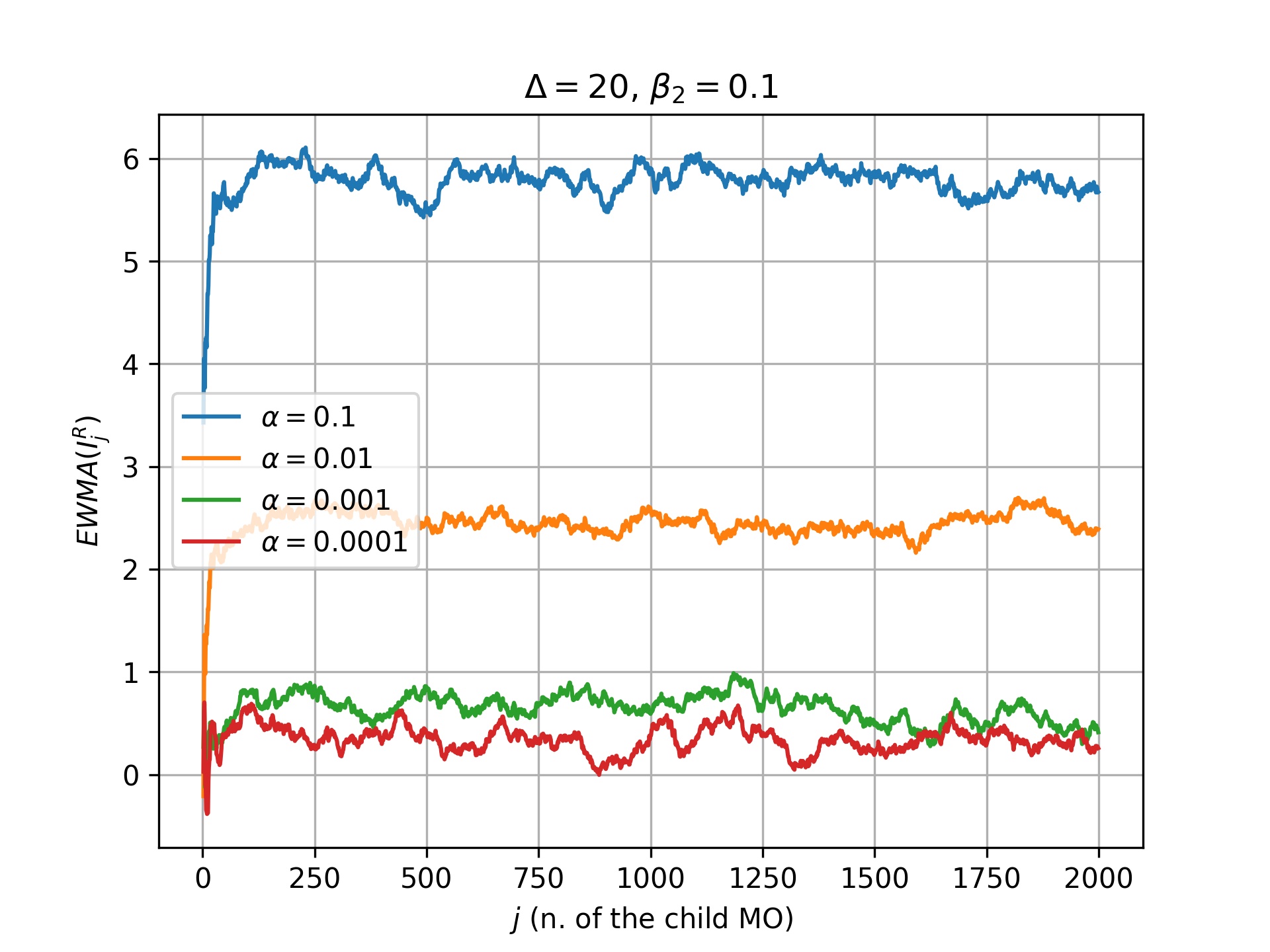}
\includegraphics[width=0.5\linewidth]{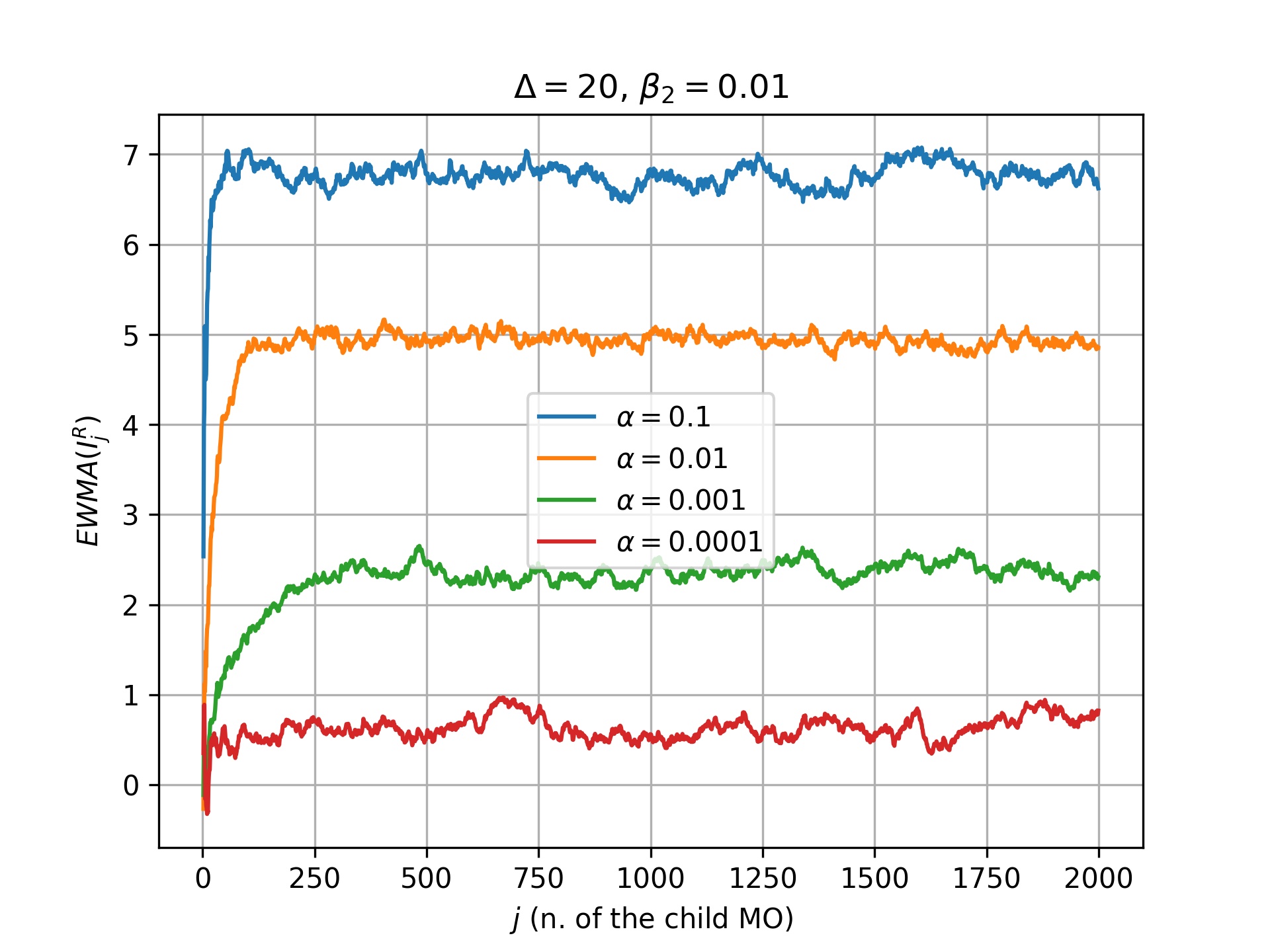}
    
    \caption{NMZI model. Evolution of the exponentially weighted moving averages (EWMAs) of the reversion component $I^R$ of the price impact. }
    \label{impact_components_reversion}
\end{figure}

\begin{figure}
\includegraphics[width=0.5\linewidth]{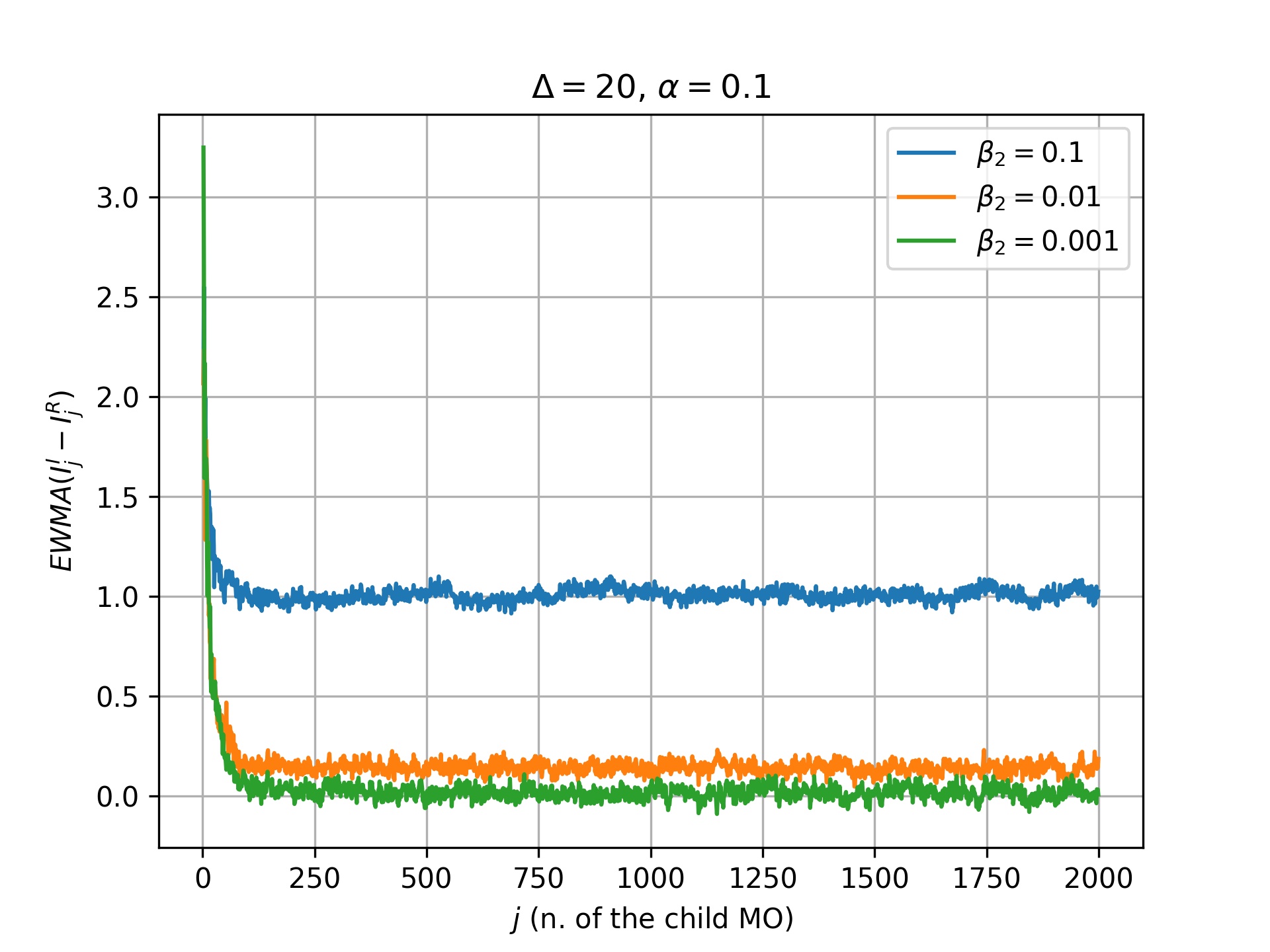}
\includegraphics[width=0.5\linewidth]{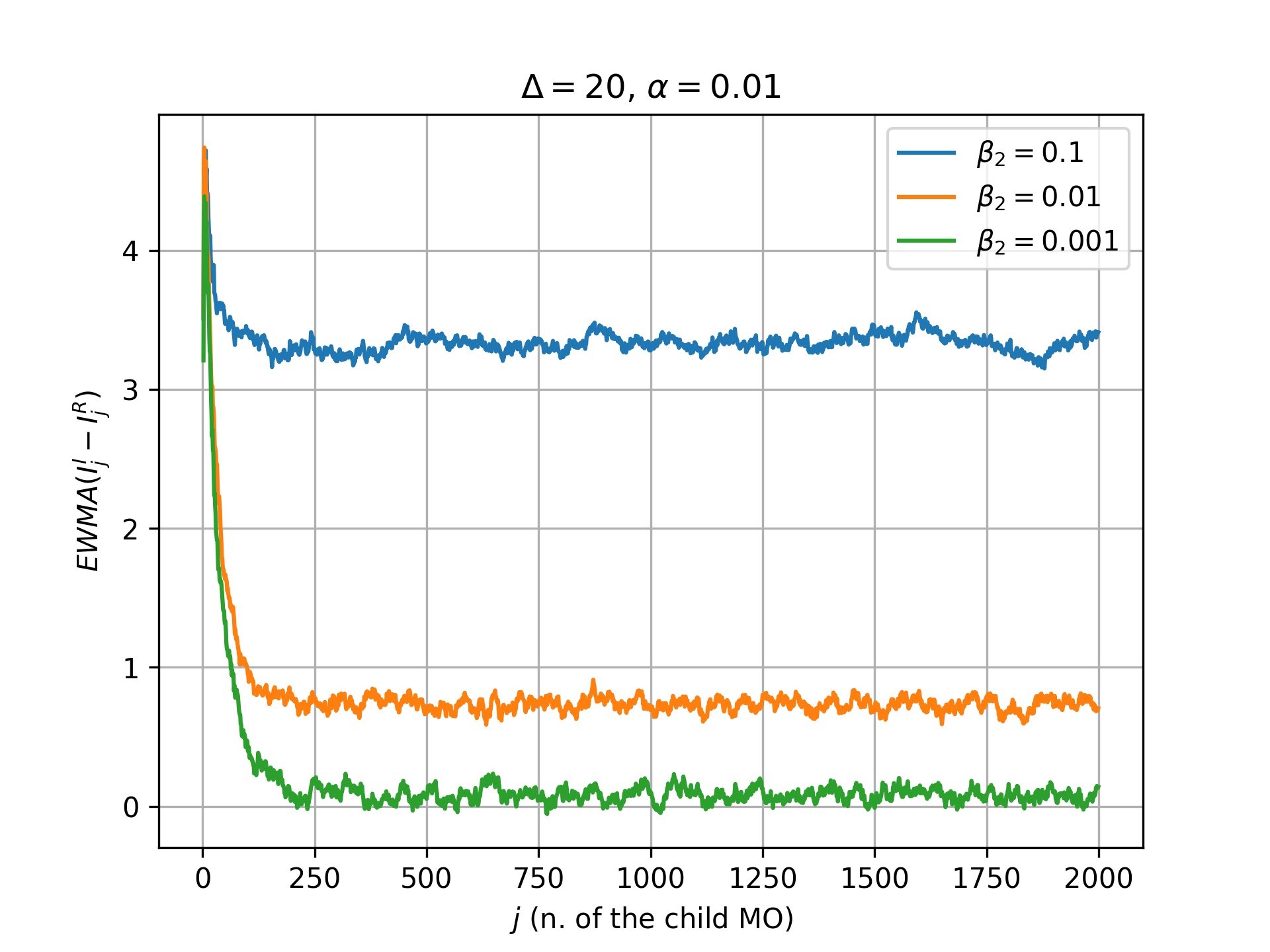}
\includegraphics[width=0.5\linewidth]{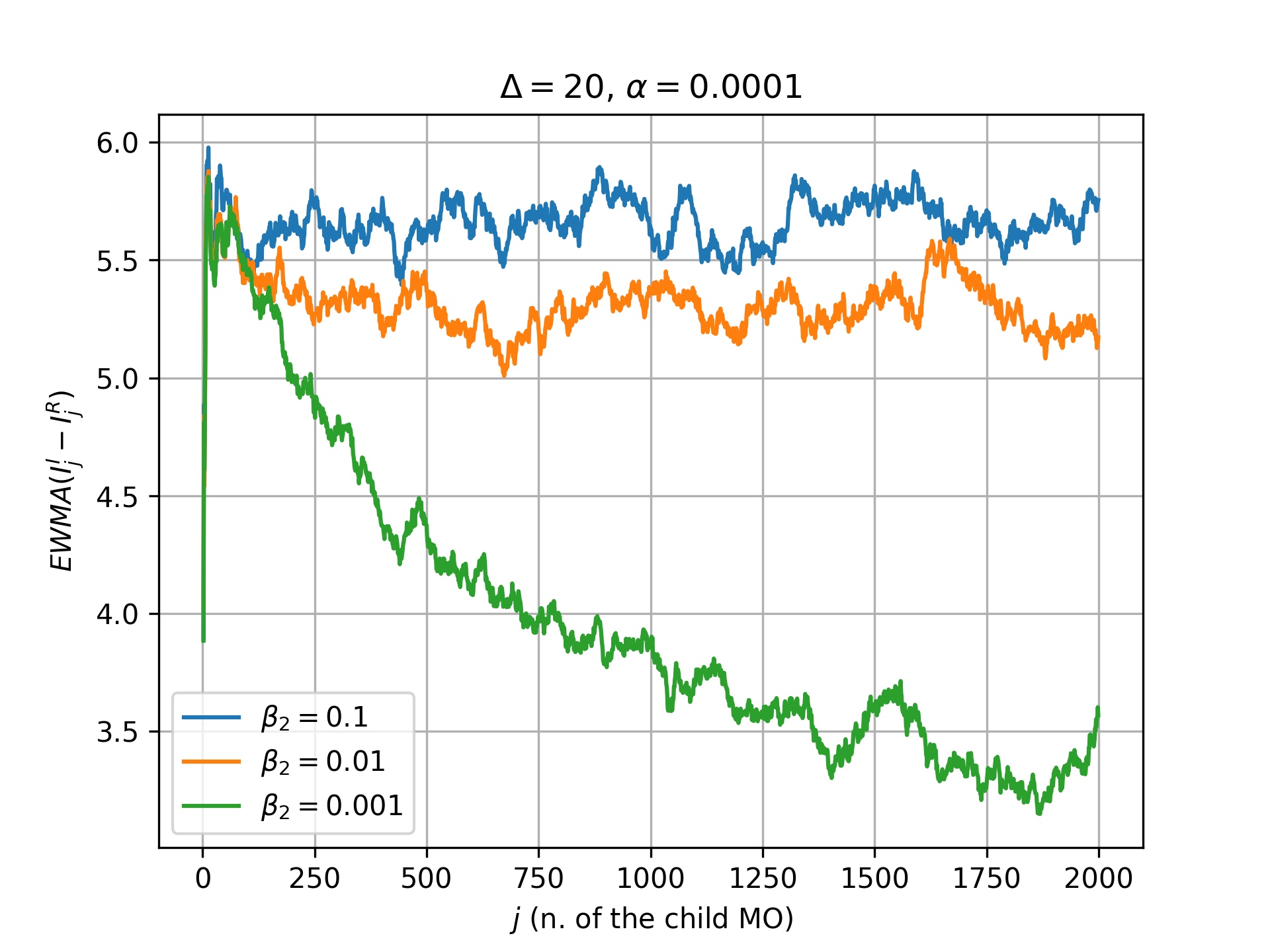}
\includegraphics[width=0.5\linewidth]{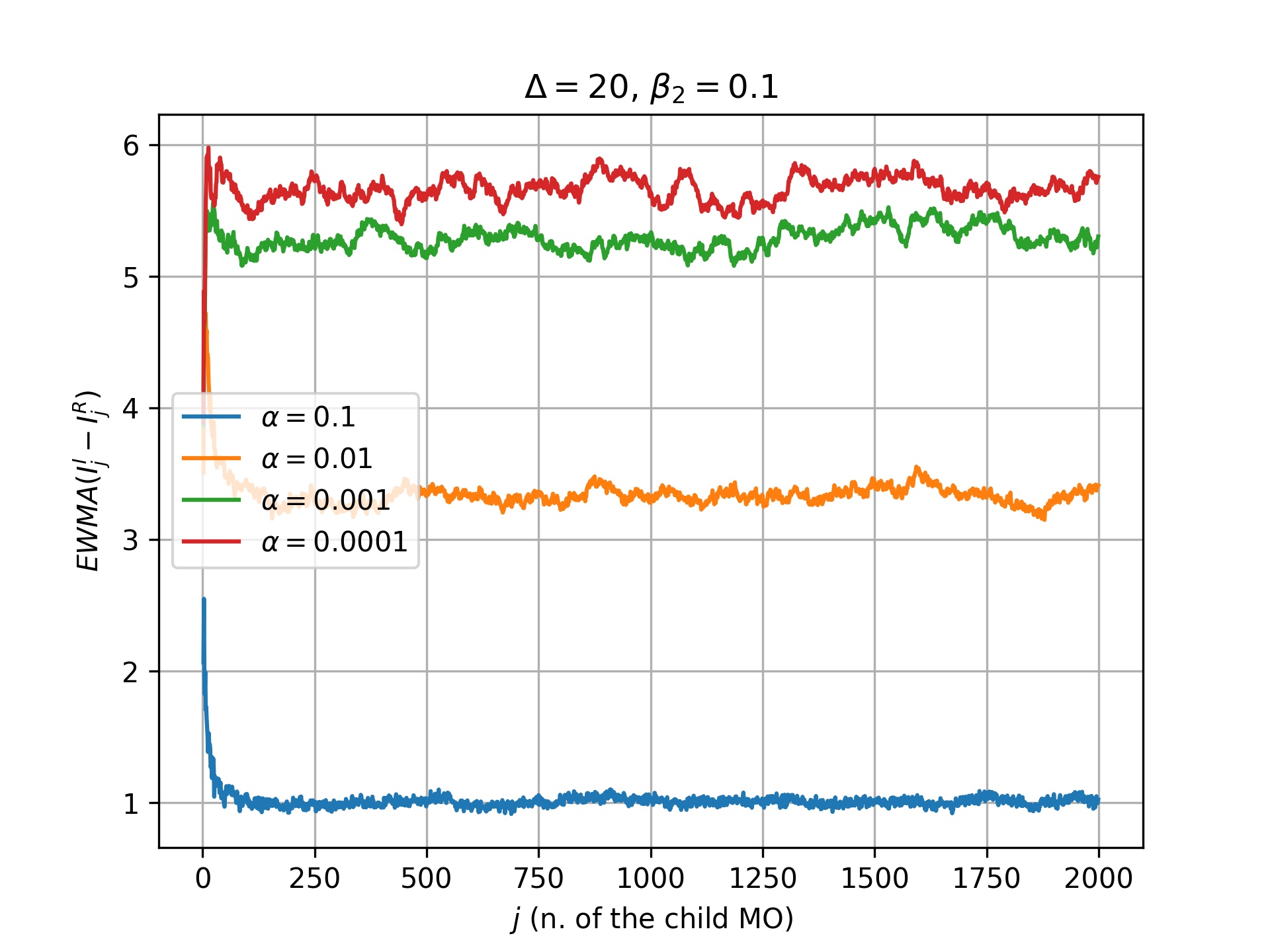}
\includegraphics[width=0.5\linewidth]{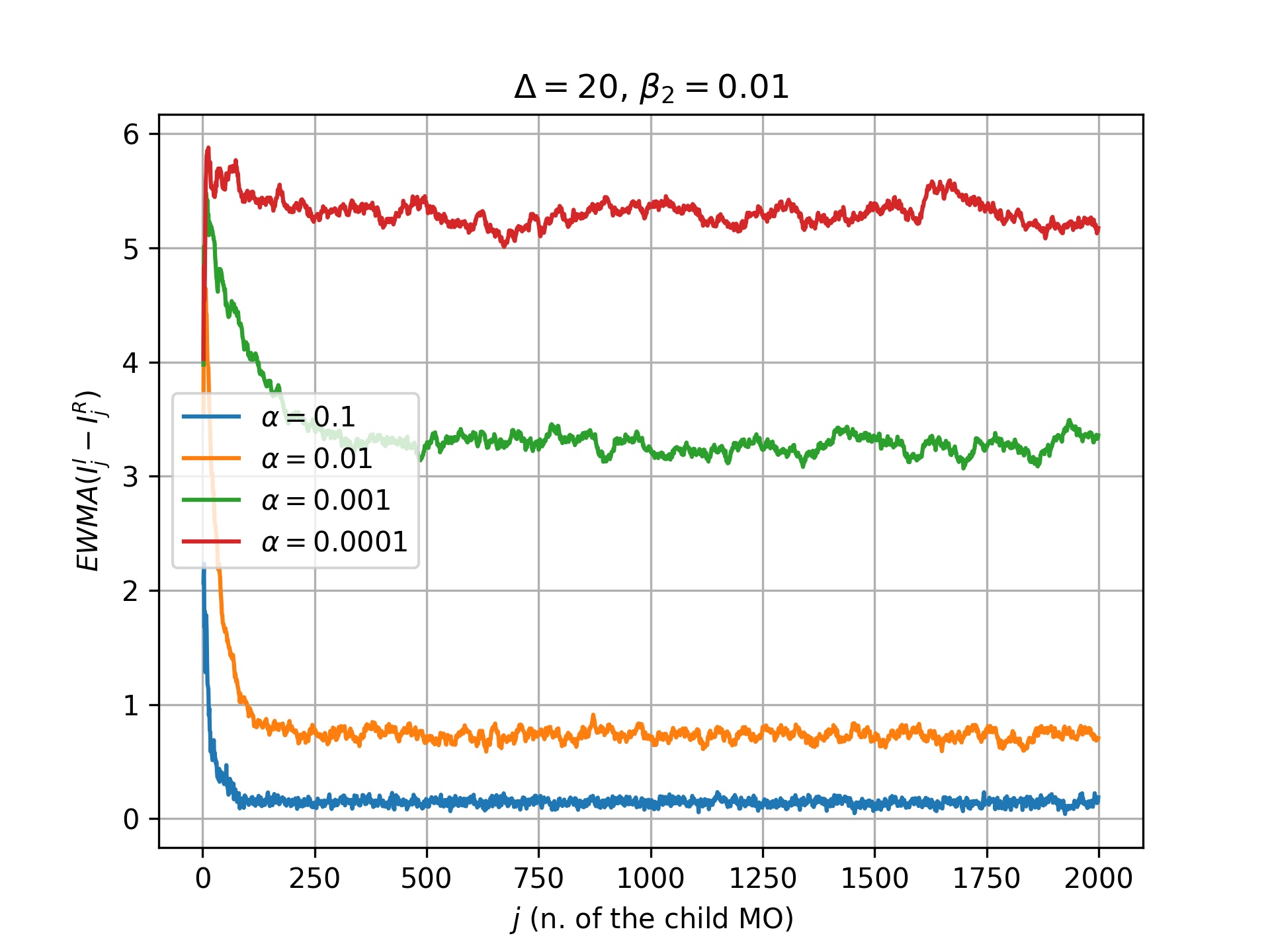}
    
    \caption{NMZI model. Evolution of the exponentially weighted moving averages (EWMAs) of the overall impact $I^I - I^R$. }
    \label{impact_components_total}
\end{figure}

\begin{figure}
\includegraphics[width=0.5\linewidth]{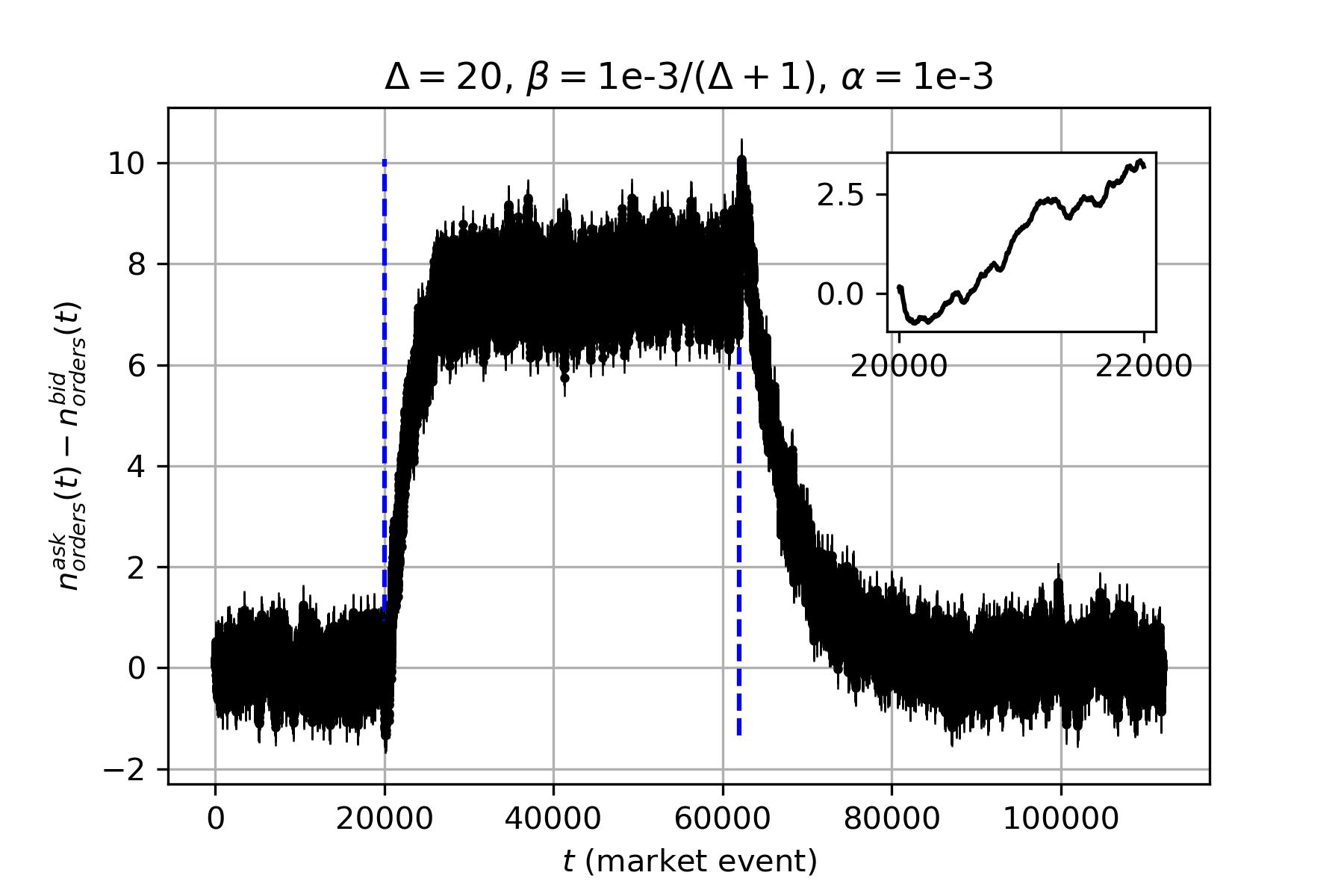}
\includegraphics[width=0.5\linewidth]{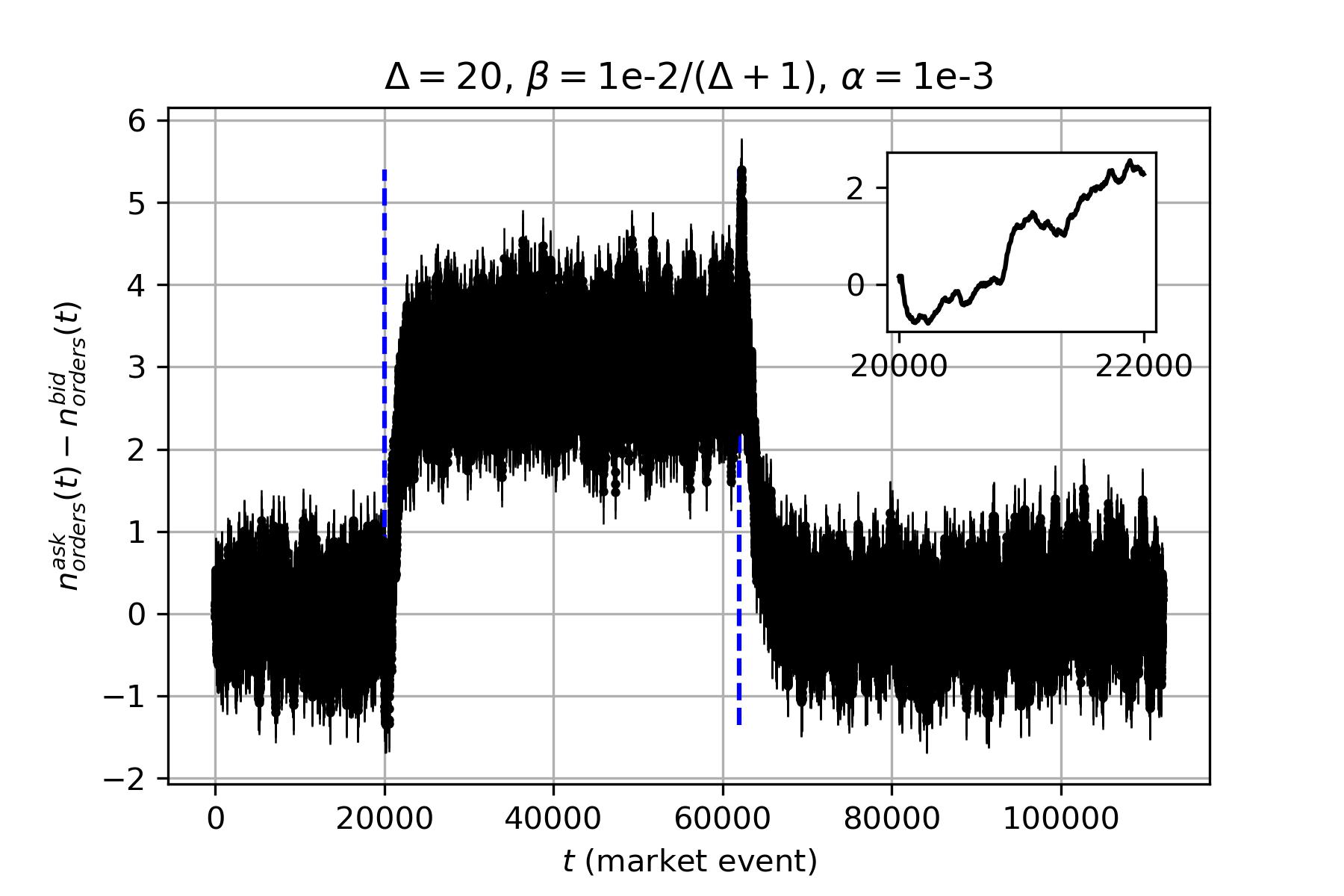}
\includegraphics[width=0.5\linewidth]{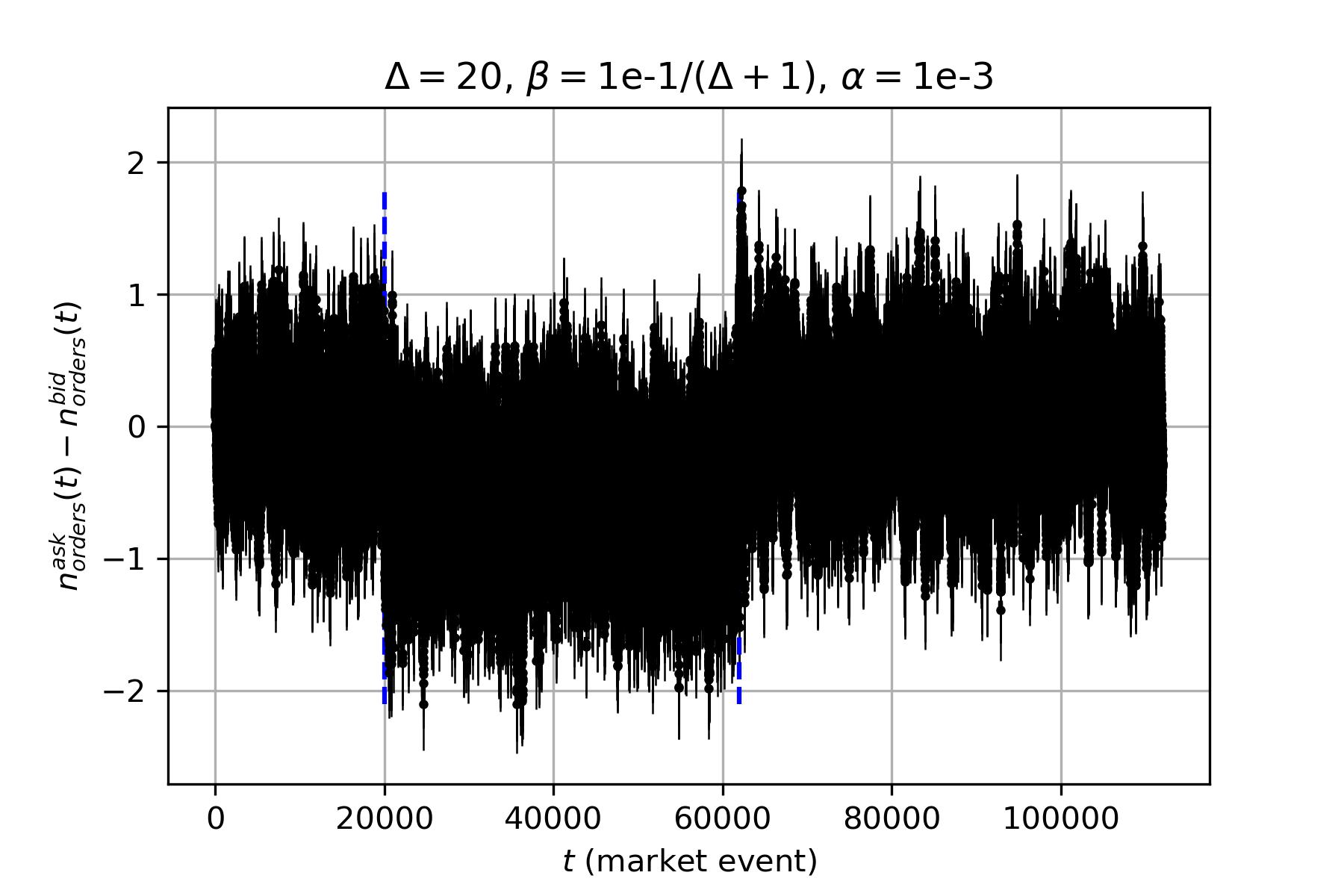}
    \caption{NMZI model. Evolution in time of the imbalance between the number of orders in the ask and bid side of the LOB. The vertical dashed blue lines are the beginning and the end of the metaorder execution. The inset plots focus on the first events after the execution starts.}
    \label{imbalance_evolution}
\end{figure}

\begin{figure}
\includegraphics[width=.5\linewidth]{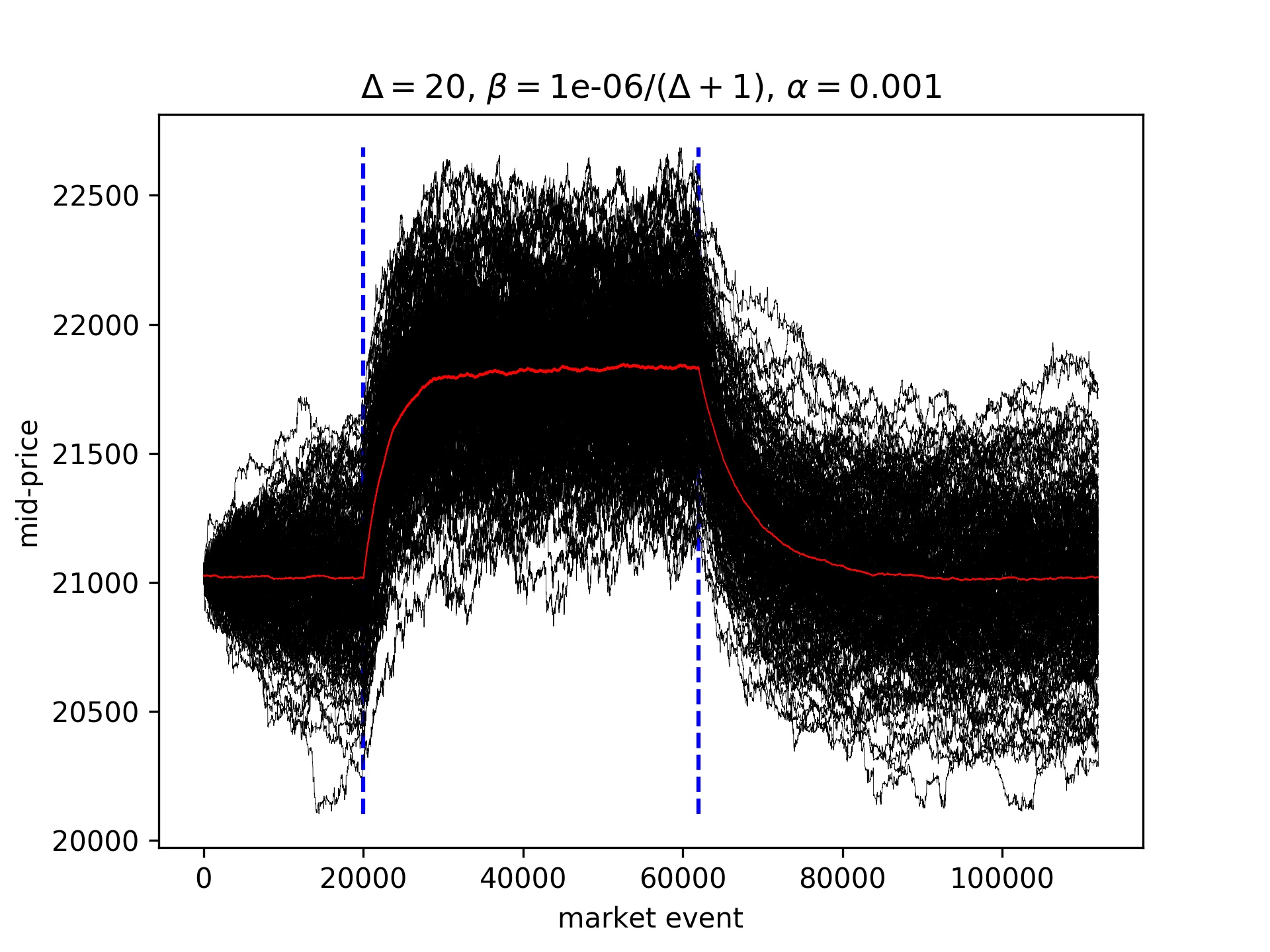}
\includegraphics[width=.5\linewidth]{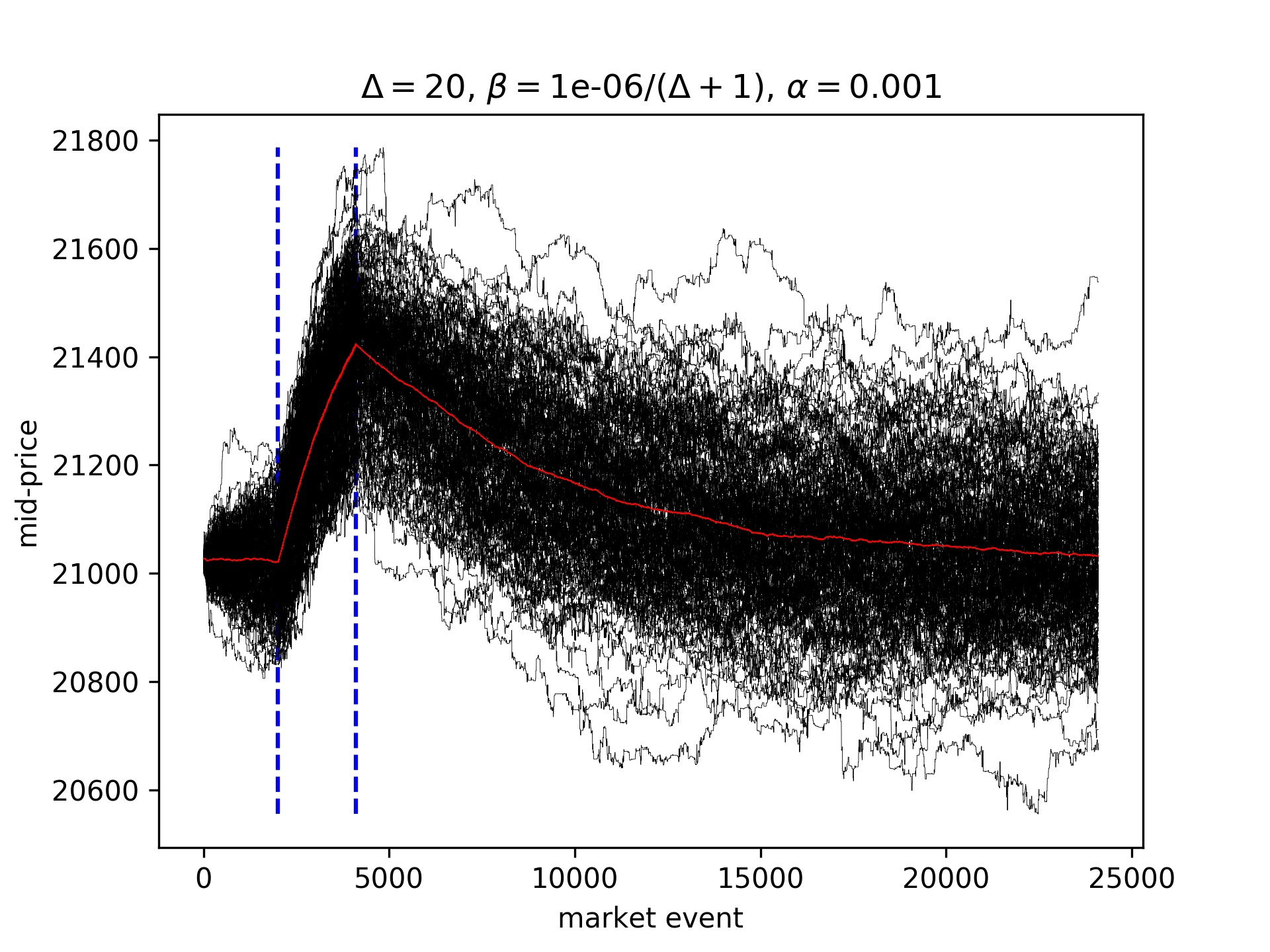}

    \caption{NMZI model. Mid-price path for an execution with total volume $Q= 2,000$ (left) and $Q= 100$ (right), buy direction and $\Delta = 20$. Each black line corresponds to a simulation. The red line is the mean path. The vertical dashed blue lines are the beginning and the end of the metaorder execution.}
    \label{fig_mid_price_paths_trading_interval_mZI_beta1e-6}
\end{figure}

\begin{figure}
    \includegraphics[width=0.5\linewidth]{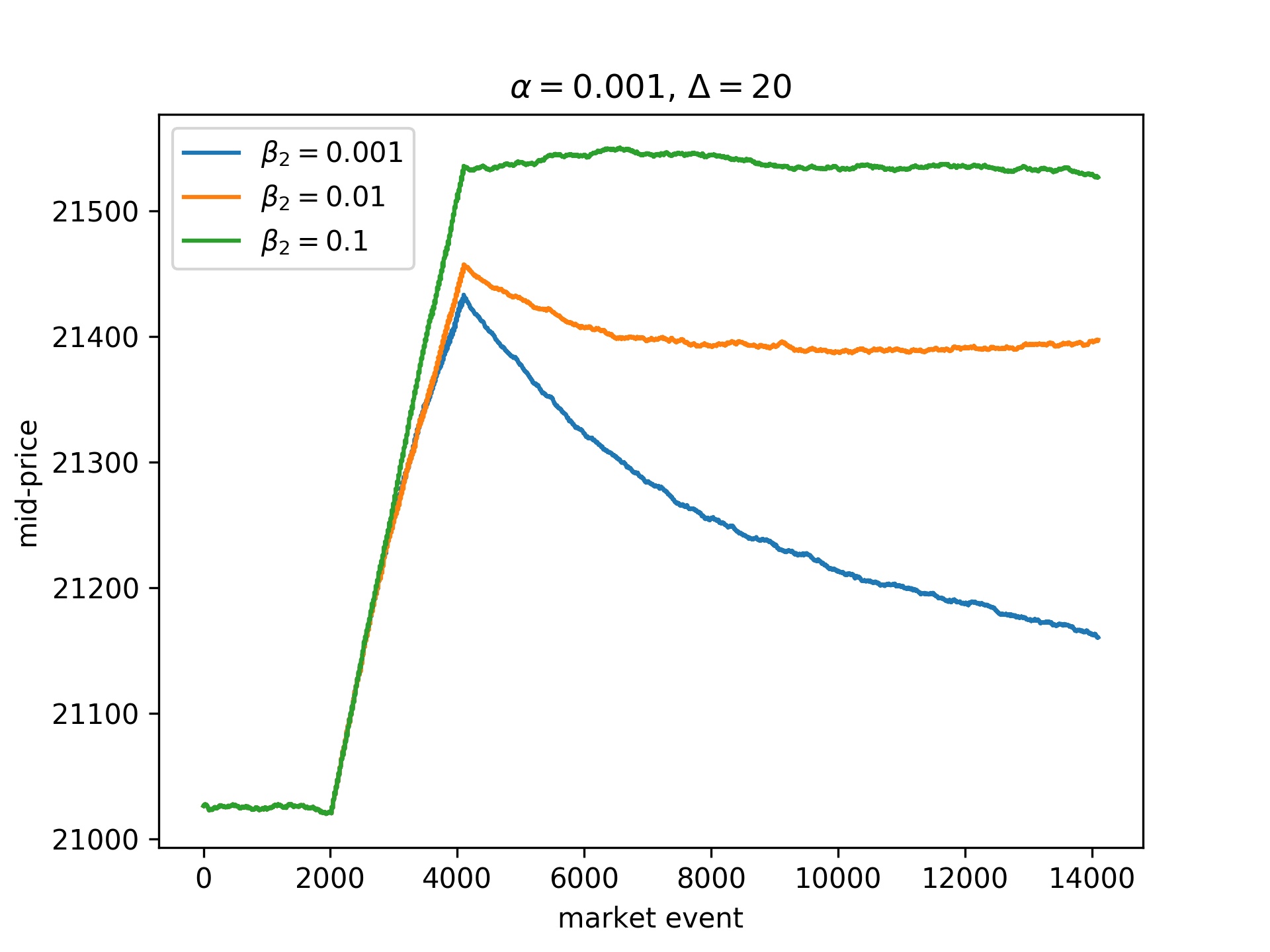}
    \includegraphics[width=0.5\linewidth]{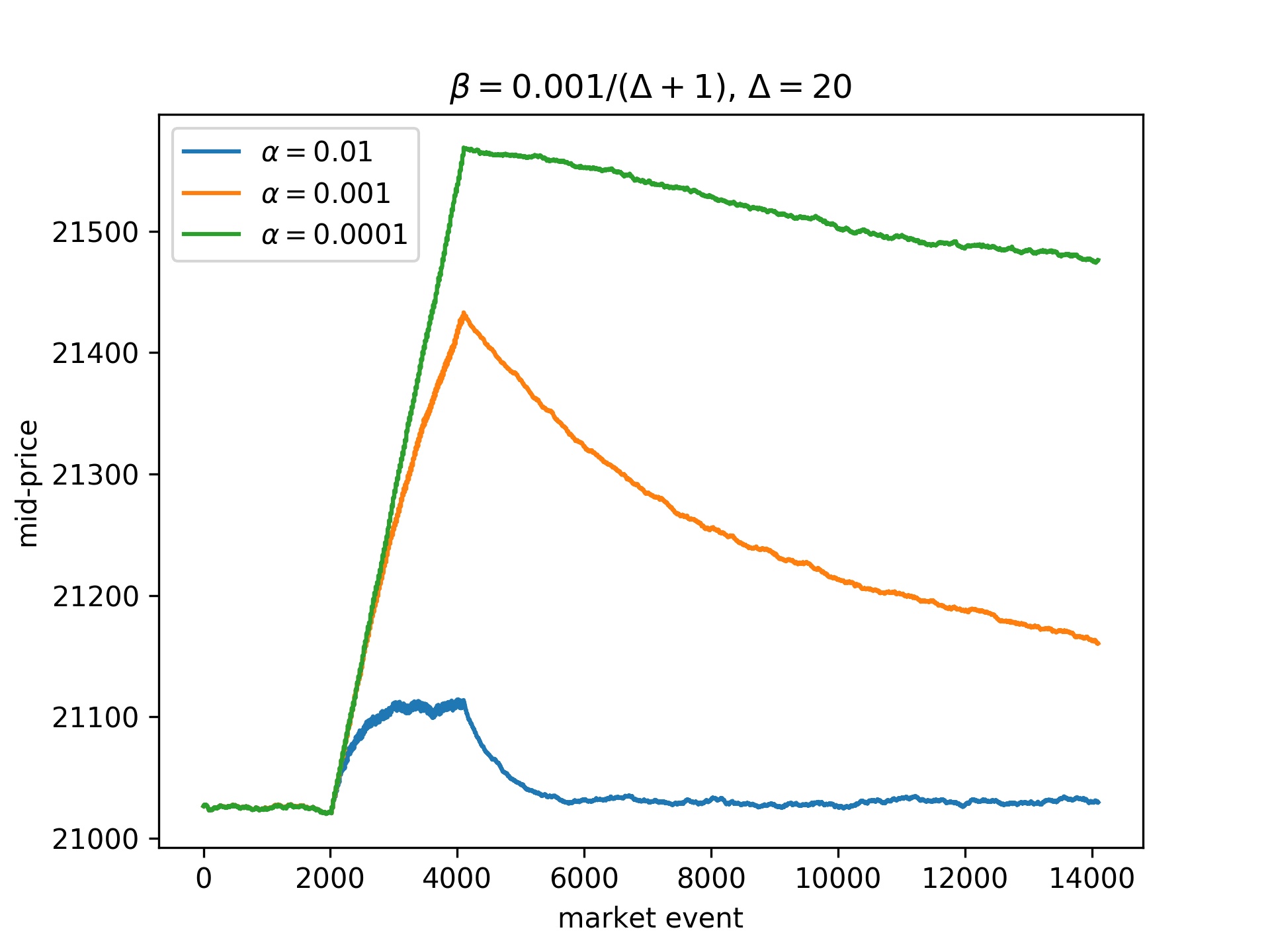}
    \caption{NMZI model. Mean mid-price path before, during and after the execution of a metaorder with $Q=100$, $\Delta = 20$ and $\alpha = 10^{-3}$ and several values of $\beta_2$ (left), $\beta_2 = 10^{-3}$ and several values of $\alpha$ (right).}
    \label{compare_midprice_beta20.001}
\end{figure}

\begin{figure}
    \includegraphics[width=0.5\linewidth]{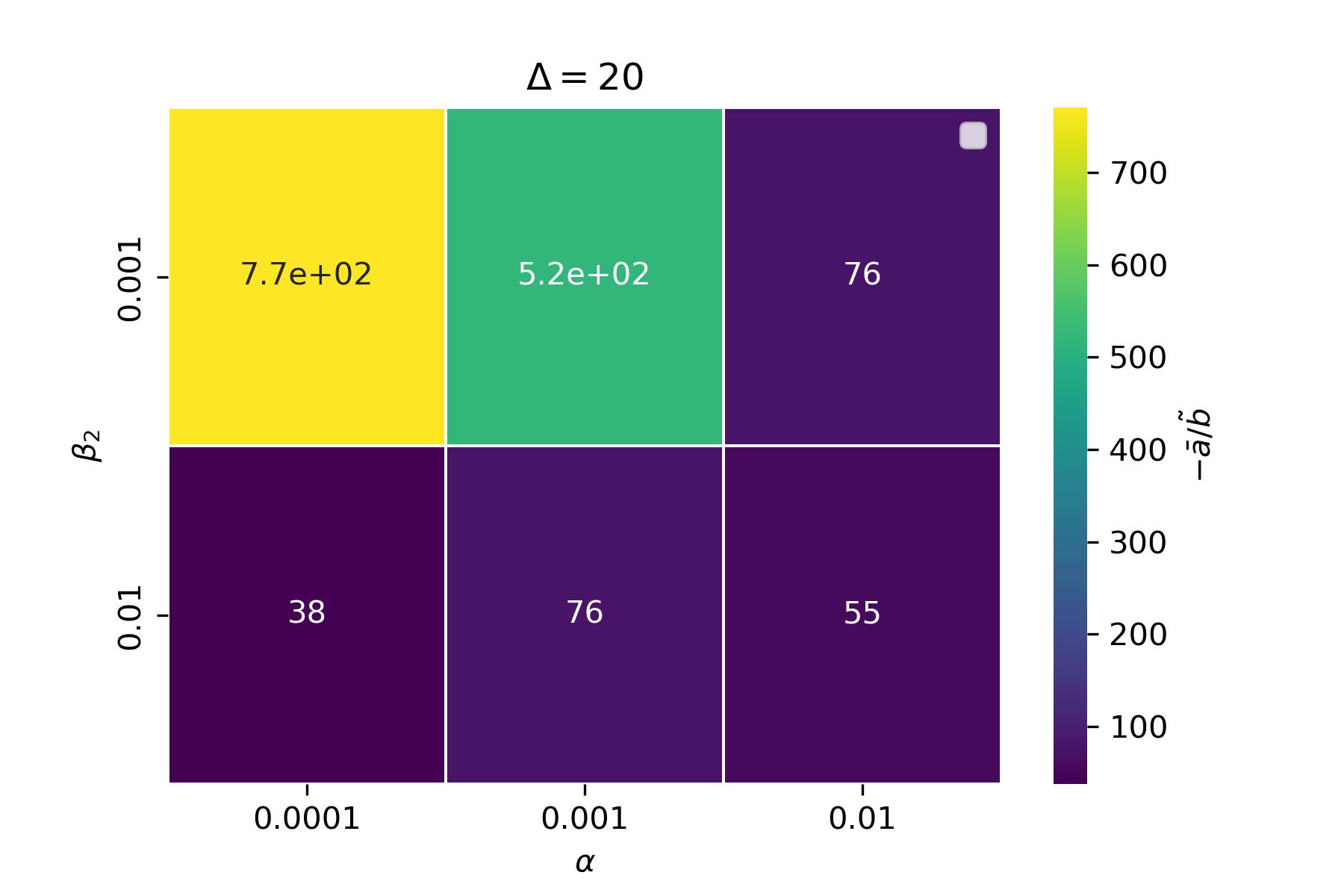}
    \includegraphics[width=0.5\linewidth]{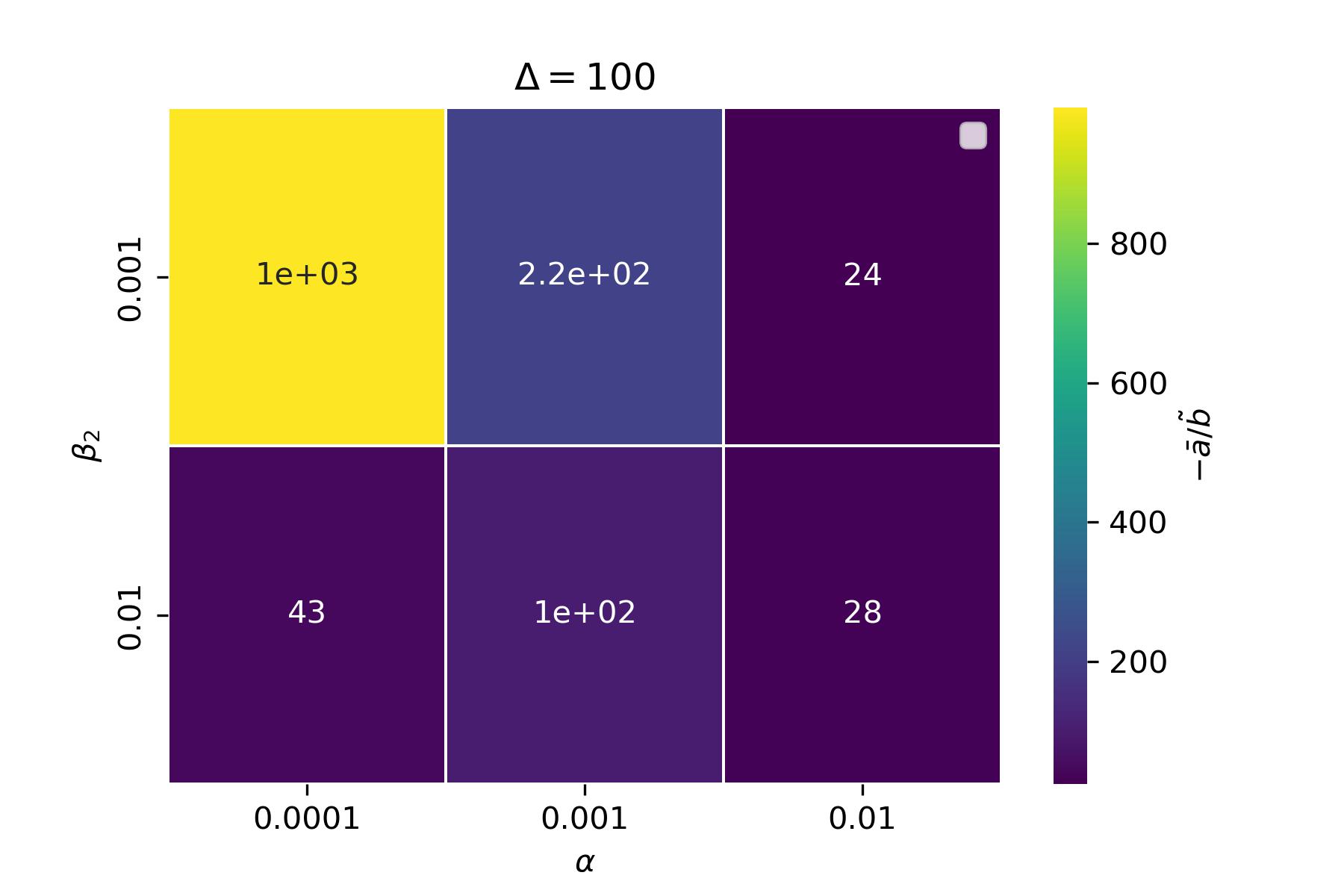}
    \includegraphics[width=0.5\linewidth]{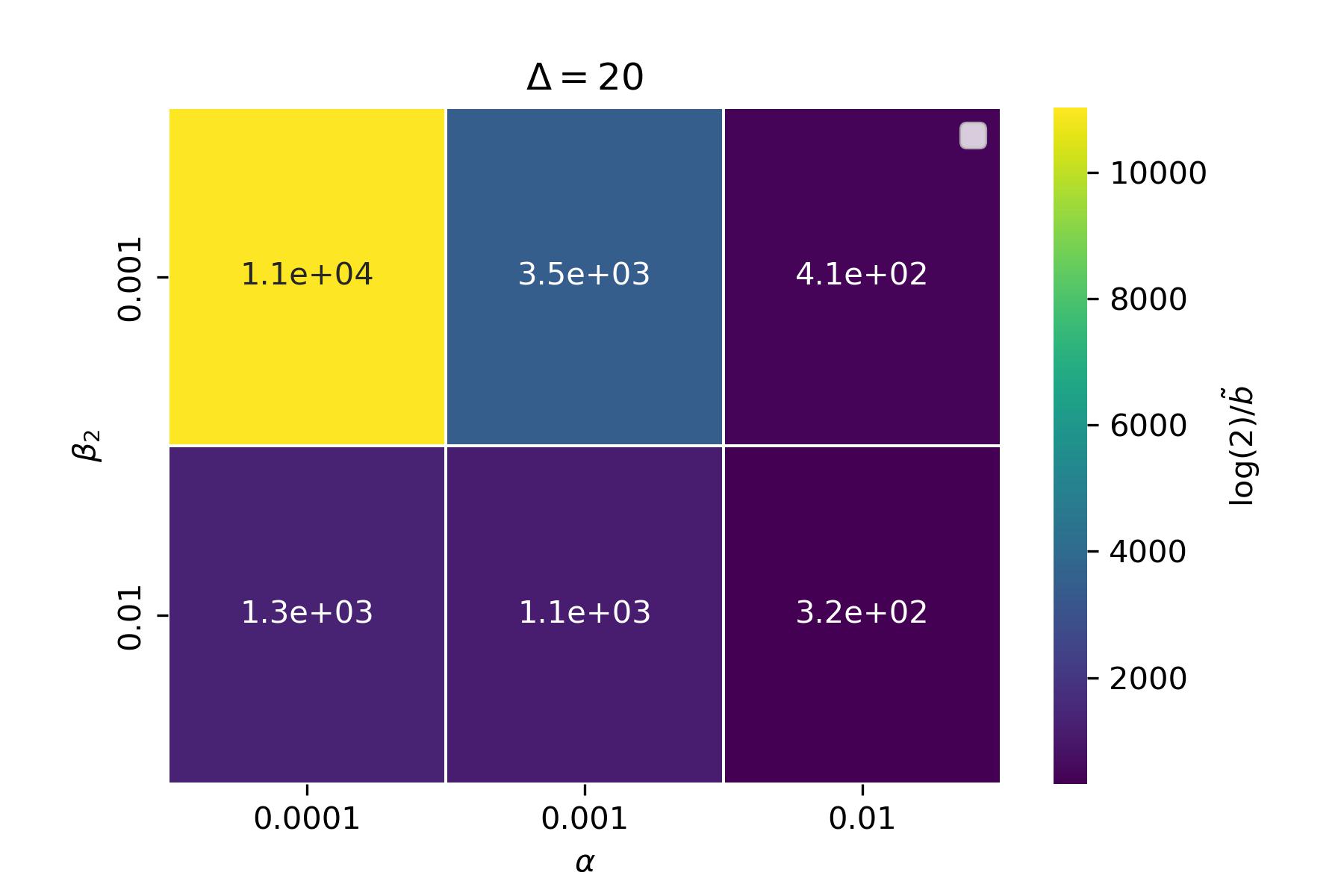}
    \includegraphics[width=0.5\linewidth]{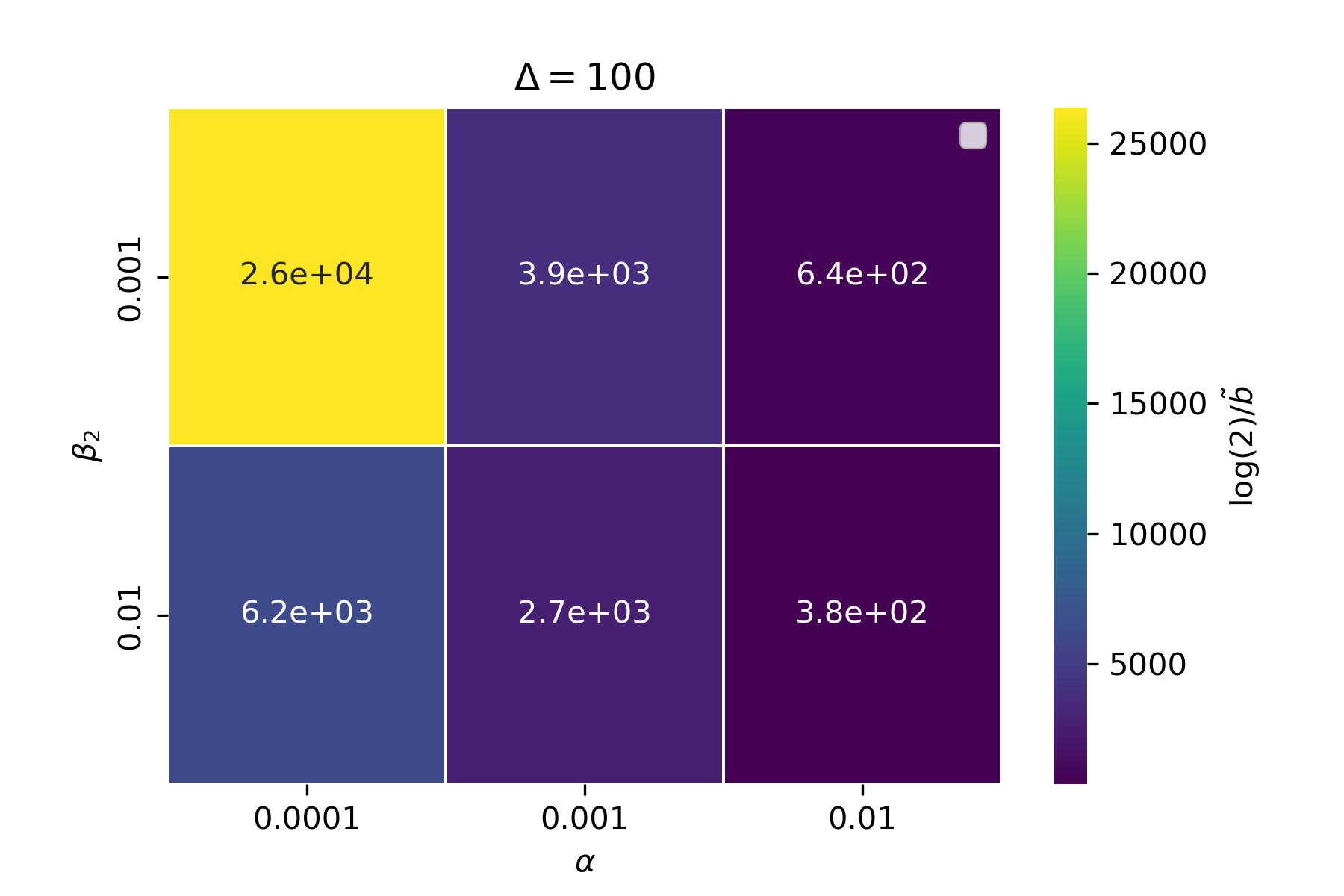}
    \caption{NMZI model. Heat maps representing $-\bar{a}/\tilde{b}$ and $\log(2)/\tilde{b}$ where $\bar{a}$ and $\tilde{b}$ are obtained by fitting $m_{t} = c - \bar{a}/\tilde{b} e^{-\tilde{b}(t - t_Q - 1)}$ to the mid-price evolution which follows the execution of a buy metaorder ending at $t_Q$.}
    \label{fig_heatmaps_after_execution}
\end{figure}

This section of the Appendix contains the Figures from \ref{fig_ewmareturns_mZI} to \ref{fig_heatmaps_after_execution}. They are related to the Non-Markovian Zero Intelligence model and are explained in the main text.

\begin{figure}
\centering
    \includegraphics[scale = 0.5]{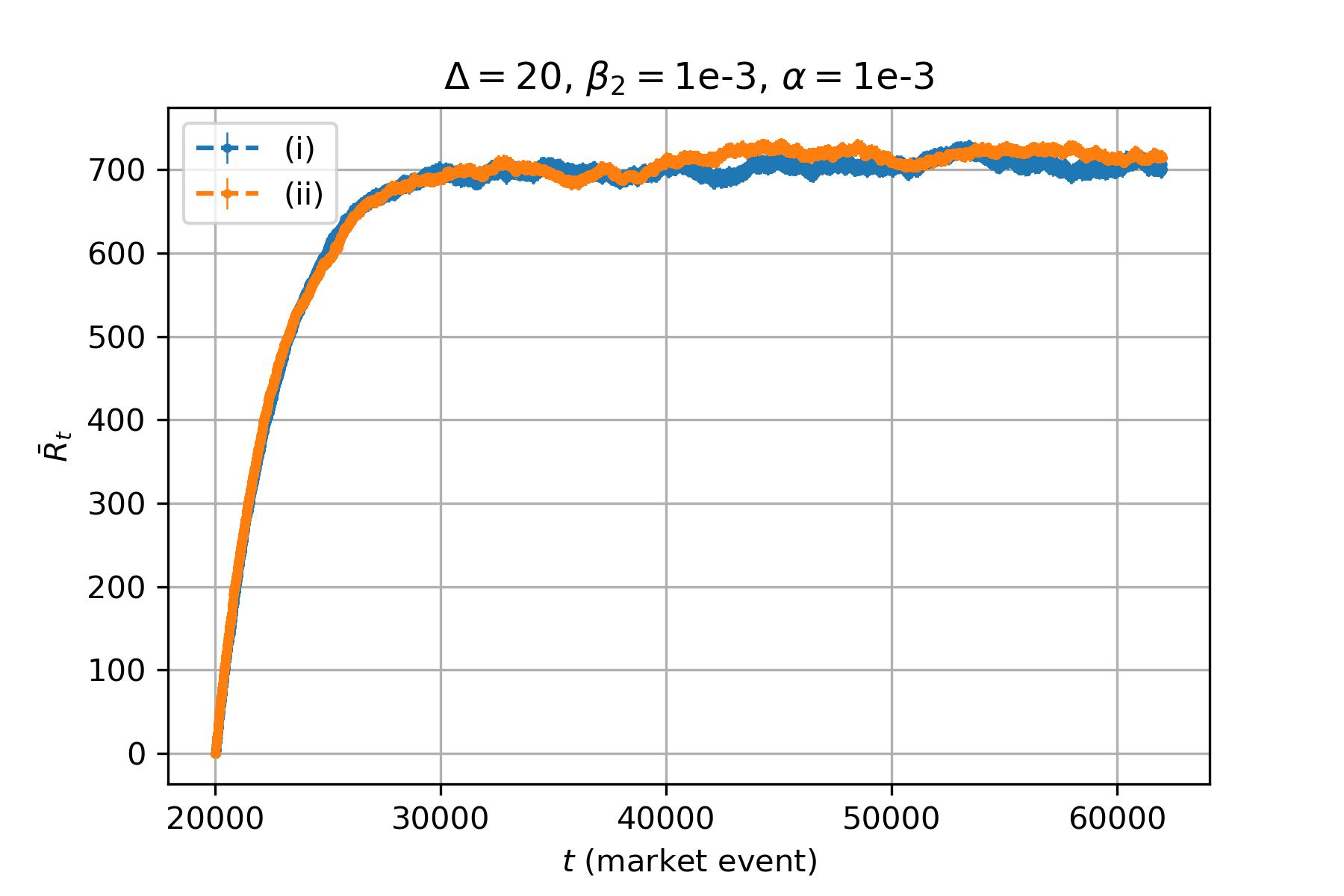}
    \includegraphics[scale = 0.5]{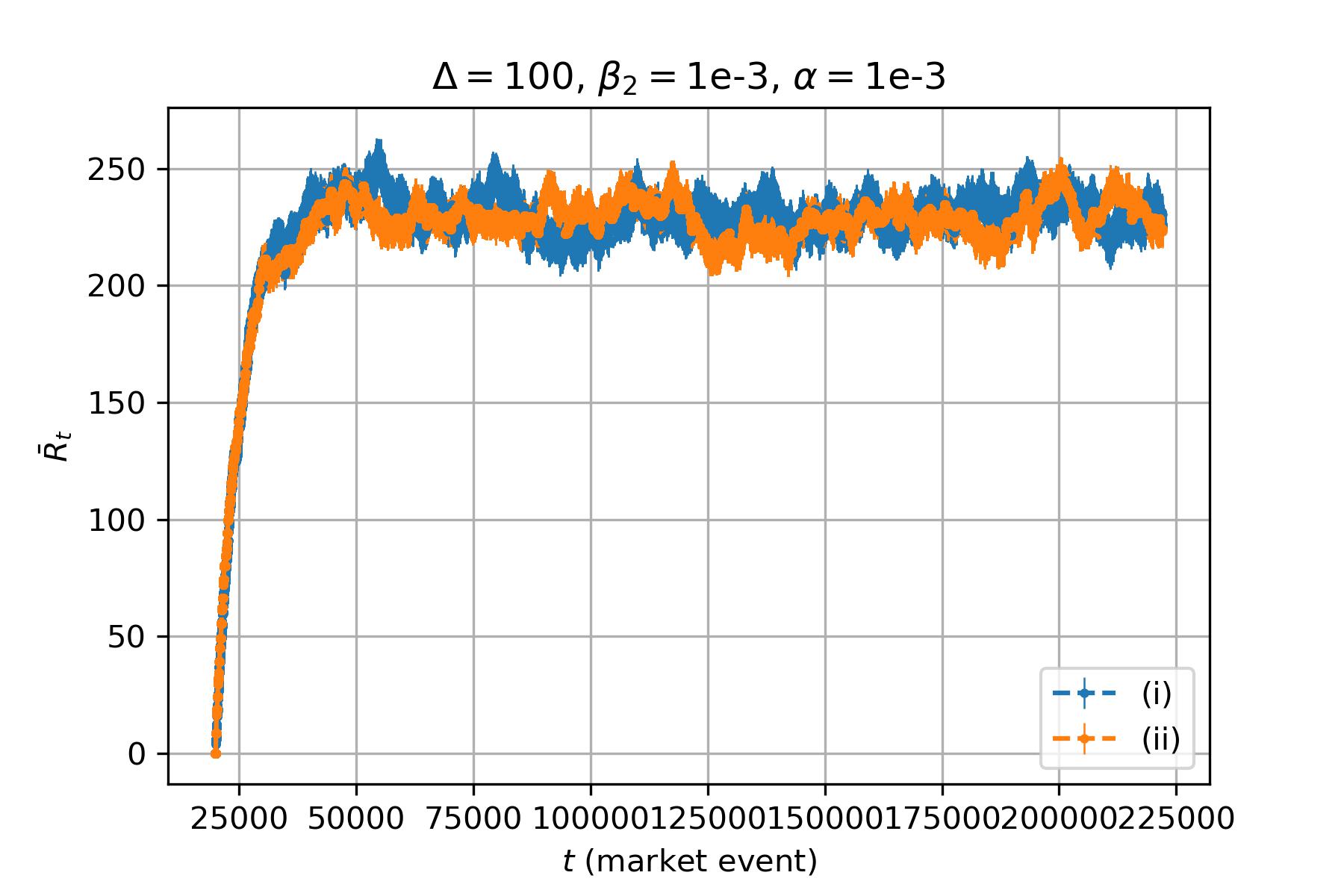}
    \includegraphics[scale = 0.5]{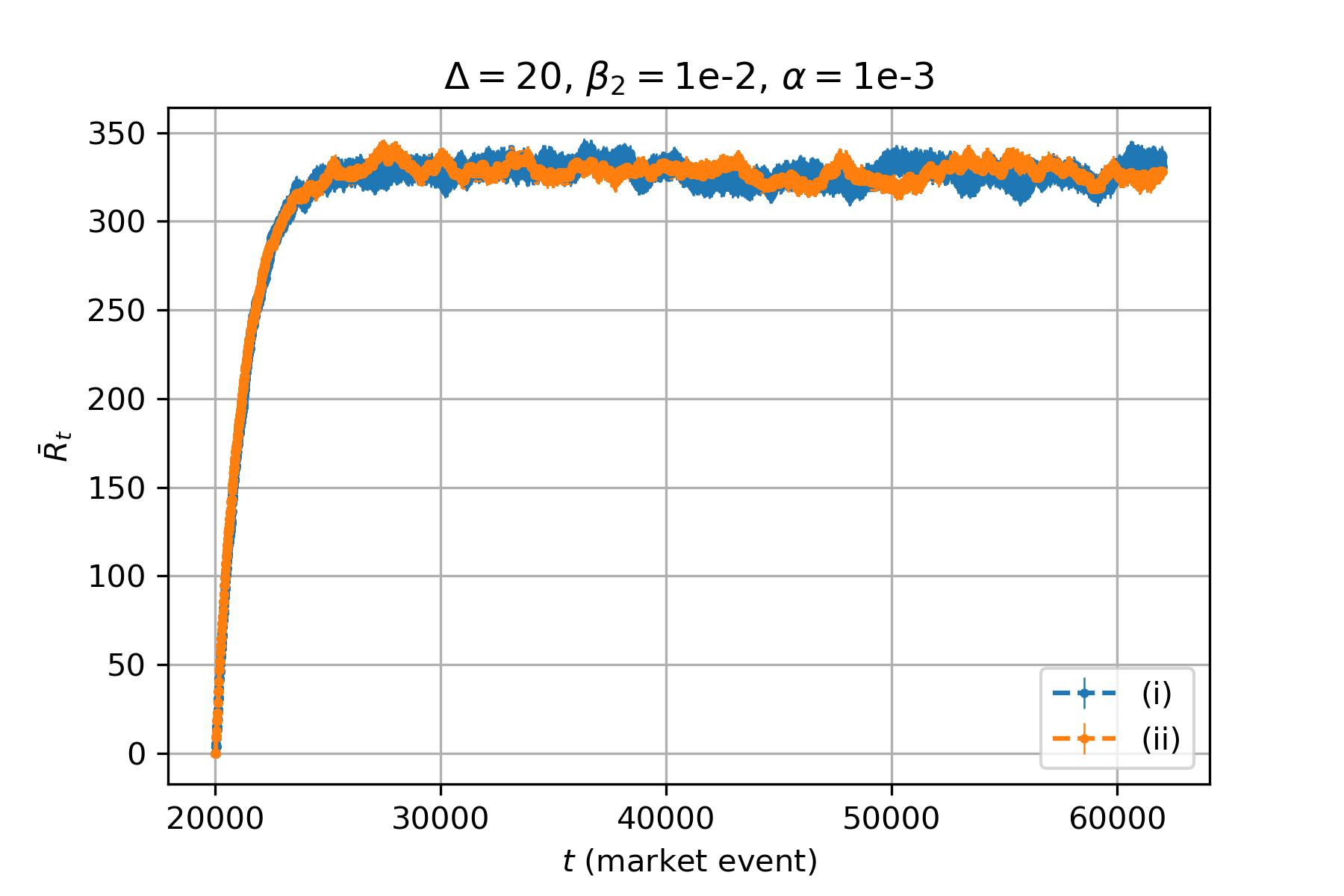}
    \includegraphics[scale = 0.5]{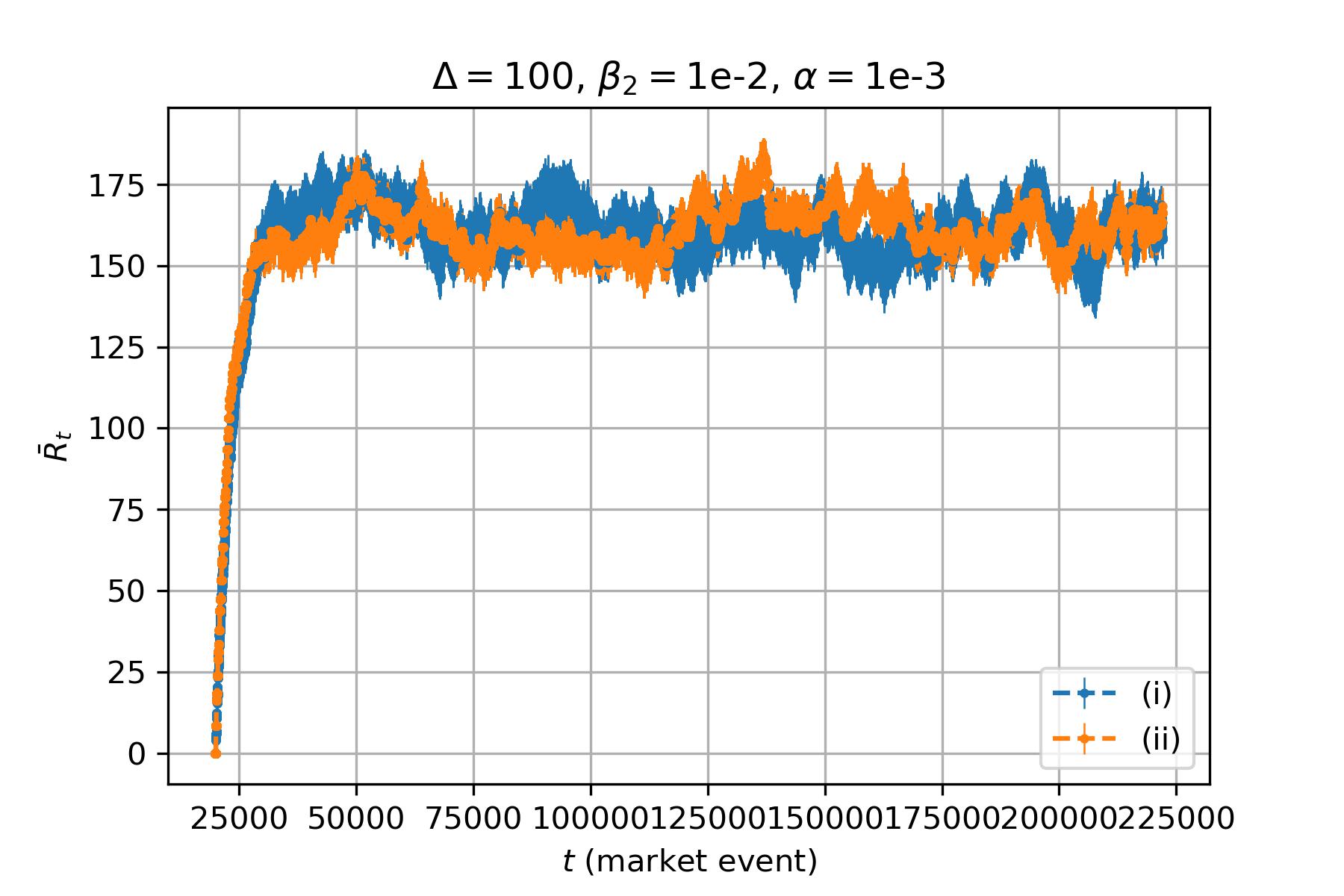}
    \caption{NMZI model. Path of the exponential weighted mid-price return $\bar{R}_t$ versus market events' times. The metaorder execution starts at time $20,000$, it has total volume $Q= 2,000$ and buy direction. In the legend, (i) and (ii) refer to the descriptions in Eq. \eqref{eq_EWMAreturns_descriptions}.}
    \label{fig_ewmareturns_mZI}
\end{figure}


\newpage

\begin{thebibliography}{99}
\bibitem{konark_review_2024}{Jain, K., Firoozye, N., Kochems, J. and Treleaven, P.. Limit Order Book Simulations: A Review, 2024. arXiv preprint: \href{https://doi.org/10.48550/arXiv.2402.17359}{
https://doi.org/10.48550/arXiv.2402.17359}.}

\bibitem{kyle1985}{Kyle, A.S.. Continuous auctions and insider trading. \textit{Econometrica: J. Econom. Soc.}, 1315–1335, 1985. \href{https://doi.org/10.2307/1913210}{https://doi.org/10.2307/1913210}.}

\bibitem{bouchaud_book_2018}{Bouchaud, J. P., Bonart, J., Donier, J. and Gould, M.. \textit{Trades, Quotes and Prices: Financial Markets Under the Microscope}, 2018 (Cambridge University Press).}

\bibitem{moro_2009}{Moro, E., Vicente, J., Moyano, L. G., Gerig, A., Farmer J. D., Vaglica, G., Lillo, F. and Mantegna, R. N.. Market impact and trading profile of hidden orders in stock markets. \textit{Phys. Rev. E}, 2009, 80(6), 066102. \href{https://doi.org/10.1103/PhysRevE.80.066102}{https://doi.org/10.1103/PhysRevE.80.066102}.}

\bibitem{brokman2015}{Brokmann, X., S\'{e}ri\'{e}, E., Kockelkoren, J. and Bouchaud, J. P.. Slow Decay of Impact in Equity Markets. \textit{Mark. Microstruct. Liquidity}, 01, 02, 1550007, 2015. \href{https://doi.org/10.1142/S2382626615500070}{https://doi.org/10.1142/S2382626615500070}.}

\bibitem{zarinelli_2015}{Zarinelli, E., Treccani, M., Farmer, J. D. and Lillo, F.. Beyond the square root: Evidence for logarithmic dependence of market impact on size and participation rate. \textit{Mark. Microstruct. Liquidity}, 1, 02, 1550004, 2015. \href{https://doi.org/10.1142/S2382626615500045}{https://doi.org/10.1142/S2382626615500045}.}

\bibitem{bucci2018}{Bucci, F., Benzaquen, M., Lillo, F. and Bouchaud, J. P.. Slow Decay of Impact in Equity Markets: Insights from the ANcerno Database. \textit{Mark. Microstruct. Liquidity}, 04, 03n04, 1950006, 2018. \href{https://doi.org/10.1142/S2382626619500060}{https://doi.org/10.1142/S2382626619500060}.}

\bibitem{daniels_2003}{Daniels, M., Farmer J., Gillemot, L., Iori G. and Smith, E.. Quantitative model of price diffusion and market friction based on trading as a mechanistic random process. \textit{Phys. Rev. Lett.}, 90:108102, 04, 2003. \href{https://doi.org/10.1103/PhysRevLett.90.108102}{https://doi.org/10.1103/PhysRevLett.90.108102}.}

\bibitem{smith_2003}{Smith, E., Farmer, J.D., Gillemot, L. and Krishnamurthy, S.. Statistical theory of the continuous double auction. \textit{Quant. Finance}, 3(6):481–514, 2003. \href{https://doi.org/10.1088/1469-7688/3/6/307}{https://doi.org/10.1088/1469-7688/3/6/307}.}

\bibitem{farmer_2005}{Farmer, J. D., Patelli, P., and Zovko, I. I. The predictive power of zero intelligence in financial markets. \textit{PNAS}, 102(6), 2254–2259, 2005. \href{https://doi.org/10.1073/pnas.0409157102}{https://doi.org/10.1073/pnas.0409157102}.}

\bibitem{bouchaud_2002}{Bouchaud, J. P., Mézard, M. and Potters, M.. Statistical properties of stock order books: empirical results and models. \textit{Quant. Finance}, 2(4), 251–256, 2002. \href{https://doi.org/10.1088/1469-7688/2/4/301}{https://doi.org/10.1088/1469-7688/2/4/301}.}

\bibitem{cont_2010}{Cont, R., Stoikov, S. and Talreja, R. A Stochastic Model for Order Book Dynamics. \textit{Oper. Res.}, 58(3), 549–563, 2010. \href{https://doi.org/10.1287/opre.1090.0780}{https://doi.org/10.1287/opre.1090.0780}.}

\bibitem{abergel_2011}{Abergel, F., Jedidi, A.. A Mathematical Approach to Order Book Modelling. In: \textit{Econophysics of Order-driven Markets. New Economic Windows.}, edited by Abergel, F., Chakrabarti, B.K., Chakraborti, A., Mitra, M., 2011. (Springer: Milano). \href{https://doi.org/10.1007/978-88-470-1766-5_7}{https://doi.org/10.1007/978-88-470-1766-5\_7}.}

\bibitem{hult_2010}{Hult, H. and Kiessling, J.. Algorithmic trading with Markov chains, 2010. Preprint: \href{https://urn.kb.se/resolve?urn=urn:nbn:se:kth:diva-25075}{https://urn.kb.se/resolve?urn=urn:nbn:se:kth:diva-25075}.}

\bibitem{huang_2013}{Huang, W., Lehalle, C.-A. and Rosenbaum, M.. Simulating and Analyzing Order Book Data: The Queue-Reactive Model. \textit{J. Am. Stat. Assoc.}, 110.509, pp. 107-122, 2015. \href{https://doi.org/10.1080/01621459.2014.982278}{https://doi.org/10.1080/01621459.2014.982278}.}

\bibitem{lu_2018b}{Lu, X. and Abergel, F.. Order-book modeling and market making strategies. \textit{Mark. Microstruct. Liquidity}, 4, 01n02, 1950003, 2018. \href{https://doi.org/10.1142/S2382626619500035}{https://doi.org/10.1142/S2382626619500035}.}

\bibitem{toke_2010}{Toke, I.M.. “Market Making” in an Order Book Model and Its Impact on the Spread. In: \textit{Econophysics of Order-driven Markets. New Economic Windows}, edited by Abergel, F., Chakrabarti, B.K., Chakraborti, A., Mitra, M., 2011 (Springer: Milano). \href{https://doi.org/10.1007/978-88-470-1766-5_4}{https://doi.org/10.1007/978-88-470-1766-5\_4}}.

\bibitem{zheng_2014}{Zheng, B., Roueff, F. and Abergel, F.. Ergodicity and scaling limit of a constrained multivariate Hawkes process, 2014. arXiv preprint: \href{https://doi.org/10.48550/arXiv.1301.5007}{https://doi.org/10.48550/arXiv.1301.5007}.}

\bibitem{bacry_2016}{Bacry, E., Jaisson, T., and Muzy, J.-F.. Estimation of
slowly decreasing hawkes kernels: application to high-frequency order book dynamics. \textit{Quant. Finance}, 16(8):1179–1201, 2016. \href{https://doi.org/10.1080/14697688.2015.1123287}{https://doi.org/10.1080/14697688.2015.1123287}.}

\bibitem{lu_2018a}{Lu, X. and Abergel, F.. High-dimensional Hawkes processes for limit order books: modelling,
empirical analysis and numerical calibration. \textit{Quant. Finance}, 18(2), 249-264, 2018. \href{https://doi.org/10.1080/14697688.2017.1403142}{https://doi.org/10.1080/14697688.2017.1403142}.}

\bibitem{mounjid_2019}{Mounjid, O., Rosenbaum, M. and Saliba, P.. From asymptotic properties of general point processes to the ranking of financial agents, 2019. arXiv preprint: \href{https://doi.org/10.48550/arXiv.1906.05420}{https://doi.org/10.48550/arXiv.1906.05420
}.}

\bibitem{lee_2022}{Lee, K. and Seo, B. K.. Modeling Bid and Ask Price Dynamics with an Extended Hawkes Process and Its Empirical Applications for High-Frequency Stock Market Data. \textit{J. Financ. Econom.}, 21(4), 1099–1142, 2023. \href{https://doi.org/10.1093/jjfinec/nbab029}{https://doi.org/10.1093/jjfinec/nbab029}.}

\bibitem{paddrik_2012}{Paddrik, M. E., Hayes, R. L., Todd, A., Yang, S. Y., Scherer, W. and Beling, P.. An Agent Based Model of the E-Mini S\&P 500 and the Flash Crash, 2012. SSRN preprint: \href{http://dx.doi.org/10.2139/ssrn.1932152}{http://dx.doi.org/10.2139/ssrn.1932152}.}

\bibitem{byrd_2019}{Byrd, D., Hybinette, M. and Hybinette Balch, T.. ABIDES: Towards High-Fidelity Market Simulation for AI Research, 2019. arXiv preprint: \href{https://doi.org/10.48550/arXiv.1904.12066}{https://doi.org/10.48550/arXiv.1904.12066}.}

\bibitem{belcak_2020}{Belcak, P., Calliess, J.P., Zohren, S.. Fast agent-based simulation framework of limit order books with applications to pro-rata markets and the study of latency effects, 2020. arXiv preprint: \href{https://doi.org/10.48550/arXiv.2008.07871}{https://doi.org/10.48550/arXiv.2008.07871}.
}

\bibitem{li_2020}{Li, J., Wang, X., Lin, Y., Sinha, A. and Wellman, M.P.. Generating Realistic Stock Market Order Streams, 2020. arXiv preprint: \href{https://doi.org/10.48550/arXiv.2006.04212}{https://doi.org/10.48550/arXiv.2006.04212}.}

\bibitem{shi_2021}{Shi, Z., Chen, Y. and Cartlidge J.. The LOB Recreation Model: Predicting the Limit Order Book from TAQ History Using an Ordinary Differential Equation Recurrent Neural Network, 2021. arXiv preprint: \href{https://doi.org/10.48550/arXiv.2103.01670}{https://doi.org/10.48550/arXiv.2103.01670}.
}

\bibitem{shi_2022}{Shi, Z. and Cartlidge, J.. The Limit Order Book Recreation Model (LOBRM): An Extended
Analysis. In: \textit{Machine Learning and Knowledge Discovery in Databases. Applied Data Science Track:
European Conference}, ECML PKDD 2021, Bilbao, Spain, September 13-17, 2021, Proc., Part IV
21. Springer, pp. 204-220. \href{https://doi.org/10.1007/978-3-030-86514-6_13}{https://doi.org/10.1007/978-3-030-86514-6\_13}.} 

\bibitem{coletta_2022}{Coletta, A., Moulin, A., Vyetrenko, S. and Balch, T.. Learning to simulate realistic limit order book markets from data as a World Agent. In \textit{Proc. Third ACM Int. Conf. AI in Finance}, pp. 428-436, 2022. \href{https://doi.org/10.1145/3533271.3561753}{https://doi.org/10.1145/3533271.3561753}.}

\bibitem{cont_2023a}{Cont, R., Cucuringu, M., Kochems, J. and Prenzel, F.. Limit Order Book Simulation with Generative Adversarial Networks, 2023. SSRN preprint: \href{http://dx.doi.org/10.2139/ssrn.4512356}{http://dx.doi.org/10.2139/ssrn.4512356}.}

\bibitem{nagy_2023}{Nagy, P., Frey, S., Sapora, S., Li, K., Calinescu, A., Zohren, S., Foerster, J.. Generative AI for End-to-End Limit Order Book Modelling: A Token-Level Autoregressive Generative Model of Message Flow Using a Deep State Space Network. In \textit{Proc. Fourth ACM Int. Conf. AI in Finance}, pp. 91–99. \href{https://doi.org/10.1145/3604237.3626898}{https://doi.org/10.1145/3604237.3626898}.}

\bibitem{kumar_2024}{Kumar, P.. Deep Hawkes process for high-frequency market making. \textit{J. Banking and Financ. Technol.}, 2024, 8, 11–28. \href{https://doi.org/10.1007/s42786-024-00049-8}{https://doi.org/10.1007/s42786-024-00049-8}.}

\bibitem{lakner_2016}{Lakner, P., Reed, J. and Stoikov, S.. High frequency asymptotics for the limit order book. \textit{Mark. Microstruct. Liquidity}, 2016, 2.01, p. 1650004. \href{https://doi.org/10.1142/S2382626616500040}{https://doi.org/10.1142/S2382626616500040}.}

\bibitem{huang_2017}{Huang, W. and Rosenbaum, M.. Ergodicity and dffusivity of Markovian order book models: a general framework. \textit{SIAM J. Financ. Math}, 2017, 8.1, pp. 874-900. \href{https://doi.org/10.1137/16M1064337}{https://doi.org/10.1137/16M1064337}.}

\bibitem{hambly_2020}{Hambly, B., Kalsi, J. and Newbury, N.. Limit order books, diffusion approximations and refected SPDEs: from microscopic to macroscopic models. \textit{Appl. Math. Finance}, 2020, 27(1–2), 132–170. \href{https://doi.org/10.1080/1350486X.2020.1758176}{https://doi.org/10.1080/1350486X.2020.1758176}.}

\bibitem{cont_degond_2023}{Cont, R., Degond, P. and Xuan L.. A mathematical framework for modelling order book dynamics. \textit{SIAM J. Financ. Math}, 2025, 16.1, pp. 123-166. \href{https://doi.org/10.1137/22M1541538}{https://doi.org/10.1137/22M1541538}.}

\bibitem{donier_2015}{Donier, J., Bonart, J., Mastromatteo, I. and Bouchaud, J. P.. A fully consistent, minimal model for non-linear market impact. \textit{Quant. Finance}, 2015, 15:7, 1109-1121. \href{https://doi.org/10.1080/14697688.2015.1040056}{https://doi.org/10.1080/14697688.2015.1040056}.}

\bibitem{fosset}{Fosset, A., Bouchaud, J. P. and Benzaquen, M.. Endogenous liquidity crises. {\it J. Stat. Mech: Theory Exp.}, 2020. \href{https://doi.org/10.1088/1742-5468/ab7c64}{https://doi.org/10.1088/1742-5468/ab7c64}.}

\bibitem{toth_2011}{Tóth, B., Lemperiere, Y., Deremble, C., De Lataillade, J., Kockelkoren, J. and Bouchaud, J. P..
Anomalous price impact and the critical nature of liquidity in financial markets. \textit{Phys. Rev. X}, 2011, 1(2), 021006. \href{https://doi.org/10.1103/PhysRevX.1.021006}{https://doi.org/10.1103/PhysRevX.1.021006}.}

\bibitem{chan1988}{Chan, K. C.. On the Contrarian Investment Strategy. \textit{J. Bus.}, 1988, 61(2), 147–63. \href{https://doi.org/10.1086/296425}{https://doi.org/10.1086/296425}.}

\bibitem{lakonishok1994}{Lakonishok, J., Shleifer, A. and Vishny, R.W.. Contrarian Investment, Extrapolation, and Risk. \textit{J. Finance}, 1994, 49, 1541-1578. \href{https://doi.org/10.1111/j.1540-6261.1994.tb04772.x}{https://doi.org/10.1111/j.1540-6261.1994.tb04772.x}.}

\bibitem{odean1998}{Odean, T.. Are investors reluctant to realize their losses? \textit{J. Finance}, 1998, 53 (5), 1775–1798. \href{https://doi.org/10.1111/0022-1082.00072}{https://doi.org/10.1111/0022-1082.00072}.}

\bibitem{grinblatt2000}{Grinblatt, M. and Keloharju, M.. The investment behavior and performance of various investor types: a study of Finland’s unique data set. \textit{J. Financ. Econ.}, 2000, 55 (1), 43–67. \href{https://doi.org/10.1016/S0304-405X(99)00044-6}{https://doi.org/10.1016/S0304-405X(99)00044-6}.}

\bibitem{badrinath2002}{Badrinath, S.G. and Wahal, S.. Momentum trading by institutions. \textit{J. Finance}, 2002, 57 (6), 2449–2478. \href{ https://doi.org/10.1111/1540-6261.00502}{ https://doi.org/10.1111/1540-6261.00502}.}

\bibitem{dehaan2011}{De Haan, L. and Kakes, J.. Momentum or contrarian investment strategies: Evidence from Dutch institutional investors. \textit{J. Banking \& Finance}, 2011, 35(9), 2245-2251. \href{https://doi.org/10.1016/j.jbankfin.2011.01.027}{https://doi.org/10.1016/j.jbankfin.2011.01.027}.}

\bibitem{che2018}{Che, L.. Investor types and stock return volatility. \textit{J. Empirical Finance}, 2018, 47 (1), 139–161. \href{https://doi.org/10.1016/j.jempfin.2018.03.005}{https://doi.org/10.1016/j.jempfin.2018.03.005}.}

\bibitem{baltzer2019}{Baltzer, M., Jank, S. and Smajlbegovic, E.. Who trades on momentum? \textit{J. Financ. Mark.}, 2019, 42 (1), 56–74. \href{https://doi.org/10.1016/j.finmar.2018.08.003}{https://doi.org/10.1016/j.finmar.2018.08.003}.}

\bibitem{grinblatt2020}{Grinblatt, M., Jostova, G., Petrasek, L. and Philipov, A.. Style and skill: hedge funds, mutual funds, and momentum. \textit{Manage. Sci.}, 2020, 66 (12), 5505–5531. \href{https://doi.org/10.1287/mnsc.2019.3433}{https://doi.org/10.1287/mnsc.2019.3433}.}

\bibitem{bradrania2023}{Bradrania, R. and Wu, W..
Foreign institutions, local investors and momentum trading. \textit{J. Empirical Finance}, 2023, 73, 40-64. \href{https://doi.org/10.1016/j.jempfin.2023.05.005}{https://doi.org/10.1016/j.jempfin.2023.05.005}.}

\end{thebibliography}
\end{document}